\documentclass[11pt,a4paper]{article}
\usepackage{graphicx}
\usepackage{mathptmx}
\usepackage{amssymb}
\usepackage{bm}
\usepackage{times}
 at 6truept 
\font\tmrms=ptmr at 6truept 
\input prepictex.tex
\input pictex.tex
\input postpictex.tex
\def\psidot{\stackrel{\hbox{\small\  .}}{\psi\hspace{0pt}}}

\def\fnote#1#2{\begingroup\def\thefootnote{#1}\footnote{#2}\addtocounter
{footnote}{-1}\endgroup}

\begin{document}
\baselineskip=12pt

\begin{center}
\begin{large}
{\bf Primordial non-Gaussianities from inflation}
\end{large}
\end{center}

\begin{center}
Hael Collins\fnote{$\dagger$}{Electronic address:  hcollins@andrew.cmu.edu} 
\vskip4truept
{\it Department of Physics}
\vskip1truept

{\it Carnegie Mellon University, Pittsburgh, Pennsylvania\/}
\vskip1truept
{\small (Dated:  29 May 2014)}
\end{center}
\vskip12truept

\begin{abstract}
These notes present a detailed introduction to Maldacena's calculation [1] of the three-point function generated by the simplest class of inflationary models:  those with a single inflaton field whose potential satisfies the slow-roll conditions and whose quantum fluctuations start in the asymptotic Bunch-Davies vacuum state.  The three-point function should be the most readily observed evidence for non-Gaussianities amongst the primordial fluctuations produced by inflation.  In these inflationary theories the non-Gaussianities are predicted to be extremely small, being naturally suppressed by the small slow-roll parameters.
\end{abstract}
\vskip24truept

\noindent
{\sc Any\/} cosmological model that seeks a deeper or more complete explanation of the universe must be able to account for the origin of the great variety of structures that appear in it today.  Most of this structural complexity in the universe appears to have arisen through processes which are largely understood from material that was once in a much simpler state than it is today.  Stars and galaxies, for example, which now stand out so clearly from their much emptier surroundings, could have grown through the slow gravitational collapse of gases which were initially distributed far more uniformly.  This basic picture suggests that the universe in its primeval stages was once extremely homogeneous, a prediction that matches well with the observations made so far.  However, this mechanism also requires that, even at the very earliest times, some spatial variation must always have existed.  In a perfectly homogeneous and isotropic universe, the process of collapse and growth, and the development of ever more complex structures, would never have begun.

The theory of inflation provides a mechanism for generating the initial spatial variations in the universe.  In the inflationary picture, space-time itself fluctuates quantum mechanically about a background that is expanding at an accelerating rate.  This extreme expansion spreads the fluctuations, which begin with a tiny spatial extent, throughout a vast region of the universe, where they eventually become small classical fluctuations in the space-time curvature---or equivalently, small spatial variations in the strength of gravity.  Since everything in the universe feels the influence of gravity, these fluctuations in the gravitational field are transferred to the matter and radiation fields, creating slightly overdense and underdense regions.  The resulting matter fluctuations then become the `initial conditions' that start the process of collapse which forms the stars and galaxies of later epochs.

To test whether this picture is correct, it is necessary to describe very accurately the properties of the pattern generated by inflation for the original, primordial fluctuations in space-time which can then be compared with what is inferred from observations.  Before entering into a detailed calculation of these primordial perturbations, it is instructive to explain first from a more general perspective how the fluctuations of a quantum field lead to a classical pattern of perturbations.  Since a quantum field is never measured directly, it is necessary also to determine how and in what form its information is available.

So to begin, consider a quantum field, which will be written as $\zeta(t,\vec x)$.  Later this field will be connected with the fluctuations of the background space-time, but for now its exact physical meaning will be left a little vague.  Although $\zeta(t,\vec x)$ is a quantum field, its influence is inferred by how it affects classical things, so the information contained in $\zeta(t,\vec x)$ needs to be converted into a classical function, or rather, into a set of classical functions.  One way to do so is by taking the expectation values of products of the field $\zeta(t,\vec x)$ at different places.  If the field is in a particular, time-dependent, quantum state, $|0(t)\rangle$, then the functions, 
$$
\langle 0(t) | \zeta(t,\vec x_1) \zeta(t,\vec x_2) \cdots \zeta(t,\vec x_n) 
|0(t)\rangle ,
$$
tell the extent to which fluctuations in different places are correlated with each other in that state, and they are accordingly called {\it $n$-point correlation functions\/}.  Usually, most of the emphasis on deriving the inflationary prediction for the initial fluctuations focuses on the field, but it is equally important to understand the state too.  Because of the dramatic expansion during inflation, the detailed assumptions made about the properties of this state---even at seemingly infinitesimal distances---can have an influence on the predictions for the pattern of the fluctuations.

Each $n$-point function potentially contains unique information that is not found in any of the others.  If the fluctuations in the very early stages of the universe are small compared with the value of the background, then it should be possible to evaluate these correlation functions perturbatively.  This fact implies further that the higher order $n$-point functions will be progressively more suppressed by this inherent smallness of $\zeta(t,\vec x)$.  So given a limited experimental accuracy, only the lowest few correlators will be observable in practice.  If the spatially independent background has been chosen correctly, then the one-point function vanishes, 
$$
\langle 0(t) | \zeta(t,\vec x) |0(t)\rangle = 0 ,
$$
meaning that the first observable measure of the primordial fluctuations is provided by the two-point function, 
$$
\langle 0(t) | \zeta(t,\vec x) \zeta(t,\vec y) |0(t)\rangle ,
$$
followed next by the three-point function, 
$$
\langle 0(t) | \zeta(t,\vec x) \zeta(t,\vec y) \zeta(t,\vec z) |0(t)\rangle , 
$$
and so on.

Calculating these correlation functions requires knowing something about the dynamics governing the field $\zeta(t,\vec x)$.  Suppose that they are determined by an action $S$ with an associated Lagrange density ${\cal L}[\zeta]$, 
$$
S = \int d^4x\, {\cal L}[\zeta(x)] .
$$
Since $\zeta$ is small, the Lagrangian can be expanded as a series in powers of the field,
$$
S = \int d^4x\, \bigl\{ 
{\cal L}_0 + {\cal L}^{(2)}[\zeta(x)] + {\cal L}^{(3)}[\zeta(x)] + {\cal L}^{(4)}[\zeta(x)] + \cdots 
\bigr\} , 
$$
where ${\cal L}^{(2)}$ is quadratic in the field, ${\cal L}^{(3)}$ is cubic, {\it etc\/}.  The zeroth piece fixes the time evolution of the background; and since this background is always implicitly a solution to the equations of motion, the ${\cal L}^{(1)}$ term vanishes.  Associating a vertex with each of these terms, with one leg for each factor of the field, 
$$
\beginpicture
\setcoordinatesystem units <1.00truept,1.00truept>
\setplotarea x from -72 to 32, y from -17 to 24
\circulararc 360 degrees from 8 0 center at 0 0
\plot  8 0   32 0 /
\plot -8 0  -32 0 /
\put {${\cal L}^{(2)}\sim$} [r] at -38 0
\setshadesymbol ({\tmrms .})
\setshadegrid span <0.9pt>
\setquadratic
\hshade -5.65 -5.65 -5.65 <,z,,>  0 -8.00 -5.65   5.65 -5.65 -5.65 /
\vshade -5.65 -5.65  5.65 <z,z,,> 0 -8.00  8.00   5.65 -5.65  5.65 /
\hshade -5.65  5.65  5.65 <z,,,>  0  5.65  8.00   5.65  5.65  5.65 /
\endpicture
\ , \quad
\beginpicture
\setcoordinatesystem units <1.00truept,1.00truept>
\setplotarea x from -60 to 21, y from -17 to 24
\circulararc 360 degrees from 8 0 center at 0 0
\plot  6.93 -4   20.78 -12 /
\plot -6.93 -4  -20.78 -12 /
\plot 0 8  0 24 /
\put {${\cal L}^{(3)}\sim$} [r] at -27 0
\setshadesymbol ({\tmrms .})
\setshadegrid span <0.9pt>
\setquadratic
\hshade -5.65 -5.65 -5.65 <,z,,>  0 -8.00 -5.65   5.65 -5.65 -5.65 /
\vshade -5.65 -5.65  5.65 <z,z,,> 0 -8.00  8.00   5.65 -5.65  5.65 /
\hshade -5.65  5.65  5.65 <z,,,>  0  5.65  8.00   5.65  5.65  5.65 /
\endpicture
\ , \quad
\beginpicture
\setcoordinatesystem units <1.00truept,1.00truept>
\setplotarea x from -60 to 21, y from -17 to 24
\circulararc 360 degrees from 8 0 center at 0 0
\plot  5.65  5.65   16.97  16.97 /
\plot  5.65 -5.65   16.97 -16.97 /
\plot -5.65 -5.65  -16.97 -16.97 /
\plot -5.65  5.65  -16.97  16.97 /
\put {${\cal L}^{(4)}\sim$} [r] at -27 0
\setshadesymbol ({\tmrms .})
\setshadegrid span <0.9pt>
\setquadratic
\hshade -5.65 -5.65 -5.65 <,z,,>  0 -8.00 -5.65   5.65 -5.65 -5.65 /
\vshade -5.65 -5.65  5.65 <z,z,,> 0 -8.00  8.00   5.65 -5.65  5.65 /
\hshade -5.65  5.65  5.65 <z,,,>  0  5.65  8.00   5.65  5.65  5.65 /
\endpicture
\ , \quad \hbox{etc.,}
$$
the first few contributions to the two-point function, in a perturbative expansion, can be diagrammatically represented by
$$
\beginpicture
\setcoordinatesystem units <1.00truept,1.00truept>
\setplotarea x from -144 to 144, y from -8 to 24
\circulararc 360 degrees from 8 0 center at 0 0
\plot  8 0   32 0 /
\plot -8 0  -32 0 /
\put {$+$} [c] at 44 0
\circulararc 360 degrees from 96 0 center at 88 0
\circulararc 282.11 degrees from 95.54 2.67 center at 88 12
\plot 56 0  80 0 /
\plot 96 0  120 0 /
\put {$+$} [c] at 132 0
\put {$\cdots$} [l] at 144 0
\put {$\langle 0(t)|\zeta(x)\zeta(y) |0(t)\rangle$} [r] at -56 0
\put {$=$} [c] at -44 0
\put {$x$} [c] at -32  -9
\put {$y$} [c] at  32 -10
\put {$x$} [c] at  56  -9
\put {$y$} [c] at 120 -10
\setshadesymbol ({\tmrms .})
\setshadegrid span <0.9pt>
\setquadratic
\hshade -5.65 -5.65 -5.65 <,z,,>  0 -8.00 -5.65   5.65 -5.65 -5.65 /
\vshade -5.65 -5.65  5.65 <z,z,,> 0 -8.00  8.00   5.65 -5.65  5.65 /
\hshade -5.65  5.65  5.65 <z,,,>  0  5.65  8.00   5.65  5.65  5.65 /
\hshade -5.65 82.35 82.35 <,z,,>  0 80.00 82.35   5.65 82.35 82.35 /
\vshade 82.35 -5.65  5.65 <z,z,,> 88 -8.00  8.00   93.65 -5.65  5.65 /
\hshade -5.65 93.65 93.65 <z,,,>  0  93.65 96.00   5.65  93.65 93.65 /
\endpicture
$$
while the three-point function is 
$$
\beginpicture
\setcoordinatesystem units <1.00truept,1.00truept>
\setplotarea x from -144 to 144, y from -8 to 24
\circulararc 360 degrees from 8 0 center at 0 0
\plot  6.93 -4   20.78 -12 /
\plot -6.93 -4  -20.78 -12 /
\plot 0 8  0 24 /
\put {$+$} [c] at 44 0
\put {$\cdots$} [l] at 56 0
\put {$\langle 0(t)|\zeta(x)\zeta(y)\zeta(z)|0(t)\rangle$} [r] at -56 0
\put {$=$} [c] at -44 0
\put {$x$} [c] at -26 -17
\put {$y$} [c] at  26 -18
\put {$z$} [c] at   0  30
\setshadesymbol ({\tmrms .})
\setshadegrid span <0.9pt>
\setquadratic
\hshade -5.65 -5.65 -5.65 <,z,,>  0 -8.00 -5.65   5.65 -5.65 -5.65 /
\vshade -5.65 -5.65  5.65 <z,z,,> 0 -8.00  8.00   5.65 -5.65  5.65 /
\hshade -5.65  5.65  5.65 <z,,,>  0  5.65  8.00   5.65  5.65  5.65 /
\endpicture
$$
plus further higher order corrections which have not been written explicitly.  In these expressions, the locations of the individual fields have been abbreviated as $x = (t,\vec x)$, $y = (t,\vec y)$ and $z = (t,\vec z)$.  Beyond cubic order, the leading contributions to the correlation functions are made up in part by powers of the lower order correlators.  At quartic order, for instance, the leading behaviour receives contributions both from the three possible pairings of two-point functions as well as from the connected graph associated with the ${\cal L}^{(4)}$ vertex, 
$$
\beginpicture
\setcoordinatesystem units <1.00truept,1.00truept>
\setplotarea x from -18 to 18, y from -18 to 18
\circulararc 360 degrees from 6  12 center at 0  12
\circulararc 360 degrees from 6 -12 center at 0 -12
\plot  6  12   18  12 /
\plot -6  12  -18  12 /
\plot  6 -12   18 -12 /
\plot -6 -12  -18 -12 /
\put {$x_1$} [r] at -21  10
\put {$x_2$} [l] at  21  10
\put {$x_3$} [r] at -21 -14
\put {$x_4$} [l] at  21 -14
\setshadesymbol ({\tmrms .})
\setshadegrid span <0.9pt>
\setquadratic
\hshade  7.76 -4.24  -4.24 <,z,,>  12 -6.00 -4.24   16.24 -4.24 -4.24 /
\vshade -4.24  7.76  16.24 <z,z,,>  0  6.00 18.00    4.24  7.76 16.24 /
\hshade  7.76  4.24   4.24 <z,,,>  12  4.24  6.00   16.24  4.24  4.24 /
\hshade -16.24 -4.24  -4.24 <,z,,>  -12 -6.00  -4.24   -7.76 -4.24  -4.24 /
\vshade -4.24 -16.24 -7.76 <z,z,,>  0 -18.00 -6.00    4.24 -16.24 -7.76 /
\hshade -16.24  4.24   4.24 <z,,,>  -12  4.24   6.00   -7.76  4.24   4.24 /
\endpicture
\quad + \quad
\beginpicture
\setcoordinatesystem units <1.00truept,1.00truept>
\setplotarea x from -18 to 18, y from -18 to 18
\circulararc 360 degrees from  18 0 center at  12 0
\circulararc 360 degrees from -18 0 center at -12 0
\plot  12  6   12  18 /
\plot  12 -6   12 -18 /
\plot -12  6  -12  18 /
\plot -12 -6  -12 -18 /
\put {$x_1$} [c] at -12  24
\put {$x_2$} [c] at  12  24
\put {$x_3$} [c] at -12 -25
\put {$x_4$} [c] at  12 -25
\setshadesymbol ({\tmrms .})
\setshadegrid span <0.9pt>
\setquadratic
\hshade -4.24  7.76  7.76 <,z,,>   0  6.00  7.76    4.24  7.76  7.76 /
\vshade  7.76 -4.24  4.24 <z,z,,> 12 -6.00  6.00   16.24 -4.24  4.24 /
\hshade -4.24 16.24 16.24 <z,,,>   0 16.24 18.00    4.24 16.24 16.24 /
\hshade  -4.24 -16.24 -16.24 <,z,,>    0 -18.00 -16.24    4.24 -16.24 -16.24 /
\vshade -16.24  -4.24   4.24 <z,z,,> -12  -6.00   6.00   -7.76  -4.24   4.24 /
\hshade  -4.24  -7.76  -7.76 <z,,,>    0  -7.76  -6.00    4.24  -7.76  -7.76 /
\endpicture
\quad + \quad
\beginpicture
\setcoordinatesystem units <1.00truept,1.00truept>
\setplotarea x from -18 to 18, y from -18 to 18
\circulararc 360 degrees from -3  9 center at -9  9
\circulararc 360 degrees from -3 -9 center at -9 -9
\plot   1.5    1.5    18    18 /
\plot  -1.5   -1.5 -4.76 -4.76 /
\plot -13.24  13.24  -18    18 /
\plot -13.24 -13.24  -18   -18 /
\plot  -4.76   4.76   18   -18 /
\put {$x_1$} [c] at -24  24
\put {$x_2$} [c] at  24  24
\put {$x_3$} [c] at -24 -24
\put {$x_4$} [c] at  24 -24
\setshadesymbol ({\tmrms .})
\setshadegrid span <0.9pt>
\setquadratic
\hshade   4.76 -13.24 -13.24 <,z,,>   9 -15.00 -13.24   13.24 -13.24 -13.24 /
\vshade -13.24   4.76  13.24 <z,z,,> -9   3.00  15.00   -4.76   4.76  13.24 /
\hshade   4.76  -4.76  -4.76 <z,,,>   9  -4.76  -3.00   13.24  -4.76  -4.76 /
\hshade -13.24 -13.24 -13.24 <,z,,>  -9 -15.00 -13.24   -4.76 -13.24 -13.24 /
\vshade -13.24 -13.24  -4.76 <z,z,,> -9 -15.00  -3.00   -4.76 -13.24  -4.76 /
\hshade -13.24  -4.76  -4.76 <z,,,>  -9  -4.76  -3.00   -4.76  -4.76  -4.76 /
\endpicture
\quad + \quad
\beginpicture
\setcoordinatesystem units <1.00truept,1.00truept>
\setplotarea x from -18 to 18, y from -18 to 18
\circulararc 360 degrees from 8 0 center at 0 0
\plot  5.65  5.65   18  18 /
\plot  5.65 -5.65   18 -18 /
\plot -5.65 -5.65  -18 -18 /
\plot -5.65  5.65  -18  18 /
\put {$x_1$} [c] at -24  24
\put {$x_2$} [c] at  24  24
\put {$x_3$} [c] at -24 -24
\put {$x_4$} [c] at  24 -24
\setshadesymbol ({\tmrms .})
\setshadegrid span <0.9pt>
\setquadratic
\hshade -5.65 -5.65 -5.65 <,z,,>  0 -8.00 -5.65   5.65 -5.65 -5.65 /
\vshade -5.65 -5.65  5.65 <z,z,,> 0 -8.00  8.00   5.65 -5.65  5.65 /
\hshade -5.65  5.65  5.65 <z,,,>  0  5.65  8.00   5.65  5.65  5.65 /
\endpicture
\quad + \quad\cdots
$$
which is equivalently written as 
\begin{eqnarray}
\langle 0(t) | \zeta(x_1) \zeta(x_2) \zeta(x_3) \zeta(x_4) |0(t)\rangle 
&\!\!\!\!=\!\!\!\!& 
\langle 0| \zeta(x_1) \zeta(x_2) |0\rangle\ 
\langle 0| \zeta(x_3) \zeta(x_4) |0\rangle 
\nonumber \\
&& 
+ \langle 0| \zeta(x_1) \zeta(x_3) |0\rangle\  
  \langle 0| \zeta(x_2) \zeta(x_4) |0\rangle 
\nonumber \\
&& 
+ \langle 0| \zeta(x_1) \zeta(x_4) |0\rangle\  
  \langle 0| \zeta(x_2) \zeta(x_3) |0\rangle 
\nonumber \\
&&
+ \langle 0(t) | \zeta(x_1) \zeta(x_2) \zeta(x_3) \zeta(x_4) |0(t)\rangle \big|_{\rm connected}
\nonumber \\
&&
+ \cdots . 
\nonumber
\end{eqnarray}

A scenario that has no higher order interactions beyond the quadratic ones,
$$
{\cal L}[\zeta(x)] = {\cal L}_0 + {\cal L}^{(2)}[\zeta(x)] ,
\qquad\qquad\hbox{(exactly)} 
$$
is said to be a {\it Gaussian\/} theory.  In such theories, all the odd-point functions vanish (since diagrammatically it is quite obvious that it is impossible to divide an odd number of points into pairs connected by ${\cal L}^{(2)}$ without always leaving one point unpaired---and this one-point function vanishes), while all the even-point functions can be decomposed entirely into products of the two-point function.  In the Gaussian case, the four-point function would then be equal to just the first three terms shown above, without the connected part.  There is no reason to suppose that the primordial pattern of fluctuations in the actual universe forms a {\it perfectly\/} Gaussian pattern.  However, inflationary models typically have cubic (and higher) interactions that are suppressed, so they usually predict a pattern that is still largely Gaussian in its character.  One test of the inflationary picture then is to see whether the actual primordial fluctuations have a non-Gaussian component that is large or small in comparison with the natural inflationary prediction.  To do so, it is important to know the explicit form of the three-point function expected by inflation.

These notes derive the {\it standard\/} inflationary prediction for the three-point function of the primordial fluctuation, the most readily observed signal of a non-Gaussian component within these fluctuations.  It is generated by the cubic terms in the action, ${\cal L}^{(3)}$.  This set of terms will be calculated for the simplest possible setting, where there is only a single scalar field participating in the inflationary expansion.

This derivation will be developed in several stages, building up to the full calculation of the three-point function.  The first stage analyses how the background evolves during inflation, and it introduces a few dimensionless parameters associated with this background, before proceeding to the calculation of the quadratic action for the fluctuations and the resulting behaviour for the two-point function.  This calculation of the two-point function defines a set of coordinates and the general method that will be used in these notes to describe the quantum fluctuations about the background.  Once the method that will be used to describe the fluctuations has been illustrated for the quadratic terms, the full calculation of the cubic terms and the three-point function that results from them will finally begin.
\vskip24truept

\noindent{\bf\large I. THE BACKGROUND}
\vskip9truept

\noindent 
To start, consider a simple inflationary model with a scalar field $\phi$ that is responsible for a period of accelerated expansion in the very early universe.  If this field has a potential energy described by $V(\phi)$, then the total action, including terms for the dynamics of $\phi$ and the space-time itself, is 
$$
S = \int d^4x\, \sqrt{-g}\, \Bigl\{
{\textstyle{1\over 2}} M_{\rm pl}^2\, R 
+ {\textstyle{1\over 2}} g^{\mu\nu} \partial_\mu\phi\partial_\nu\phi 
- V(\phi) \Bigr\} , 
$$
where $M_{\rm pl} = (8\pi G)^{-1/2}$ is the reduced Planck mass, $G$ being Newton's constant.  $R$ is the scalar curvature associated with the space-time, whose geometry is described by the metric $g_{\mu\nu}$.  All of the fields can be placed on the same footing by choosing units where $M_{\rm pl} = 1$; since $M_{\rm pl}$ only accompanies the gravitational field and not $\phi$, it is usually a simple matter to restore it whenever it might be needed.

Varying the action with respect to a small change in the metric, $g_{\mu\nu}\to g_{\mu\nu} + \delta g_{\mu\nu}$ produces Einstein's equation, which relates how the presence of matter and energy---in this case, that produced by the scalar field $\phi$---affects the geometry of the space-time where they exist, 
$$
R_{\mu\nu} - {\textstyle{1\over 2}} g_{\mu\nu} R = T_{\mu\nu} . 
$$
$T_{\mu\nu}$ is the energy-momentum tensor of the field $\phi$, 
$$
T_{\mu\nu} = \partial_\mu\phi\partial_\nu\phi 
- {\textstyle{1\over 2}} g_{\mu\nu} 
g^{\lambda\sigma} \partial_\lambda\phi\partial_\sigma\phi 
+ g_{\mu\nu} V . 
$$
Varying with respect to the scalar field $\phi\to\phi+\delta\phi$ yields the equation of motion for the scalar field,
$$
\nabla^2\phi + {\delta V\over\delta\phi} = 0 . 
$$

Before attempting to explain the structure of the quantum fluctuations produced by the inflationary era, it is good to begin by following how the classical background evolves.  Empirically, at very large scales or at very early times the universe appears to be quite uniform spatially.  Inflation assumes this spatial uniformity holds over a large enough patch of the universe, from the beginning of inflation to its end, that the classical part of the metric can be treated as though it depended only on the time coordinate.  Such a metric can be put into the following standard form,\footnote{This choice of the metric implicitly assumes that space-time is spatially flat.  Empirically the early universe does indeed appear to be spatially flat---or is at least very nearly so---so throughout these notes such a background will be assumed.}
$$
ds^2 = dt^2 - e^{2\rho(t)}\, \delta_{ij}\, dx^i dx^j .
$$
The rate at which $\rho(t)$ is changing corresponds to the Hubble scale,
$$
H(t) = \dot\rho(t) \equiv {d\rho\over dt} , 
$$
which is the characteristic energy scale associated with the gravitational evolution.  Evaluating the components of the Einstein equation for this metric yields 
\begin{eqnarray}
3\dot\rho^2 &\!\!\!=\!\!\!& {1\over 2} \dot\phi^2 + V 
\nonumber \\
-2\ddot\rho - 3\dot\rho^2 &\!\!\!=\!\!\!& {1\over 2} \dot\phi^2 - V , 
\nonumber
\end{eqnarray}
while the equation for the field itself is 
$$
\ddot\phi + 3\dot\rho\dot\phi + {\delta V\over\delta\phi} = 0 . 
$$

So far, aside from restricting to the case of a single scalar field and a particular assumption about the approximate homogeneity of the metric, the setting has not been otherwise constrained.  Most importantly, the potential energy of the field $V(\phi)$ has been left unspecified.  Inflationary models, if they are to produce a sufficient amount of expansion, typically must be in a `slow-roll' phase where the value of the field is not changing too rapidly---that is, $\dot\phi$ should be `small'.  Further, to maintain this condition over a sufficiently long period to produce a necessary amount of inflation, the field's acceleration ($\ddot\phi$) must also be tiny.  These two conditions can be put a little more precisely by requiring the following dimensionless parameters to be small,
$$
\epsilon = {d\over dt} {1\over H} = - {\ddot\rho\over\dot\rho^2} ,
\qquad\qquad
\delta = {1\over H} {\ddot\phi\over\dot\phi} 
= {\ddot\phi\over\dot\rho\dot\phi} .  
$$
In the limit where both $\epsilon\ll 1$ and $\delta\ll 1$, the derivatives of these parameters are even still smaller, $\dot\epsilon, \dot\delta \ll H\epsilon, H\delta$, so $\epsilon$ and $\delta$ will usually be treated as though they were constants.  By applying the background equations of motion to solve for $\ddot\rho = - {1\over 2}\dot\phi^2$, $\epsilon$ can also be expressed as 
$$
\epsilon = {1\over 2} {\dot\phi^2\over\dot\rho^2} . 
$$
This form for $\epsilon$ makes what is meant by a `small' value for $\dot\phi$ more precise:  the value of the field should change slowly when compared with the rate at which the background itself is changing, $\dot\rho = H$.  Restoring $M_{\rm pl}$ for a moment, the value for $\epsilon$ is more genuinely
$$
\epsilon = {1\over 2} {1\over M_{\rm pl}^2} {\dot\phi^2\over\dot\rho^2} , 
$$
which shows that it is indeed dimensionless.

When $\epsilon=0$ exactly, then $\dot\phi=0$; in this case, both $V(\phi)$ and $\dot\rho = H$ are constants as well.  A space-time with a constant, positive ($V>0$) vacuum energy density is called {\it de Sitter space\/}.  Therefore, $\epsilon$ can also be regarded as a parameter that characterises by how much an inflationary space-time departs from a purely de Sitter background.
\vskip24truept

\noindent{\bf\large II. THE QUADRATIC ACTION}
\vskip9truept

\noindent 
One very appealing property of inflation is that a tiny amount of spatial dependence is inevitable.  These tiny, primordial variations in the space-time, which are naturally present in inflation, provide the initial inhomogeneities needed to explain the beginnings of the structures that are observed today.  Their origin in the theory lies in the quantum behaviour of both the field and the space-time.  Therefore, in addition to the classical quantities $\phi(t)$ and $g_{\mu\nu}$ associated with the background, consider a quantum correction to each as well, 
\begin{eqnarray}
g_{\mu\nu}(t) 
&\!\!\!\!\to\!\!\!\!& g_{\mu\nu}(t) + \delta g_{\mu\nu}(t,\vec x)
\nonumber \\
\phi(t) 
&\!\!\!\!\to\!\!\!\!& \phi(t) + \delta\phi(t,\vec x) . 
\nonumber
\end{eqnarray}
Since $\delta g_{\mu\nu}$ and $\delta\phi$ are quantum mechanical, they are always fluctuating, and they introduce some spatial dependence into what would otherwise be a featureless background.  The extreme expansion during inflation stretches these fluctuations, which would otherwise remain of a tiny spatial extent, to vast sizes.  This process happens so rapidly that the fluctuations are soon frozen into the space-time and thereafter remain beyond any further causal influence while the inflationary stage lasts.  This is the basic mechanism that fills the universe with a pattern of tiny primordial perturbations according to inflation.  This section derives the leading, quadratic component of this pattern as a preliminary step before analysing the leading non-Gaussian component.

One difficulty in describing these fluctuations is that general relativity contains a fair amount of redundancy.  Although a coordinate system must be chosen in order to compare a prediction with what is observed, nothing that depends in detail on this coordinate choice corresponds to a genuine physical effect.  These notes will only be considering the scalar fluctuations, which are responsible for the density and the temperature fluctuations seen in the matter of the early universe.  A scalar fluctuation is one that transforms as a scalar under the unbroken {\it spatial\/} rotational and translational symmetries of the background rather than as a scalar under a general four-dimensional coordinate transformation.  To count the number of distinct scalar functions that are possible in a general perturbation $\delta g_{\mu\nu}(t,\vec x)$ to the classical background, first divide these fluctuations into blocks, 
$$
\delta g_{\mu\nu}(t,\vec x) = 
\pmatrix{\delta g_{00}(t,\vec x)  &\delta g_{0i}(t,\vec x)\cr 
         \delta g_{i0}(t,\vec x)  &\delta g_{ij}(t,\vec x)\cr} . 
$$
$\delta g_{00}$ itself provides one scalar field.  Another is generated by making a three-vector from a scalar field by taking its spatial derivative; thus $\delta g_{0i} \sim \partial_i B$ contains a scalar field.  $\delta g_{ij}$ contains two scalar fields---they are its trace and one generated by taking two derivatives of a scalar function, $\partial_i\partial_j\xi$.  Including a final fluctuation for the actual scalar field $\delta\phi$, all told there are five separate scalar fields.

Most of these fields have no independent physical meaning.  Two of them, for example, are determined by how the coordinates are chosen.  Dividing a general coordinate transformation, $x^\mu \to x^\mu + \delta x^\mu$, into its temporal and spatial pieces, there is one scalar function in $\delta x^0$ and it is again possible to form a spatial vector by taking a derivative of a second, $\delta x^i \sim \delta^{ij}\partial_j f$.  Additionally, it will later become apparent that two more of these scalar fields are non-propagating degrees of freedom fixed by two constraints.  Thus, the five potential scalar fields are reduced by four, leaving a single physical field.  The following calculation shows how this reduction proceeds, isolating this one physical scalar field and then analysing how its dynamics lead to a prediction for the two-point function.

To study the fluctuations about the simple, spatially invariant inflationary background, write the metric in the following general form, 
\begin{eqnarray}
ds^2 &\!\!\!=\!\!\!& 
N^2\, dt^2 - h_{ij} \bigl( N^i\, dt + dx^i \bigr)\bigl( N^j\, dt + dx^j \bigr)
\nonumber \\
&\!\!\!=\!\!\!& 
(N^2 - N_iN^i)\, dt^2 - 2N_i\, dtdx^i - h_{ij}\, dx^idx^j .
\nonumber 
\end{eqnarray}
This metric was originally introduced by Arnowitt, Deser and Misner [2] to analyse gravity from a Hamiltonian perspective.  In this framework, $N(t,\vec x)$ is called the {\it lapse function\/} while $N^i(t,\vec x)$ is the {\it shift vector\/}.  Note that $N_i$ is defined to be $N_i \equiv h_{ij}N^j$.  The components of the inverse metric are 
$$
g^{00} = {1\over N^2}, \qquad
g^{0i} = - {1\over N^2} N^i, \qquad
g^{ij} = - {1\over N^2} \bigl[ N^2 h^{ij} - N^i N^j \bigr] . 
$$
The spatial components, $h_{ij}$, can be used to define a metric for the three-dimensional spatial hypersurfaces of the full space-time.  To distinguish the curvature and covariant derivatives calculated using this metric, $h_{ij}$, from those evaluated with the full space-time metric, $g_{\mu\nu}$, the former will be written with a caret---as $\hat\nabla_i$ or $\hat R$, for example.  The full action of the theory, rewritten in terms of this metric, then becomes 
$$
S = {1\over 2} \int d^4x\, \sqrt{h}\, 
\Bigl\{ N\hat R + {1\over N} (E_{ij}E^{ij} - E^2) 
+ {1\over N}(\dot\phi - N^i\partial_i\phi)^2 
- N h^{ij}\partial_i\phi\partial_j\phi - 2NV(\phi) \Bigr\} . 
$$
In this expression, a new spatial tensor $E_{ij}$ has been introduced.  It is defined by
\begin{eqnarray}
E_{ij} 
&\!\!\!=\!\!\!& {\textstyle{1\over 2}} \bigl[ 
\dot h_{ij} - \hat\nabla_i N_j - \hat\nabla_j N_i \bigr] 
\nonumber \\
&\!\!\!=\!\!\!& {\textstyle{1\over 2}} \bigl[ 
\dot h_{ij} - h_{ik}\partial_j N^k - h_{jk}\partial_i N^k 
- N^k\partial_k h_{ij} \bigr] , 
\nonumber
\end{eqnarray}
and it is closely related to the extrinsic curvature associated with how the spatial surfaces are embedded in the full space-time.  The spatial indices are still implicitly being contracted using the metric $h_{ij}$; so, for example,
$$
E_{ij}E^{ij} - E^2 
= \bigl[ h^{ik} h^{jl} - h^{ij} h^{kl} \bigr] E_{ij}E_{kl} . 
$$

The fields $N$ and $N^i$ are both Lagrange multipliers, with no underlying dynamics.  Their equations of motion produce two constraints which reduce the number of independent scalar degrees of freedom by two.  Varying with respect to $N\to N + \delta N$ yields 
$$
\hat R - N^{-2} (E_{ij}E^{ij} - E^2) 
- N^{-2}(\dot\phi - N^i\partial_i\phi)^2 
- h^{ij}\partial_i\phi\partial_j\phi - 2V = 0 , 
$$
while $N^i\to N^i + \delta N^i$ gives 
$$
\hat\nabla_j \bigl[ N^{-1} (E_i^{\ j} - \delta_i^{\ j} E) \bigr]
= N^{-1} (\dot\phi - N^j\partial_j\phi) \partial_i\phi .
$$

Since the behaviour of the gravitational part of the action is fairly complicated, and since the actual quantum fluctuations of the field and the metric are small when compared with the classical background values, the fluctuations can be studied by expanding the action to the necessary order in the fluctuations for the quantity being analysed.  At first, when calculating the two-point function, it will be only necessary to keep those terms in the action that are quadratic in the fluctuations; but later, when evaluating the three-point function, the leading signal of a non-Gaussian pattern, the cubic terms will be kept as well.

A general parametrisation of the scalar fluctuations about the background in the metric is provided by 
$$
N = 1 + 2\Phi(t,\vec x) , \qquad
N^i = \delta^{ij} \partial_j B(t,\vec x) , \qquad
h_{ij} = e^{2\rho(t)} \bigl[ (1+ 2\zeta(t,\vec x))\, \delta_{ij} + \partial_i\partial_j\xi \bigr] , 
$$
and in the field's fluctuations by
$$
\phi = \phi_0(t) + \delta\phi(t,\vec x) . 
$$
This particular parametrisation is not left unchanged under a small change of the coordinates; but the calculation here uses the freedom to choose a particular set of coordinates to simplify certain parts of the analysis, so it is not necessary to write an explicitly coordinate-invariant form  for the fluctuations from the start.  One final raising and lowering convention that will be followed is that the indices of derivatives acting on any one of the scalar fluctuations---$\zeta$, $\Phi$, $B$, or $\xi$---will always be implicitly raised or contracted with a Kronecker $\delta^{ij}$ (rather than $h_{ij}$) as was done above in the definition of $N^i$ as $\delta^{ij}\partial_jB$.

As was described in a general sense before, redefining the time coordinate and shifting the spatial coordinate through the derivative of a scalar function, $x^i \to x^i + \delta^{ij} \partial_j f(t,\vec x)$, removes two of the five scalar functions in the general parametrisation.  This freedom can be used to choose the coordinates so that the fluctuations of the scalar field vanish, $\delta\phi=0$, and so that $\xi=0$, leaving just three scalar fields,
$$
N = 1 + 2\Phi(t,\vec x) , \qquad
N^i = \delta^{ij} \partial_j B(t,\vec x) , \qquad
h_{ij} = e^{2\rho(t)+2\zeta(t,\vec x)}\, \delta_{ij} , \qquad
\phi = \phi(t) .
$$
Notice that $\zeta$ has been slightly redefined so that it now appears in the exponent.  At first order these coordinates are the same as the initial definition above; however, this form will be much more useful when analysing the three-point function later.  This choice for the coordinates already simplifies the constraint equations quite a bit, since $\phi$ has no spatial derivatives, 
\begin{eqnarray}
& \hat R - N^{-2} (E_{ij}E^{ij} - E^2) - N^{-2} \dot\phi^2 - 2V = 0 &
\nonumber \\
& \hat\nabla_j \bigl[ N^{-1} (E_i^{\ j} - \delta_i^{\ j} E) \bigr] = 0 . &
\nonumber
\end{eqnarray}
The potential energy $V(\phi)$ has still been left fairly general, other than the assumption that it satisfies the slow-roll conditions.  It can be removed entirely from the first constraint equation by applying the $tt$ component of the Einstein equation, 
$$
\hat R - N^{-2} \bigl( E_{ij}E^{ij} - E^2 \bigr) - 6\dot\rho^2 
+ \bigl( 1 - N^{-2} \bigr) \dot\phi^2 = 0 . 
$$

Now these constraints are ready to be solved to first order in the set of coordinates that was chosen.  The scalar curvature $\hat R$ associated with the spatial metric $h_{ij}$ is 
$$
\hat R = e^{-2\rho-2\zeta} \bigl[ 
- 4 \partial_k\partial^k\zeta - 2 \partial_k\zeta\partial^k\zeta 
\bigr] 
= - 4 e^{-2\rho} \partial_k\partial^k\zeta + {\cal O}(\zeta^2) ,
$$
while $E_{ij}$ is 
$$
E_{ij} = e^{2\rho} \bigl[ \dot\rho(1+2\zeta)\delta_{ij} 
+ \dot\zeta\delta_{ij} - \partial_i\partial_j B \bigr] + \cdots
$$
so that 
$$
E_{ij}E^{ij} - E^2 = - 6 \dot\rho^2 - 12 \dot\rho\dot\zeta 
+ 4 \dot\rho\partial_i\partial_j B  + \cdots . 
$$
The constraint equation for $N$ then becomes---again, to first order in the fluctuations---
$$
- 3 \dot\rho \bigl[ 2 \dot\rho \Phi - \dot\zeta \bigr]
- \partial_k\partial^k \bigl[ \dot\rho B + e^{-2\rho}\zeta \bigr] 
+ \dot\phi^2 \Phi = 0 . 
$$
Similarly expanding the constraint for $N^i$ to first order yields,
$$
2\partial_i \bigl[ 2\dot\rho\Phi - \dot\zeta \bigr] = 0 . 
$$
This equation removes one of the scalar degrees of freedom by fixing $\Phi$, 
$$
\Phi = {1\over 2} {\dot\zeta\over\dot\rho} . 
$$
Although this constraint would seemingly have allowed the addition of an arbitrary constant as well, that constant was implicitly chosen to be zero so that the original background metric would be restored when the fluctuations are removed.  When this result is inserted into the constraint for $N$, it similarly completely fixes another one of the scalar fields, $B$, 
$$
B = - {e^{-2\rho}\over\dot\rho} \zeta + \chi
\qquad{\rm with}\qquad 
\partial_k\partial^k\chi = {1\over 2}{\dot\phi^2\over\dot\rho^2} \dot\zeta . 
$$

Having solved the constraint equations and thereby eliminated two of the scalar fields, $\Phi$ and $B$, by expressing them in terms of the one remaining field $\zeta$, it is time to determine the quadratic action for this remaining scalar field.  First, substitute the background equation, $V = 3\dot\rho^2 - {1\over 2}\dot\phi^2$, into the action once again, still using the coordinate system where the fluctuation of the field $\delta\phi=0$ vanishes, to obtain
$$
S = {1\over 2} \int d^4x\, \sqrt{h}\, 
\Bigl\{ N\hat R + N^{-1} (E_{ij}E^{ij} - E^2) - 6N\dot\rho^2 
+ (N + N^{-1}) \dot\phi^2 \Bigr\} . 
$$
To calculate the two-point function of $\zeta$, expand this integrand to second order in the small fluctuations.  At this order, 
$$
\hat R = e^{-2\rho} \bigl\{ 
- 4 \partial_k\partial^k\zeta + 8 \zeta\partial_k\partial^k\zeta 
- 2 \partial_k\zeta\partial^k\zeta + \cdots \bigr\}
$$
and 
\begin{eqnarray}
E_{ij}E^{ij} - E^2 
&\!\!\!=\!\!\!& 
- 6 \dot\rho^2 
- 12 \dot\rho\dot\zeta 
+ 4 \dot\rho\partial_k\partial^k B 
- 6 \dot\zeta^2 
+ 12 \dot\rho \partial_k\zeta \partial^k B 
+ 4 \dot\zeta\partial_k\partial^k B 
\nonumber \\
&&
+ (\partial_i\partial_j B) (\partial^i\partial^j B) 
- (\partial_k\partial^k B)^2 
+ \cdots
\nonumber
\end{eqnarray}
and 
$$
N + {1\over N} = 2 + {\dot\zeta^2\over\dot\rho^2} + \cdots . 
$$
The first term in the action is responsible for the spatial part of the kinetic term for $\zeta$, 
\begin{eqnarray}
\sqrt{h} N \hat R &\!\!\!=\!\!\!& 
2 e^{\rho} {\ddot\rho\over\dot\rho^2} \partial_k\zeta \partial^k \zeta
- \partial_0 \biggl\{ 2e^{\rho}{1\over\dot\rho} \zeta\partial_k\partial^k \zeta 
\biggr\}
\nonumber \\
&&
- 2 e^\rho \partial_k \biggl\{
\biggl( 2 + \zeta + {\ddot\rho\over\dot\rho^2}\zeta 
+ {\dot\zeta\over\dot\rho} \biggr) \partial^k\zeta 
- {1\over\dot\rho} \zeta\partial^k\dot\zeta 
\biggr\} + \cdots ; 
\nonumber 
\end{eqnarray}
using the background equations to replace $\ddot\rho = - {1\over 2} \dot\phi^2$, only one term remains 
$$
\sqrt{h} N \hat R = 
- e^{\rho} {\dot\phi^2\over\dot\rho^2} \partial_k\zeta \partial^k \zeta
+ \cdots , 
$$
up to derivative terms which have no dynamical effect.  Taken together, the rest of the terms to second order are 
\begin{eqnarray}
&&\!\!\!\!\!\!\!\!\!\!\!\!\!\!\!\!\!\!\!\!\!\!\!\!\!\!
\!\!\!\!\!\!\!\!\!\!\!\!\!\!\!\!\!\!\!\!\!\!\!\!\!\!
\sqrt{h} \biggl\{ {1\over N} \bigl( E_{ij}E^{ij} - E^2 \bigr)
- 6 N\dot\rho^2 + \biggl( N + {1\over N} \biggr) \dot\phi^2 \biggr\} 
\nonumber \\
&\!\!\!=\!\!\!& 
e^{3\rho} {\dot\phi^2\over\dot\rho^2} \dot\zeta^2  
+ e^{3\rho} \bigl[ 2\ddot\rho + \dot\phi^2 \bigr] 
\bigl[ 2 + 6\zeta + 9\zeta^2 \bigr]
\nonumber \\
&&
- 2 \partial_0 \bigl\{ e^{3\rho} \dot\rho \bigl[ 2 + 6\zeta + 9\zeta^2 \bigr] 
\bigr\}
\nonumber \\
&&
+ e^{3\rho} \partial_k \bigl\{
\bigl( 4\dot\rho + 12\dot\rho\zeta - \partial_j\partial^j B \bigr) \partial^k B 
+ (\partial_j B) (\partial^j\partial^k B) 
\bigr\} + \cdots . 
\nonumber
\end{eqnarray}
Using the background equation $\dot\phi^2 = - 2\ddot\rho$ once again leaves just one term that is not a total derivative,
$$
\sqrt{h} \biggl\{ {1\over N} \bigl( E_{ij}E^{ij} - E^2 \bigr)
- 6 N\dot\rho^2 + \biggl( N + {1\over N} \biggr) \dot\phi^2 \biggr\} 
= e^{3\rho} {\dot\phi^2\over\dot\rho^2} \dot\zeta^2 + \cdots . 
$$
So what is left after applying the background equations and ignoring total derivative terms is a remarkably simple expression for the quadratic part of the action for the fluctuations,
$$
S = \int dt\, {\dot\phi^2\over\dot\rho^2} \int d^3\vec x\, 
e^{3\rho} \Bigl\{
{\textstyle{1\over 2}} \dot\zeta^2
- {\textstyle{1\over 2}} e^{-2\rho} \partial_k\zeta \partial^k \zeta
+ \cdots \Bigr\} . 
$$
Notice that the action is directly proportional to $\dot\phi^2/\dot\rho^2$.  Until now, the slow-roll approximation has not been used at all; but replacing the background functions with the appropriate parameters introduced earlier, the fluctuations only survive as long as 
$$
{\dot\phi^2\over\dot\rho^2} = 2\epsilon M_{\rm pl}^2 \not= 0 . 
$$
In a purely de Sitter background, the second order action vanishes entirely and no primordial fluctuations are generated.  
\vskip24truept

\noindent{\bf\large III. THE TWO-POINT FUNCTION}
\vskip9truept

\noindent 
The fluctuations that are being considered are essentially quantum mechanical in their nature.  The next step is to rewrite the field so that its action more closely resembles the standard form used for a quantum field in a flat space-time.  Once this has been done, many of the results of ordinary quantum field theory can be applied to this inflationary setting, with a few cautionary notes.  Flat space is a tamer environment than the accelerating background of an inflating universe.  The energy scale that characterises this expansion is often assumed to be fairly close to the Planck scale.  At the Planck scale, the quantum description of the gravitational fluctuations that is being used in inflation becomes non-perturbative.  So while it will not be addressed further here, it is important to keep in mind that some of the usual assumptions about the behaviour of $\zeta$ that are appropriate for flat space no longer apply at such length or energy scales.

Quantum field theory was originally established for a flat space-time.  A flat background is invariant under a ten-dimensional set of Poincar\'e transformations---translations, rotations and boosts.  The canonical form of the action for a scalar field in flat space has a kinetic term which is normalised with a canonical factor of one-half:  ${1\over 2} (\dot\varphi^2 - \partial_k\varphi \partial^k\varphi)$.  In order to state some of the predictions for inflation, it will be easier if the quadratic action is first put into a form that resembles the canonical one.   To do so, first rescale the fluctuation $\zeta$ by  
$$
\varphi(t,\vec x) = e^\rho {\dot\phi\over\dot\rho}\, \zeta(t,\vec x) . 
$$
The time-derivative of $\zeta$, in terms of $\varphi$ and the slow-roll parameters, is 
$$
\dot\zeta
= e^{-\rho} {\dot\rho\over\dot\phi} 
\bigl[ \dot\varphi - \dot\rho (1+\epsilon + \delta) \varphi \bigr] . 
$$
The Lagrangian for the field $\varphi$ then begins to resemble its canonical form, 
\begin{eqnarray}
{1\over 2} e^{3\rho}{\dot\phi^2\over\dot\rho^2} \dot\zeta^2
- {1\over 2} e^{\rho}{\dot\phi^2\over\dot\rho^2} \partial_k\zeta\partial^k\zeta
&\!\!\!=\!\!\!&
{1\over 2} e^\rho \dot\varphi^2 
- {1\over 2} e^{-\rho} \partial_k\varphi\partial^k\varphi
- {1\over 2} e^{-\rho} m^2 \varphi^2 
\nonumber \\
&&
- {1\over 2} \partial_0 
\bigl\{ e^\rho \dot\rho (1 + \epsilon + \delta) \varphi^2 \bigr\} , 
\nonumber
\end{eqnarray}
where an effective, {\it time-dependent\/}, mass has appeared for the field $\varphi(t,\vec x)$,
$$
m^2(t) = - e^{2\rho} \dot\rho^2 (2 + \delta) (1 + \epsilon + \delta) 
- e^{2\rho} \dot\rho (\dot\epsilon + \dot\delta) . 
$$

So far, the slow-roll parameters have been included without actually assuming that they are small.  But now looking in the slow-roll limit, where $\epsilon^2, \epsilon\delta, \delta^2, \dot\epsilon/\dot\rho$, and $\dot\delta/\dot\rho$ are each much smaller than $\epsilon$ and $\delta$, the mass to leading order is more simply given by
$$
m^2 = - e^{2\rho} \dot\rho^2 (2 + 2\epsilon + 3\delta) + \cdots . 
$$
The time and space coordinates are still weighted with different powers of the scale factor $e^{\rho(t)}$; to put them on a similar footing, introduce a {\it conformal time\/} coordinate, given by 
$$
\eta(t) \equiv \int dt\, e^{-\rho(t)} , 
$$
in terms of which the quadratic action for $\varphi(\eta(t),\vec x)$ becomes
$$
S = \int d\eta d^3\vec x\, \bigl\{
{\textstyle{1\over 2}} \varphi^{\prime\, 2} 
- {\textstyle{1\over 2}} \vec\nabla\varphi\cdot\vec\nabla\varphi
- {\textstyle{1\over 2}} m^2 \varphi^2 
\bigr\} . 
$$
Hereafter, a prime denotes a derivative with respect to the conformal time and a standard vector notation has been introduced where $\vec a\cdot\vec b = \delta_{ij}\, a^i b^j$.  The conformal time coordinate is assumed to be negative, $\eta\in (-\infty,0]$, since for this choice time runs forwards; the coordinate could also have been chosen to be positive, but then as time {\it advances\/} the coordinate {\it diminishes\/}, which is why the negative branch is the one that is more often used.

Varying of $S$ with respect to $\varphi$ determines the equation of motion for $\varphi$, 
$$
\varphi^{\prime\prime} - \vec\nabla\cdot\vec\nabla\varphi + m^2 \varphi = 0 . 
$$
Superficially, this equation appears to be exactly that of a free massive field in flat space; however, it is not fully covariant under Poincar\'e transformations.  The mass is not a constant, but depends rather on the conformal time $\eta$, 
$$
m^2 = - \rho^{\prime\, 2} (2 + 2\epsilon + 3\delta) + \cdots . 
$$
Since the background is still invariant under purely {\it spatial\/} transformations, the field $\varphi(t,\vec x)$ can be expanded, as usual, in operators that create or annihilate plane waves, 
$$
\varphi(\eta,\vec x) = \int {d^3\vec k\over (2\pi)^3}\, 
\bigl\{ \varphi_k(\eta) e^{i\vec k\cdot\vec x} a_{\vec k}
+ \varphi_k^*(\eta) e^{-i\vec k\cdot\vec x} a_{\vec k}^\dagger \bigr\} . 
$$
The time-dependent part of the eigenmodes is then the solution to a Klein-Gordon equation with a time-dependent mass,
$$
\varphi^{\prime\prime}_k + (k^2 + m^2) \varphi_k = 0 . 
$$
To solve this equation requires knowing the behaviour of $m(\eta)$---at least to leading order in the slow-roll parameters $\epsilon$ and $\delta$.  In terms of the conformal time, the $\epsilon$ parameter is defined by 
$$
\epsilon = {d\over dt} {1\over H} 
= e^{-\rho(\eta)} {d\over d\eta} {e^{\rho(\eta)}\over\rho'}
= 1 + {d\over d\eta} {1\over\rho'}
\qquad\Rightarrow\qquad
d \biggl( {1\over\rho'} \biggr) = - (1 - \epsilon)\, d\eta . 
$$
This equation can be easily integrated and when the order $\epsilon^2$ corrections are neglected it becomes
$$
\rho' = - {1+\epsilon\over\eta} + \cdots . 
$$
The constant of integration has been fixed so that the standard result, $e^{-\rho(\eta)} = - H\eta$, is recovered in the de Sitter limit ($\epsilon\to 0$), remembering that the Hubble scale is a constant in de Sitter space.  This result allows $\rho'$ in the expression for the mass to be replaced with its explicit time dependence,
$$
m^2 = - {1\over\eta^2} (2 + 6\epsilon + 3\delta) + \cdots , 
$$
which in turn determines the behaviour of the modes  
$$
\varphi^{\prime\prime}_k 
+ k^2 \biggl[ 1 - {2+3(2\epsilon+\delta)\over k^2\eta^2}
\biggr] \varphi_k = 0 , 
$$
at least to leading order in $\epsilon$ and $\delta$.

The solution of this differential equation is best expressed as a linear combination of Hankel functions, 
$$
\varphi_k(\eta) 
= N_k (-k\eta)^{1/2}\, 
\bigl[ H_\nu^{(1)}(-k\eta) + \theta_k\, H_\nu^{(2)}(-k\eta) \bigr] , 
$$
where their common index $\nu$ is 
$$
\nu = {\textstyle{3\over 2}} \sqrt{1+{\textstyle{4\over 3}}(2\epsilon+\delta)} 
=  {\textstyle{3\over 2}} + 2\epsilon + \delta + \cdots . 
$$
As a second order differential equation, any particular solution requires two further conditions to determine the constants of integration, $N_k$ and $\theta_k$, to specify it completely.  One of these conditions is automatically provided by quantum field theory.  The canonical commutation relation between the field $\varphi$ and its conjugate momentum $\pi$ for a local, causal field should be 
$$
\bigl[ \varphi(\eta,\vec x), \pi(\eta,\vec y) \bigr] 
= i\delta^3(\vec x-\vec y)
\qquad\qquad
\pi = {\delta{\cal L}\over\delta\varphi'} = \varphi' . 
$$
This relation is the field-theoretic analogue of the quantum mechanical commutator between the position ($\hat X$) and momentum ($\hat P$) operators, $[\hat X,\hat P] = i$.  In terms of the modes $\varphi_k$, this relation implies that 
$$
\varphi_k \varphi_k^{\prime *} - \varphi_k^* \varphi_k^{\prime} = i , 
$$
which fixes one of the constants of integration, 
$$
N_k = - {\sqrt{\pi}\over 2\sqrt{k}} {1\over\sqrt{1-\theta_k\theta_k^*}} . 
$$
A further condition is needed to fix $\theta_k$.

The second property ordinarily assumed is that space-time becomes locally flat at arbitrarily small separations.  One of the postulates of general relativity is that it is always possible to choose a locally flat frame for any space-time point.  And since quantum field theory was developed for a {\it globally\/} flat background, it is very tempting to impose local flatness as a principle applicable in an arbitrary background.  However, what is meant by a short distance in an inflationary background is not an absolute statement, for it depends on precisely {\it when\/} this condition is being imposed.  Wavelengths that at one time might have been small compared with the curvature of the background will later no longer be so, having been stretched along with the expansion of the space.  

What is still a bit more troubling is that the dynamical scale for the inflationary expansion, the Hubble scale $H$, is usually chosen to be an appreciable fraction of the Planck scale.  At distances smaller than this Planck threshold, a description of nature that simultaneously applies the principles of both quantum field theory and general relativity so far does not seem to be consistent.  The standard prescription is to defined the modes so that in the infinite past, $t\to-\infty$, they match with the positive energy modes of flat space.  This prescription defines what is called the `Bunch-Davies' state:  the vacuum defined in the infinite past that is associated with the free, or quadratic, part of the action.

Applying this condition fixes the remaining constant of integration in $\varphi_k(\eta)$.  Put a little more precisely, this condition requires that the modes at very short distances---or equivalently at very large spatial momenta ($k\to\infty$)---should match the functional form of the standard modes for a quantum theory in a flat space-time.  Expanding the solution for $\varphi_k$ in this limit, produces 
$$
\lim_{k\to\infty}\varphi_k(\eta) 
= {1\over\sqrt{1-\theta_k\theta_k^*}}\, 
\biggl[ 
e^{-i{\pi\over 2}(2\epsilon+\delta)} {e^{-ik\eta}\over\sqrt{2k}} 
+ \theta_k\,e^{i{\pi\over 2}(2\epsilon+\delta)} {e^{ik\eta}\over\sqrt{2k}} 
\biggr] . 
$$
For a massless quantum field in flat space, the positive energy vacuum modes are those for which $\varphi_k(\eta) = e^{-ik\eta}/\sqrt{2k}$, so the Bunch-Davies prescription requires that $\theta_k = 0$.  These two constraints on the state---the canonical commutation relation and the matching with the flat-space vacuum at short distances---thus completely determine the momentum modes
$$
\varphi_k(\eta) 
=  - {\sqrt{\pi}\over 2} \sqrt{-\eta}\, H_\nu^{(1)}(-k\eta) , 
$$
which in turn defines the metric fluctuations, $\zeta(\eta,\vec x)$.

With this exact expression for the modes, it is at last possible to evaluate the two-point correlation function for the fluctuations mentioned at the beginning, 
$$
\langle 0(\eta) | \zeta(\eta,\vec x) \zeta(\eta,\vec y) |0(\eta)\rangle . 
$$
This two-point function is often expressed in terms of a {\it power spectrum\/}, which is its Fourier transform with a few conventional factors (sometimes) extracted for convenience, 
$$
\langle 0(\eta) | \zeta(\eta,\vec x) \zeta(\eta,\vec y) |0(\eta)\rangle 
= \int {d^3\vec k\over (2\pi)^3}\, e^{i\vec k\cdot(\vec x-\vec y)} 
{2\pi^2\over k^3} P_k(\eta) . 
$$
By applying the rescaling that connects $\zeta$ with $\varphi$, the power spectrum for the fluctuation $\zeta$ is 
$$
P_k(\eta) = e^{-2\rho} {\dot\rho^2\over\dot\phi^2} 
{k^3\over 2\pi^2} \varphi_k\varphi_k^* . 
$$
Then replacing $\dot\rho^2/\dot\phi^2$ with $1/(2M_{\rm pl}^2\epsilon)$ in this expression, while at the same time substituting the factor $e^{-2\rho(\eta)}$ with 
$$
e^{-2\rho} = {(-H\eta)^2\over(1+\epsilon)^2} + \cdots , 
$$
the leading form for the power spectrum is found to be
$$
P_k(\eta) = {H^2\over M_{\rm pl}^2} {(-k\eta)^3\over 16\pi} 
{H_\nu^{(1)}(-k\eta)\, H_\nu^{(2)}(-k\eta)\over \epsilon(1+\epsilon)^2} 
+ \cdots . 
$$
Since the interest here is to learn how the power spectrum changes with the scale $k$, rather than with the conformal time $\eta$, the time dependence of $H(\eta)$ is not explicitly shown, since $H$ has no dependence on $k$ at all, whereas in the other factors where it occurs together with $k$ as $-k\eta$.

Inflation is intended to increase the size of a causally connected patch of the universe far beyond what it would have been in a purely radiation- or matter-dominated universe.  The idea is that what is seen at large distance-scales today began as fluctuations with a very tiny spatial extent at the time of inflation.  From the perspective of the more recent ages of the universe, the fluctuations laid down by inflation are, for a longer or shorter while, beyond the causal reach of any process until the horizon $1/\dot\rho$ catches up with the physical size of a fluctuation, which is also growing with the expansion of the universe, though not as fast as the horizon.  During the inflationary era, this relation is reversed; during inflation, the horizon changes only very slowly---$1/\dot\rho=1/H$ is nearly constant---while the physical spatial sizes of the fluctuations are stretched very dramatically.  In the de Sitter limit, the horizon size would in fact be constant; it is called a horizon since for any observer, anything at a distance greater than $1/\dot\rho$ is unobservable, and time-like Killing vectors tip over and become space-like as they are continued beyond the horizon, just as for the horizon of a black hole in the standard Schwarzschild coordinates.  However, the de Sitter horizon is a little different---in de Sitter space, each observer sees his or her own horizon and it is not associated with any space-time singularity, and of course de Sitter space does not asymptotically approach Minkowski space as is the case far from a black hole horizon.

So for inflation, the observationally important values of $k$ are those associated with modes that have been stretched well outside this Hubble horizon by the end of inflation.  Writing the {\it physical\/} $k$, which is stretched by the scale factor over time, as $k_{\rm phys}=e^{-\rho(\eta)}k$, the relevant modes are those for which 
$$
e^{-\rho(\eta)}k \ll H 
\qquad\Rightarrow\qquad 
- {H\eta\over 1+\epsilon}\, k \ll H 
\qquad\Rightarrow\qquad 
- k\eta \ll 1 .
$$
Expanding the power spectrum in this limit, $-k\eta\to 0$, then yields 
$$
P_k(\eta) = {1\over 8\pi^2} {1\over\epsilon} {H^2\over M_{\rm pl}^2} 
(-k\eta)^{-4\epsilon-2\delta} + \cdots . 
$$
This power spectrum can be equivalently written in the form 
$$
P_k(\eta) = {1\over 8\pi^2} {1\over\epsilon} {H^2\over M_{\rm pl}^2} 
(-k_*\eta)^{-4\epsilon-2\delta} 
\biggl( {k\over k_*} \biggr)^{-4\epsilon-2\delta} + \cdots 
$$
by introducing a reference scale $k_*$, which is useful when comparing with experiments.\footnote{In the WMAP experiment [3], for example, the data are normalized with respect to a $k_*$ of 2 Gpc$^{-1}$.}

This minimal inflationary picture has influenced how experimental results are typically analysed; fits to observations usually choose a power-law form for the power spectrum, which is compatible with this basic scaling-form predicted by inflation,
$$
P_k(\eta) \equiv \Delta_\zeta^2(k_*)  
\biggl( {k\over k_*} \biggr)^{n_s-1} . 
$$
$\Delta_\zeta^2$ is called the {\it amplitude\/} of the power spectrum and $n_s$ is its {\it tilt\/}.  The predictions for these parameters for the simple inflationary model that has been described here are
$$
\Delta_\zeta^2(k_*) \approx 
{1\over 8\pi^2} {1\over\epsilon} {H^2\over M_{\rm pl}^2} 
\qquad\hbox{and}\qquad
n_s = 1 - 4\epsilon - 2\delta . 
$$
\vskip24truept

\noindent{\bf\large IV. THE CUBIC ACTION}
\vskip9truept

\noindent 
The three-point function is determined by the cubic interactions, which means that the action next needs to be expanded a further order in the small fluctuations.  Much of the difficulty in extracting these cubic terms occurs because a straightforward expansion of the action to third order in $\zeta$ produces a set of interactions where some of the coefficients obscure the true size of their physical effect.  For example, some of the terms that emerge at third-order superficially appear not to be suppressed at all, though in fact the three-point function is suppressed by a factor of $\epsilon^2$.  With other coordinates for describing the fluctuations, this suppression would have been obvious.  Unfortunately, these coordinates, which are presented in the appendix, are not as useful for describing the fluctuations once they have been stretched well outside the horizon.  For this purpose it is $\zeta(t,\vec x)$ that is ideal---it becomes essentially constant far outside the horizon and it is closely related to the classical fluctuations in the spatial curvature which are commonly used for treating the inhomogeneities of the universe after inflation has ended.  Unfortunately, there is no perfect set of coordinates that is at once suited to all possible purposes.

Most of this section is devoted to showing how to convert the cubic interactions---those that emerge when the action is expanded directly---into a simpler form where only the $\epsilon^2$ suppressed terms are left.  This latter form will be used to calculate the inflationary prediction for the three-point function in the next section.  Going from one set of interactions to the other is a bit tedious.  Only at the very end, when all of the separate calculations have been put back together, will the goal of finding a simpler set of cubic interactions finally be attained.

Before embarking on this calculation, it is useful to explain its basic logic.  The first step is just to expand the action to cubic order in the fluctuations.  Thereafter follows a lengthy series of integrations by parts, the results of which will be sifted into three distinct groups.  One set contains total derivatives, together with terms that vanish when the background equation, $\ddot\rho = -{1\over 2} \dot\phi^2$, is imposed.  These terms have no dynamical effect, although for thoroughness they will all be carefully retained.  The second set is directly proportional to the linear equation of motion for $\zeta$,
$$
{d\over dt} \biggl\{ e^{3\rho} {\dot\phi^2\over\dot\rho^2} \dot\zeta \biggr\} 
- e^\rho {\dot\phi^2\over\dot\rho^2} \partial_k\partial^k\zeta . 
$$
This set does contribute in part to the three-point function, but it will be treated separately when computing the three-point function by means of a nonlinear shift in the field.  The final set of terms are those that do not fall into either of these two classes.  This set will be seen to be clearly order $\epsilon^2$ and are therefore small in the slow-roll limit.

To begin, start once more with the coordinates that were introduced during the derivation of the quadratic action, 
$$
N = 1 + {\dot\zeta\over\dot\rho}, \qquad
N^i = \delta^{ij}\partial_j B, \qquad
h_{ij} = e^{2\rho+2\zeta}\, \delta_{ij},\qquad
\phi = \phi(t) ,
$$
where one of the Lagrange constraints has already been used to define $N$ in terms of $\zeta$.  As before, the action for this system is 
$$
S = {1\over 2} \int d^4x\, \sqrt{h}\, \bigl\{ N\hat R 
+ N^{-1} \bigl( E_{ij}E^{ij} - E^2 \bigr) - 6N\dot\rho^2 
+ \bigl( N + N^{-1} \bigr) \dot\phi^2 \bigr\} 
$$
once the background equation has been imposed to remove the explicit appearance of the potential $V(\phi)$.

The exact expressions for $\hat R$ and $E_{ij}$ in these coordinates are 
$$
\hat R = - 4 e^{-2\rho} e^{-2\zeta} \partial_k\partial^k\zeta 
-2 e^{-2\rho} e^{-2\zeta} \partial_k\zeta \partial^k\zeta
$$
and 
$$
E_{ij} = e^{2\rho+2\zeta} \bigl\{ [\dot\rho + \dot\zeta 
- \partial_k\zeta\partial^k B ] \delta_{ij}
- \partial_i\partial_j B \bigr\} , 
$$
so that 
\begin{eqnarray}
E_{ij}E^{ij} - E^2 
&\!\!\!=\!\!\!& 
- 6 \bigl( \dot\rho + \dot\zeta - \partial_k\zeta\partial^k B \bigr)^2
+ 4 \bigl( \dot\rho + \dot\zeta - \partial_k\zeta\partial^k B \bigr) 
\bigl( \partial_l\partial^l B \bigr)
\nonumber \\
&& 
+ \bigl( \partial_k\partial_l B \bigr) \bigl( \partial^k\partial^l B \bigr)
- \bigl( \partial_k\partial^k B \bigr)^2 . 
\nonumber 
\end{eqnarray}
The rest of the factors need also to be expanded to third order, 
\begin{eqnarray}
\sqrt{h} &\!\!\!=\!\!\!& e^{3\rho} e^{3\zeta} 
= e^{3\rho} \bigl\{ 
1 + 3\zeta + {\textstyle{9\over 2}} \zeta^2 + {\textstyle{9\over 2}} \zeta^3
+ \cdots \bigr\} 
\nonumber \\
{1\over N} &\!\!\!=\!\!\!& 
1 - {\dot\zeta\over\dot\rho} + {\dot\zeta^2\over\dot\rho^2} 
- {\dot\zeta^3\over\dot\rho^3} + \cdots .
\nonumber 
\end{eqnarray}
Expanding, for example, the first term in the Lagrangian produces the following set of third-order terms, 
$$
\sqrt{h}\, N\hat R \bigr|^{(3)} = 
- 2 e^{\rho} \zeta^2 \partial_k\partial^k\zeta 
- 2 e^{\rho} \zeta \partial_k\zeta \partial^k\zeta 
- 4 e^{\rho} {1\over\dot\rho} \dot\zeta \zeta \partial_k\partial^k\zeta 
- 2 e^{\rho}  {1\over\dot\rho} \dot\zeta \partial_k\zeta \partial^k\zeta . 
$$
The goal is to convert the action into a form where many of the leading contributions vanish because they are total derivatives.  To start this process, convert this set of four terms into a single (nonderivative) term by noticing that the last two terms resemble the first two, once they have been integrated by parts to remove the time derivatives from the $\zeta$ fields,
\begin{eqnarray}
&&\!\!\!\!\!\!\!\!\!\!\!\!\!\!\!\!\!\!\!\!\!\!\!\!\!\!\!\!\!\!\!\!\!
- 4 e^{\rho} {1\over\dot\rho} \dot\zeta \zeta \partial_k\partial^k\zeta 
- 2 e^{\rho}  {1\over\dot\rho} \dot\zeta \partial_k\zeta \partial^k\zeta 
\nonumber \\
&\!\!\!=\!\!\!& 
2 e^{\rho} \zeta^2 \partial_k\partial^k\zeta
+ 2 e^{\rho} \zeta \partial_k\zeta \partial^k\zeta
- 2 e^{\rho} {\ddot\rho\over\dot\rho^2} \zeta^2 \partial_k\partial^k\zeta
- 2 e^{\rho} {\ddot\rho\over\dot\rho^2} \zeta \partial_k\zeta \partial^k\zeta
\nonumber \\
&& 
- 2{d\over dt} \biggl\{ 
e^{\rho} {1\over\dot\rho} \zeta^2 \partial_k\partial^k\zeta
+ e^{\rho} {1\over\dot\rho} \zeta \partial_k\zeta \partial^k\zeta \biggr\}
+ \partial_k \biggl\{ 
2 e^{\rho} {1\over\dot\rho} \zeta^2 \partial^k\zeta \biggr\} . 
\nonumber 
\end{eqnarray}
Inserted back into the previous equation, and integrating it by parts once more, yields a single term that is not a total derivative, 
\begin{eqnarray}
\sqrt{h}\, N\hat R \bigr|^{(3)} 
&\!\!\!=\!\!\!& 
2 e^{\rho} {\ddot\rho\over\dot\rho^2} \zeta \partial_k\zeta \partial^k\zeta
\nonumber \\
&& 
- {d\over dt} \biggl\{ 
2 e^\rho {1\over\dot\rho} \zeta
\partial_k \bigl[ \zeta\partial^k\zeta \bigr] 
\biggr\}
+ \partial_k \biggl\{ 
2 e^{\rho} {1\over\dot\rho^2} \zeta^2 
\bigl[ \dot\rho \partial^k\dot\zeta - \ddot\rho \partial^k\zeta \bigr] 
\biggr\} . \qquad
\nonumber 
\end{eqnarray}
The background equation, $\ddot\rho = - {1\over 2} \dot\phi^2$, will be assumed frequently in this calculation; for example, the first term in this expression can be rewritten as
\begin{eqnarray}
\sqrt{h}\, N\hat R \bigr|^{(3)} 
&\!\!\!=\!\!\!& 
- e^{\rho} {\dot\phi^2\over\dot\rho^2} \zeta \partial_k\zeta \partial^k\zeta 
+ e^{\rho} {1\over\dot\rho^2} \bigl[ 2 \ddot\rho + \dot\phi^2 \bigr] 
\zeta \partial_k\zeta \partial^k\zeta 
\nonumber \\
&& 
- {d\over dt} \biggl\{ 
2 e^\rho {1\over\dot\rho} \zeta
\partial_k \bigl[ \zeta\partial^k\zeta \bigr] 
\biggr\}
+ \partial_k \biggl\{ 
2 e^{\rho} {1\over\dot\rho^2} \zeta^2 
\bigl[ \dot\rho \partial^k\dot\zeta - \ddot\rho \partial^k\zeta \bigr] 
\biggr\} . \qquad
\nonumber 
\end{eqnarray}
This step is useful; it makes the scaling of the initial term with the slow-roll parameter a little clearer, since the ratio $\dot\phi^2/\dot\rho^2$ is directly proportional to $\epsilon$.

The remaining terms in the Lagrangian contain the following cubic terms,
\begin{eqnarray}
&&\!\!\!\!\!\!\!\!\!\!\!\!\!\!\!
\sqrt{h} \biggl\{ {1\over N} \bigl( E_{ij}E^{ij} - E^2 \bigr)
- 6 N\dot\rho^2 + \biggl( N + {1\over N} \biggr) \dot\phi^2 \biggr\}^{(3)} 
\nonumber \\
&\!\!\!=\!\!\!& 
e^{3\rho} {\dot\phi^2\over\dot\rho^2} \biggl( 
3 \zeta - {\dot\zeta\over\dot\rho} 
\biggr) \dot\zeta^2 
\nonumber \\
&&
+ e^{3\rho} \biggl[ 
\biggl( 3 \zeta - {\dot\zeta\over\dot\rho} \biggr) 
\bigl( \partial_k\partial_l B \bigr) \bigl( \partial^k\partial^l B \bigr) 
- \biggl( 3 \zeta - {\dot\zeta\over\dot\rho} \biggr) 
\bigl( \partial_k\partial^k B \bigr)^2 
- 4 \bigl( \partial_k\zeta \partial^k B \bigr) 
\bigl( \partial_l\partial^l B \bigr) 
\biggr] 
\nonumber \\
&&
- {d\over dt} \bigl[ 18 e^{3\rho} \dot\rho \zeta^3 \bigr] 
+ \partial_k \bigl[ 18 e^{3\rho} \dot\rho \zeta^2\partial^k B \bigr]
+ 9 e^{3\rho} \bigl[ 2 \ddot\rho + \dot\phi^2 \bigr] \zeta^3 . 
\nonumber  
\end{eqnarray}
If the full Lagrangian for the cubic terms in the fluctuations is defined to be 
$$
2{\cal L}^{(3)} = 
\sqrt{h} \biggl\{ N\hat R + {1\over N} \bigl( E_{ij}E^{ij} - E^2 \bigr)
- 6 N\dot\rho^2 + \biggl( N + {1\over N} \biggr) \dot\phi^2 \biggr\} \biggr|^{(3)} , 
$$
then 
\begin{eqnarray}
2{\cal L}^{(3)}
&\!\!\!=\!\!\!& 
e^{3\rho} {\dot\phi^2\over\dot\rho^2} 
\biggl( 3 \zeta - {\dot\zeta\over\dot\rho} \biggr) \dot\zeta^2 
- e^{\rho} {\dot\phi^2\over\dot\rho^2} \zeta \partial_k\zeta \partial^k\zeta 
\nonumber \\
&&
+ e^{3\rho} \biggl\{ 
\biggl( 3 \zeta - {\dot\zeta\over\dot\rho} \biggr) 
\Bigl[
\bigl( \partial_k\partial_l B \bigr) \bigl( \partial^k\partial^l B \bigr) 
- \bigl( \partial_k\partial^k B \bigr)^2 
\Bigr] 
- 4 \bigl( \partial_k\zeta \partial^k B \bigr) 
\bigl( \partial_l\partial^l B \bigr) 
\biggr\}
\nonumber \\
&&
+ {\cal D}_0 , 
\nonumber  
\end{eqnarray}
where all of the non-dynamical terms generated so far have been abbreviated by ${\cal D}_0$, which is given by 
\begin{eqnarray}
{\cal D}_0 
&\!\!\!=\!\!\!& 
- {d\over dt} \biggl[ 
18 e^{3\rho} \dot\rho \zeta^3 
+ 2 e^\rho {1\over\dot\rho} \zeta
\partial_k \bigl[ \zeta\partial^k\zeta \bigr] 
\biggr]
\nonumber \\
&&
+ \partial_k \biggl[ 
18 e^{3\rho} \dot\rho \zeta^2\partial^k B 
+ 2 e^{\rho} {1\over\dot\rho^2} \zeta^2 
\bigl[ \dot\rho \partial^k\dot\zeta - \ddot\rho \partial^k\zeta \bigr] 
\biggr] 
\nonumber \\
&&
+ 9 e^{3\rho} \bigl[ 2\ddot\rho + \dot\phi^2 \bigr] \zeta^3 
+ e^{\rho} {1\over\dot\rho^2} 
\bigl[ 2\ddot\rho + \dot\phi^2 \bigr]
\zeta \partial_k\zeta \partial^k\zeta . 
\nonumber 
\end{eqnarray}

While this form for ${\cal L}^{(3)}$, the cubic part of the action, is perfectly correct, it does not immediately convey the true size of the non-Gaussian component of the primordial perturbations.  The terms on the first line are superficially suppressed by $\epsilon$, while those of the second line are not obviously suppressed at all.  However, the true size of these interactions is smaller than what either line would appear to suggest, since the actual suppression is $\epsilon^2$.  To establish this fact requires a much longer calculation.

To proceed, replace the $B$ field with the $\zeta$ field through the second of the constraints derived earlier, 
$$
B = - {e^{-2\rho}\over\dot\rho} \zeta + \chi 
\qquad{\rm with}\qquad 
\partial_k\partial^k\chi = {1\over 2} {\dot\phi^2\over\dot\rho^2} \dot\zeta ,
$$
and sort the resulting terms according to the power of their $e^\rho$ prefactors,
\begin{eqnarray}
2{\cal L}^{(3)}
&\!\!\!=\!\!\!& 
e^{3\rho} {\dot\phi^2\over\dot\rho^2} 
\biggl( 3 \zeta - {\dot\zeta\over\dot\rho} \biggr) \dot\zeta^2 
- {1\over 4} e^{3\rho} {\dot\phi^4\over\dot\rho^4} 
\biggl( 3 \zeta - {\dot\zeta\over\dot\rho} \biggr) \dot\zeta^2 
\nonumber \\
&&
- 2 e^{3\rho} {\dot\phi^2\over\dot\rho^2} \dot\zeta \partial_k\zeta \partial^k\chi
+ e^{3\rho} \biggl( 3 \zeta - {\dot\zeta\over\dot\rho} \biggr) 
\partial_k\partial_l\chi\partial^k\partial^l\chi
\nonumber \\
&&
- e^{\rho} {\dot\phi^2\over\dot\rho^2} 
\biggl( \zeta - 2 {\dot\zeta\over\dot\rho} \biggr) 
\partial_k\zeta \partial^k\zeta 
+ e^\rho {\dot\phi^2\over\dot\rho^3} \biggl( 3 \zeta - {\dot\zeta\over\dot\rho} \biggr) 
\dot\zeta \partial_k\partial^k\zeta 
\nonumber \\
&&
- 2 e^\rho {1\over\dot\rho} \biggl( 3 \zeta - {\dot\zeta\over\dot\rho} \biggr) 
\partial_k\partial_l\zeta \partial^k\partial^l\chi
+ 4 e^\rho {1\over\dot\rho} \partial_k\zeta \partial^k\chi\partial_l\partial^l\zeta 
\nonumber \\
&&
+ e^{-\rho}{1\over\dot\rho^2} \biggl( 3 \zeta - {\dot\zeta\over\dot\rho} \biggr) 
\Bigl[ 
\partial_k\partial_l\zeta \partial^k\partial^l\zeta 
- \bigl( \partial_k\partial^k\zeta \bigr)^2 
\Bigr]
- 4 e^{-\rho} {1\over\dot\rho^2} \partial_k\zeta \partial^k\zeta \partial_l\partial^l\zeta 
\nonumber \\
&&
+ {\cal D}_0 . 
\nonumber 
\end{eqnarray}

This form for the cubic action is expressed solely in terms of the scalar field $\zeta$, since $\chi$ is immediately determined by $\zeta$.  Both of the Lagrange constraints have been applied, and this cubic action is ready to be analysed term by term.  Because the process of converting this expression for ${\cal L}^{(3)}$ into one where the $\epsilon^2$ suppression is made manifest is rather lengthy, the calculation will be divided into four parts---one step for each set of terms that shares the same power for the $e^\rho$ prefactor, except for one of the $e^{3\rho}$ terms which will be treated by itself.
\vskip18truept

\noindent{\it IV.A. The $e^{-\rho}$-terms}
\vskip6truept
Since there are the fewest of them, the $e^{-\rho}$ terms will be analysed first, 
\begin{eqnarray}
2{\cal L}^{(3)}\bigr|_{e^{-\rho}}
&\!\!\!=\!\!\!& 
3 e^{-\rho}{1\over\dot\rho^2} \zeta 
\partial_k\partial_l\zeta \partial^k\partial^l\zeta 
- 3 e^{-\rho}{1\over\dot\rho^2} \zeta \bigl( \partial_k\partial^k\zeta \bigr)^2 
- 4 e^{-\rho} {1\over\dot\rho^2} \partial_k\zeta \partial^k\zeta \partial_l\partial^l\zeta 
\nonumber \\
&& 
- e^{-\rho} {1\over\dot\rho^3} \dot\zeta
\partial_k\partial_l\zeta \partial^k\partial^l\zeta 
+ e^{-\rho} {1\over\dot\rho^3} \dot\zeta \bigl( \partial_k\partial^k\zeta \bigr)^2 . 
\nonumber  
\end{eqnarray}
The two terms on the second line differ from those on the first since they both have a time derivative of the $\zeta$ field.  Integrating by parts as many times as is needed to remove all of the time derivatives from the $\zeta$'s eventually produces 
\begin{eqnarray}
&&\!\!\!\!\!\!\!\!\!\!\!\!\!\!\!\!\!\!\!\!\!\!\!\!\!\!\!
- e^{-\rho} {1\over\dot\rho^3} \dot\zeta
\partial_k\partial_l\zeta \partial^k\partial^l\zeta 
+ e^{-\rho} {1\over\dot\rho^3} \dot\zeta \bigl( \partial_k\partial^k\zeta \bigr)^2 
\nonumber \\
&\!\!\!=\!\!\!& 
- {1\over 3} e^{-\rho} {1\over\dot\rho^2} \zeta \bigl[ 
\partial_k\partial_l\zeta \partial^k\partial^l\zeta 
- \bigl( \partial_k\partial^k\zeta \bigr)^2 
\bigr] 
+ {1\over 2} e^{-\rho}{\dot\phi^2\over\dot\rho^4} \zeta \bigl[ 
\partial_k\partial_l\zeta \partial^k\partial^l\zeta 
- \bigl( \partial_k\partial^k\zeta \bigr)^2 
\bigr] 
\nonumber \\
&&
+ {1\over 3} {d\over dt} \biggl\{ 
e^{-\rho}{1\over\dot\rho^3} \zeta \bigl[ 
\bigl( \partial_k\partial^k\zeta \bigr)^2 
- \partial_k\partial_l\zeta \partial^k\partial^l\zeta 
\bigr] \biggr\}
\nonumber \\
&&
+ {2\over 3} \partial_k \biggl\{ 
e^{-\rho}{1\over\dot\rho^3} \dot\zeta \bigl[ 
\partial^k\zeta \partial_l\partial^l\zeta 
- \partial^l\zeta \partial_l\partial^k\zeta 
\bigr] 
- e^{-\rho}{1\over\dot\rho^3} \zeta \bigl[ 
\partial^k\dot\zeta \partial_l\partial^l\zeta 
- \partial^l\dot\zeta \partial_l\partial^k\zeta 
\bigr] \biggr\}
\nonumber \\
&&
+ e^{-\rho} {1\over\dot\rho^4} \biggl[ \ddot\rho + {1\over 2}\dot\phi^2 \biggr]
\zeta \bigl[ 
\bigl( \partial_k\partial^k\zeta \bigr)^2 
- \partial_k\partial_l\zeta \partial^k\partial^l\zeta 
\bigr] . 
\nonumber  
\end{eqnarray}
The first of these terms---superficially the dominant one in the slow-roll limit---when added to the rest of the order $e^{-\rho}$ terms produces an expression that coalesces into a further set of total derivatives, leaving just one term remaining, 
$$
2{\cal L}^{(3)}\bigr|_{e^{-\rho}}
= {1\over 2} e^{-\rho}{\dot\phi^2\over\dot\rho^4} \zeta \bigl[ 
\partial_k\partial_l\zeta \partial^k\partial^l\zeta 
- \bigl( \partial_k\partial^k\zeta \bigr)^2 
\bigr] 
+ \cdots ,  
$$
where the unwritten terms are those that have no dynamical effect.

It might seem that what remains is only suppressed by $\epsilon$ and not by $\epsilon^2$ after all.  However, this term is actually a part of a larger set that is ultimately proportional to the variation of the quadratic action for $\zeta$.  As will be shown in the next section, all the terms in this set that are not second order in the slow-roll parameters $\epsilon$ and $\delta$ will be seen to contribute negligibly to the three-point function in a late-time limit.  This term will therefore be written as 
$$
{\cal F}_A
= {1\over 2} e^{-\rho}{\dot\phi^2\over\dot\rho^4} \bigl[ 
\partial_k\zeta \partial^k\zeta \partial_l\partial^l\zeta 
- \zeta \partial_k\partial_l \bigl( \partial^k\zeta \partial^l\zeta \bigr)
\bigr] , 
$$
after a few more spatial integrations, in anticipation of things to come.  In the end, the $e^{-\rho}$ terms in the action are just left with 
$$
2{\cal L}^{(3)}\bigr|_{e^{-\rho}}
= {\cal F}_A + {\cal D}_A ; 
$$
${\cal D}_A$, like ${\cal D}_0$ before it, contains all the total derivative terms, 
\begin{eqnarray}
{\cal D}_A
&\!\!\!=\!\!\!& 
{1\over 3} {d\over dt} \biggl\{ 
- e^{-\rho}{1\over\dot\rho^3} \zeta \bigl[ 
\partial_k\partial_l\zeta \partial^k\partial^l\zeta 
- \bigl( \partial_k\partial^k\zeta \bigr)^2 
\bigr] \biggr\}
\nonumber \\
&&
- e^{-\rho} {1\over\dot\rho^4} \biggl[ \ddot\rho + {1\over 2}\dot\phi^2 \biggr]
\zeta \bigl[ 
\partial_k\partial_l\zeta \partial^k\partial^l\zeta 
- \bigl( \partial_k\partial^k\zeta \bigr)^2 
\bigr] 
\nonumber \\
&&
+ {2\over 3} \partial_k \biggl\{ 
e^{-\rho}{1\over\dot\rho^3} \dot\zeta \bigl[ 
\partial^k\zeta \partial_l\partial^l\zeta 
- \partial^l\zeta \partial_l\partial^k\zeta 
\bigr] 
- e^{-\rho}{1\over\dot\rho^3} \zeta \bigl[ 
\partial^k\dot\zeta \partial_l\partial^l\zeta 
- \partial^l\dot\zeta \partial_l\partial^k\zeta 
\bigr] \biggr\}
\nonumber \\
&&
+ {4\over 3} \partial_k \biggl\{ 
e^{-\rho}{1\over\dot\rho^2} \bigl[ 
2 \zeta \partial_l\zeta \partial^k\partial^l\zeta 
- 2 \zeta \partial^k\zeta \partial_l\partial^l\zeta 
-  \partial^k\zeta \partial_l\zeta \partial^l\zeta 
\bigr] 
\biggr\}
\nonumber \\
&&
+ {1\over 2} \partial_k \biggl\{ e^{-\rho}{\dot\phi^2\over\dot\rho^4} 
\bigl[ 
2 \zeta \partial^k\partial^l\zeta \partial_l\zeta 
- \partial^k\zeta \partial^l\zeta \partial_l\zeta 
\bigr] \biggr\} . 
\nonumber 
\end{eqnarray}
\vskip18truept

\noindent{\it IV.B. The $e^\rho$-terms}
\vskip6truept

The next group to be examined is the set of $e^\rho$ terms.  This part of the action encompasses a total of seven interactions, 
\begin{eqnarray}
2{\cal L}^{(3)}\bigr|_{e^\rho}
&\!\!\!=\!\!\!& 
- e^\rho {\dot\phi^2\over\dot\rho^2} \zeta \partial_k\zeta \partial^k\zeta
+ 2 e^\rho {\dot\phi^2\over\dot\rho^3} \dot\zeta \partial_k\zeta \partial^k\zeta
+ 3 e^\rho {\dot\phi^2\over\dot\rho^3} \dot\zeta \zeta \partial_k\partial^k\zeta
- e^\rho {\dot\phi^2\over\dot\rho^4} \dot\zeta^2 \partial_k\partial^k\zeta
\nonumber \\
&&
- 6 e^\rho {1\over\dot\rho} \zeta \partial_k\partial_l\zeta \partial^k\partial^l\chi
+ 2 e^\rho {1\over\dot\rho^2} \dot\zeta \partial_k\partial_l\zeta \partial^k\partial^l\chi
+ 4 e^\rho {1\over\dot\rho} \partial_k\zeta \partial^k\chi \partial_l\partial^l\zeta . 
\nonumber 
\end{eqnarray}
Though the number of these terms is not much greater than that of the $e^{-\rho}$ set just considered, their analysis is far more complicated.

The strategy for treating this set will be to concentrate on the second line first---the three terms that contain a $\chi$ field.  The middle term differs from the other two since it contains a time derivative acting on one of the $\zeta$ fields, which will be integrated by parts.  After doing so, some of the spatial derivatives will also be integrated, as the spatial derivatives are acting on different fields in the case of each of the terms.  Only once the $\chi$ terms have been thoroughly analysed and put into a more useful form for later will the first line of the ${\cal L}^{(3)}\bigr|_{e^\rho}$ Lagrangian be included among the rest.

So beginning with the $\chi$ term that has the $\dot\zeta$ factor, integrate it by parts until none of the terms with {\it four\/} spatial derivatives contain a factor of $\dot\zeta$, 
\begin{eqnarray}
2 e^\rho {1\over\dot\rho^2} \dot\zeta \partial_k\partial_l\zeta \partial^k\partial^l\chi
&\!\!\!=\!\!\!& 
- {e^\rho\over\dot\rho^2} \zeta \partial_k\partial_l\zeta \partial^k\partial^l\dot\chi 
- {e^\rho\over\dot\rho} \zeta \partial_k\partial_l\zeta \partial^k\partial^l\chi 
- e^\rho {\dot\phi^2\over\dot\rho^3} \zeta \partial_k\partial_l\zeta \partial^k\partial^l\chi 
\nonumber \\
&&
+ {1\over 2} e^\rho {\dot\phi^2\over\dot\rho^4} \zeta \partial_k\dot\zeta \partial^k\dot\zeta  
- {1\over 2} e^\rho {\dot\phi^2\over\dot\rho^4} \dot\zeta \partial_k\zeta \partial^k\dot\zeta 
\nonumber \\
&& 
+ {d\over dt} \biggl\{ 
{e^\rho\over\dot\rho^2} \zeta \partial_k\partial_l\zeta \partial^k\partial^l\chi \biggr\}
+ 2 e^\rho {1\over\dot\rho^3} \biggl[ \ddot\rho + {1\over 2} \dot\phi^2 \biggr]  \zeta \partial_k\partial_l\zeta \partial^k\partial^l\chi 
\nonumber \\
&& 
+ \partial_k \biggl\{ 
e^\rho {1\over\dot\rho^2} \bigl[ 
\dot\zeta \partial_l\zeta - \zeta \partial_l\dot\zeta 
\bigr] \partial^k\partial^l\chi \biggr\} . 
\nonumber 
\end{eqnarray}
In deriving this expansion, the constraint condition,
$$
\partial_k\partial^k\chi = {1\over 2} {\dot\phi^2\over\dot\rho^2} \dot\zeta , 
$$
has been used quite freely.  The last term of the first line of this expansion can itself be rewritten as 
\begin{eqnarray}
- e^\rho {\dot\phi^2\over\dot\rho^3} \zeta \partial_k\partial_l\zeta \partial^k\partial^l\chi 
&\!\!\!=\!\!\!& 
- e^\rho {\dot\phi^2\over\dot\rho^3} \bigl[ 
\partial_k\zeta \partial^k\chi \partial_l\partial^l\zeta 
- \zeta \partial_k\partial_l \bigl( \partial^k\zeta \partial^l\chi \bigr)
\bigr]
- {e^\rho\over 2} {\dot\phi^4\over\dot\rho^5} \dot\zeta \zeta \partial_k\partial^k\zeta 
\nonumber \\
&& 
- \partial_k \biggl\{ 
e^\rho {\dot\phi^2\over\dot\rho^3} \bigl[ 
\zeta \partial_l \bigl( \partial^k\zeta \partial^l\chi \bigr) 
+ \zeta \partial^k\partial^l\zeta \partial_l\chi 
- \partial_l \bigl( \zeta \partial^l\zeta \bigr) \partial^k\chi 
\bigr] \biggr\} . 
\nonumber 
\end{eqnarray}
Returning to the $e^\rho$ Lagrangian, its last term can be put in a form that more closely resembles the others by reordering its spatial derivatives,
\begin{eqnarray}
4 e^\rho {1\over\dot\rho} \partial_k\zeta \partial^k\chi \partial_l\partial^l\zeta 
&\!\!\!=\!\!\!& 
4 e^\rho {1\over\dot\rho} \zeta \partial_k\partial_l\zeta \partial^k\partial^l\chi 
+ 2 e^\rho {\dot\phi^2\over\dot\rho^3} \zeta \partial_k\zeta \partial^k\dot\zeta 
+ e^\rho {\dot\phi^2\over\dot\rho^3} \dot\zeta \partial_k\zeta \partial^k\zeta 
\nonumber \\
&& 
+ \partial_k \biggl\{ 
2 e^\rho {1\over\dot\rho} \bigl[ 
2 \partial^k\zeta \partial_l\zeta \partial^l\chi 
- 2 \zeta \partial_l\zeta \partial^k\partial^l\chi 
- \partial_l\zeta \partial^l\zeta \partial^k\chi 
\bigr] \biggr\} . 
\nonumber 
\end{eqnarray}
Assembling all the expanded versions of the $\chi$-terms found so far produces a rather lengthy expression,
\begin{eqnarray}
&&\!\!\!\!\!\!\!\!\!\!\!\!\!\!\!\!\!\!\!\!\!\!\!\!\!\!\!\!\!\!\!
-6  e^\rho {1\over\dot\rho} \zeta \partial_k\partial_l\zeta \partial^k\partial^l\chi 
+ 2 e^\rho {1\over\dot\rho^2} \dot\zeta \partial_k\partial_l\zeta \partial^k\partial^l\chi
+ 4 e^\rho {1\over\dot\rho} \partial_k\zeta \partial^k\chi \partial_l\partial^l\zeta 
\nonumber \\
&\!\!\!=\!\!\!& 
- e^\rho {1\over\dot\rho^2} \zeta \partial_k\partial_l\zeta \partial^k\partial^l\dot\chi 
- 3 e^\rho {1\over\dot\rho} \zeta \partial_k\partial_l\zeta \partial^k\partial^l\chi 
\nonumber \\
&& 
- e^\rho {\dot\phi^2\over\dot\rho^3} \bigl[ 
\partial_k\zeta \partial^k\chi \partial_l\partial^l\zeta 
- \zeta \partial_k\partial_l \bigl( \partial^k\zeta \partial^l\chi \bigr)
\bigr]
\nonumber \\
&& 
+ 2 e^\rho {\dot\phi^2\over\dot\rho^3} \zeta \partial_k\zeta \partial^k\dot\zeta 
+ e^\rho {\dot\phi^2\over\dot\rho^3} \dot\zeta \partial_k\zeta \partial^k\zeta 
- {1\over 2} e^\rho {\dot\phi^4\over\dot\rho^5} \dot\zeta \zeta \partial_k\partial^k\zeta 
\nonumber \\
&&
+ {1\over 2} e^\rho {\dot\phi^2\over\dot\rho^4} \zeta \partial_k\dot\zeta \partial^k\dot\zeta  
- {1\over 2} e^\rho {\dot\phi^2\over\dot\rho^4} \dot\zeta \partial_k\zeta \partial^k\dot\zeta 
\nonumber \\
&& 
+ {d\over dt} \biggl\{ 
e^\rho {1\over\dot\rho^2} \zeta \partial_k\partial_l\zeta \partial^k\partial^l\chi \biggr\}
+ 2 e^\rho {1\over\dot\rho^3} \biggl[ \ddot\rho + {1\over 2} \dot\phi^2 \biggr]  \zeta \partial_k\partial_l\zeta \partial^k\partial^l\chi 
\nonumber \\
&& 
- \partial_k \biggl\{ 
e^\rho {\dot\phi^2\over\dot\rho^3} \bigl[ 
\zeta \partial_l \bigl( \partial^k\zeta \partial^l\chi \bigr) 
+ \zeta \partial^k\partial^l\zeta \partial_l\chi 
- \partial_l \bigl( \zeta \partial^l\zeta \bigr) \partial^k\chi 
\bigr] \biggr\} 
\nonumber \\
&& 
+ \partial_k \biggl\{ 
2 e^\rho {1\over\dot\rho} \bigl[ 
2 \partial^k\zeta \partial_l\zeta \partial^l\chi 
- 2 \zeta \partial_l\zeta \partial^k\partial^l\chi 
- \partial_l\zeta \partial^l\zeta \partial^k\chi 
\bigr] \biggr\} 
\nonumber \\
&& 
+ \partial_k \biggl\{ 
e^\rho {1\over\dot\rho^2} \bigl[ 
\dot\zeta \partial_l\zeta - \zeta \partial_l\dot\zeta 
\bigr] \partial^k\partial^l\chi \biggr\} . 
\nonumber  
\end{eqnarray}
Before including the rest of the order $e^\rho$ Lagrangian, integrate the terms on the first line of this equation by parts until they assume the following form,
\begin{eqnarray}
&&\!\!\!\!\!\!\!\!\!\!\!\!\!\!\!\!\!\!\!\!\!\!\!\!\!\!\!\!\!\!\!
- e^\rho {1\over\dot\rho^2} \zeta \partial_k\partial_l\zeta \partial^k\partial^l\dot\chi 
- 3 e^\rho {1\over\dot\rho} \zeta \partial_k\partial_l\zeta \partial^k\partial^l\chi 
\nonumber \\
&\!\!\!=\!\!\!& 
- e^\rho {1\over\dot\rho^2} \partial_k\zeta \partial^k\zeta 
\partial_l\partial^l\bigl[ \dot\chi + 3 \dot\rho \chi \bigr] 
+ e^\rho {1\over\dot\rho^2} \bigl[ \dot\chi + 3 \dot\rho \chi \bigr] 
\partial_k\partial_l \bigl( \partial^k\zeta \partial^l\zeta \bigr) 
\nonumber \\
&& 
- e^\rho {1\over\dot\rho^2} \zeta 
\partial_k\partial^k\zeta \partial_l\partial^l\dot\chi 
- {3\over 2} e^\rho {\dot\phi^2\over\dot\rho^3} \dot\zeta \zeta
\partial_k\partial^k\zeta 
\nonumber \\
&& 
- \partial_k \biggl\{ 
e^\rho {1\over\dot\rho^2} \bigl[ 
\zeta \partial_l\zeta \partial^k\partial^l\dot\chi 
- \partial^k\zeta \partial_l\zeta \partial^l\dot\chi 
+ \dot\chi \partial_l \bigl( \partial^k\zeta \partial^l\zeta \bigr) 
\bigr] \biggr\} 
\nonumber \\
&& 
- \partial_k \biggl\{ 
3 e^\rho {1\over\dot\rho} \bigl[ 
\zeta \partial_l\zeta \partial^k\partial^l\chi 
- \partial^k\zeta \partial_l\zeta \partial^l\chi 
+ \chi \partial_l \bigl( \partial^k\zeta \partial^l\zeta \bigr) 
\bigr] \biggr\} 
\nonumber \\
&& 
+ \partial_k \biggl\{ e^\rho {1\over\dot\rho^2} \zeta 
\partial^k\zeta \partial_l\partial^l\dot\chi \biggr\}
+ \partial_k \biggl\{ {3\over 2} e^\rho {\dot\phi^2\over\dot\rho^3} 
\dot\zeta \zeta \partial^k\zeta \biggr\} . 
\nonumber  
\end{eqnarray}
Again, the constraint equation has been applied to rewrite $\partial_k\partial^k\chi$ in terms of $\dot\zeta$.  Next, formally invert the spatial Laplacian operator, $\partial^{-2}(\partial_k\partial^k) = 1$, in the constraint so that it becomes
$$
\chi = {1\over 2} {\dot\phi^2\over\dot\rho^2} \partial^{-2}\dot\zeta . 
$$
This form of the constraint allows the first line of the equation before the last to be written as  
\begin{eqnarray}
&&\!\!\!\!\!\!\!\!\!\!\!\!\!\!\!\!\!\!\!\!\!\!\!
- e^\rho {1\over\dot\rho^2} \partial_k\zeta \partial^k\zeta 
\partial_l\partial^l\bigl[ \dot\chi + 3 \dot\rho \chi \bigr] 
+ e^\rho {1\over\dot\rho^2} \bigl[ \dot\chi + 3 \dot\rho \chi \bigr] 
\partial_k\partial_l \bigl( \partial^k\zeta \partial^l\zeta \bigr) 
\nonumber \\
&\!\!\!=\!\!\!&  
- {1\over 2} {e^{-2\rho}\over\dot\rho^2} 
{d\over dt} \biggl\{ e^{3\rho} {\dot\phi^2\over\dot\rho^2} \dot\zeta \biggr\} 
\bigl[ \partial_k\zeta \partial^k\zeta 
- \partial^{-2} \partial_k\partial_l \bigl( \partial^k\zeta \partial^l\zeta \bigr) \bigr] , 
\nonumber 
\end{eqnarray}
after applying the identity, 
$$
\biggl( {d\over dt} + 3 \dot\rho \biggr) 
\biggl[ {\dot\phi^2\over\dot\rho^2} \dot\zeta \biggr] 
= e^{-3\rho} {d\over dt} 
\biggl\{ e^{3\rho} {\dot\phi^2\over\dot\rho^2} \dot\zeta \biggr\} , 
$$
and formally integrating the $\partial^{-2}$ operator `by parts.'  What this means can be best seen by considering two general functions $f(\vec x)$ and $g(\vec x)$ and expanding them in Fourier modes.  When acting on one of them with $\partial^{-2}$ and integrating over all spatial positions, the result is 
\begin{eqnarray}
\int d^3\vec x\, f(\vec x)\partial^{-2} g(\vec x) 
&\!\!\!\!\!\!=\!\!\!\!\!\!& 
\int d^3\vec x \int {d^3\vec k_1\over (2\pi)^3} {d^3\vec k_2\over (2\pi)^3}\, 
e^{i\vec k_1\cdot\vec x} f_{\vec k_1} 
\partial^{-2} \bigl( e^{i\vec k_2\cdot\vec x} g_{\vec k_2} \bigr)
\nonumber \\
&\!\!\!\!\!\!=\!\!\!\!\!\!& 
- \int {d^3\vec k_1\over (2\pi)^3} {d^3\vec k_2\over (2\pi)^3} \int d^3\vec x\,
e^{i(\vec k_1+\vec k_2)\cdot\vec x} f_{\vec k_1} 
{1\over k_2^2} g_{\vec k_2} 
\nonumber \\
&\!\!\!\!\!\!=\!\!\!\!\!\!& 
- \int {d^3\vec k_1\over (2\pi)^3} {d^3\vec k_2\over (2\pi)^3}\, 
(2\pi)^3 \delta^3(\vec k_1+\vec k_2)\, {1\over k_2^2} f_{\vec k_1}g_{\vec k_2} . 
\nonumber
\end{eqnarray}
The result of acting on the $f(\vec x)$ instead yields 
$$
\int d^3\vec x\, \bigl(\partial^{-2}f(\vec x)\bigr) g(\vec x) 
= - \int {d^3\vec k_1\over (2\pi)^3} {d^3\vec k_2\over (2\pi)^3}\, 
(2\pi)^3 \delta^3(\vec k_1+\vec k_2)\, {1\over k_1^2} f_{\vec k_1}g_{\vec k_2} . 
$$
But the $\delta$-function means that $k_1^2=k_2^2$.  Thus, 
$$
f(\vec x)\partial^{-2} g(\vec x) 
= \bigl(\partial^{-2}f(\vec x)\bigr) g(\vec x) , 
$$
where the integration over all spatial positions is implied.

At last it is possible to assemble all the rest of the order $e^\rho$ terms (those that did not contain the $\chi$ field) together with what was derived so far to find
\begin{eqnarray}
2{\cal L}^{(3)}\bigr|_{e^\rho}
&\!\!\!=\!\!\!& 
- {1\over 2} {e^{-2\rho}\over\dot\rho^2} 
{d\over dt} \biggl\{ e^{3\rho} {\dot\phi^2\over\dot\rho^2} \dot\zeta \biggr\} 
\bigl[ \partial_k\zeta \partial^k\zeta 
- \partial^{-2} \partial_k\partial_l \bigl( \partial^k\zeta \partial^l\zeta \bigr) \bigr]
\nonumber \\
&&
- e^\rho {\dot\phi^2\over\dot\rho^3} \bigl[ 
\partial_k\zeta \partial^k\chi \partial_l\partial^l\zeta 
- \zeta \partial_k\partial_l \bigl( \partial^k\zeta \partial^l\chi \bigr)
\bigr]
\nonumber \\
&& 
- e^\rho {\dot\phi^2\over\dot\rho^2} \zeta \partial_k\zeta \partial^k\zeta
+ 3 e^\rho {\dot\phi^2\over\dot\rho^3} \dot\zeta \partial_k\zeta \partial^k\zeta
+ {3\over 2} e^\rho {\dot\phi^2\over\dot\rho^3} \dot\zeta \zeta
\partial_k\partial^k\zeta 
\nonumber \\
&&
+ 2 e^\rho {\dot\phi^2\over\dot\rho^3} \zeta \partial_k\zeta \partial^k\dot\zeta 
- {1\over 2} e^\rho {\dot\phi^4\over\dot\rho^5} \dot\zeta \zeta \partial_k\partial^k\zeta 
- e^\rho {1\over\dot\rho^2} \zeta 
\partial_k\partial^k\zeta \partial_l\partial^l\dot\chi 
\nonumber \\
&&
- e^\rho {\dot\phi^2\over\dot\rho^4} \dot\zeta^2 \partial_k\partial^k\zeta
+ {1\over 2} e^\rho {\dot\phi^2\over\dot\rho^4} \zeta \partial_k\dot\zeta \partial^k\dot\zeta  
- {1\over 2} e^\rho {\dot\phi^2\over\dot\rho^4} \dot\zeta \partial_k\zeta \partial^k\dot\zeta 
+ \cdots , 
\nonumber  
\end{eqnarray}
plus a great many total derivative terms which will be shown explicitly at the end.  After integrating everything on the last three lines by parts, one learns that the order $e^\rho$ part of the Lagrangian contains one dynamical term that is manifestly of order $\epsilon^2$,
$$
2{\cal L}^{(3)}\bigr|_{e^\rho}
= {1\over 2} e^\rho {\dot\phi^4\over\dot\rho^4} 
\zeta \partial_k\zeta \partial^k\zeta 
+ {\cal F}_B + {\cal D}_B , 
$$
another dynamical contribution that will appear in the terms proportional to the linear equation of motion,   
\begin{eqnarray}
{\cal F}_B 
&\!\!\!=\!\!\!& 
- {1\over 2} {e^{-2\rho}\over\dot\rho^2} 
{d\over dt} \biggl\{ e^{3\rho} {\dot\phi^2\over\dot\rho^2} \dot\zeta \biggr\} 
\bigl[ \partial_k\zeta \partial^k\zeta 
- \partial^{-2} \partial_k\partial_l \bigl( \partial^k\zeta \partial^l\zeta \bigr) \bigr]
\nonumber \\
&& 
- e^\rho {\dot\phi^2\over\dot\rho^2} \biggl[ 
{\ddot\phi\over\dot\phi\dot\rho} \zeta^2 
+ {1\over 2} {\dot\phi^2\over\dot\rho^2} \zeta^2 
+ {2\over\dot\rho} \dot\zeta \zeta 
\biggr] \partial_k\partial^k\zeta
\nonumber \\ 
&& 
- e^\rho {\dot\phi^2\over\dot\rho^3} \bigl[ 
\partial_k\zeta \partial^k\chi \partial_l\partial^l\zeta 
- \zeta \partial_k\partial_l \bigl( \partial^k\zeta \partial^l\chi \bigr)
\bigr] , 
\nonumber  
\end{eqnarray}
and many, many non-dynamical ones,
\begin{eqnarray}
{\cal D}_B
&\!\!\!=\!\!\!& 
{d\over dt} \biggl\{ 
- {1\over 2} e^\rho {\dot\phi^2\over\dot\rho^4} 
\dot\zeta \zeta \partial_k\partial^k\zeta 
+ {1\over 2} e^\rho {\dot\phi^2\over\dot\rho^3} 
\zeta^2 \partial_k\partial^k\zeta 
+ e^\rho {1\over\dot\rho^2} \zeta \partial_k\partial_l\zeta \partial^k\partial^l\chi \biggr\}
\nonumber \\
&& 
+ e^\rho \biggl[ \ddot\rho + {1\over 2} \dot\phi^2 \biggr] 
\biggl\{ 
- {\dot\phi^2\over\dot\rho^5} \dot\zeta \zeta \partial_k\partial^k\zeta 
+ {3\over 2} {\dot\phi^2\over\dot\rho^4} \zeta^2 \partial_k\partial^k\zeta 
+ {2\over\dot\rho^3} \zeta \partial_k\partial_l\zeta \partial^k\partial^l\chi 
\biggr\} 
\nonumber \\
&& 
- \partial_k \biggl\{ 
{1\over 2} e^\rho {\dot\phi^2\over\dot\rho^2} \zeta^2 \partial^k\zeta \biggr\} 
- \partial_k \biggl\{ 
{1\over 4} e^\rho {\dot\phi^4\over\dot\rho^4} \zeta^2 \partial^k\zeta \biggr\} 
\nonumber \\
&& 
+ \partial_k \biggl\{ {9\over 2} e^\rho {\dot\phi^2\over\dot\rho^3} 
\dot\zeta \zeta \partial^k\zeta \biggr\} 
- \partial_k \biggl\{ 
{1\over 2} e^\rho {\dot\phi^2\over\dot\rho^3} \zeta^2 \partial^k\dot\zeta 
\biggr\} 
\nonumber \\
&& 
+ \partial_k \biggl\{ 
{1\over 2} e^\rho {\dot\phi^2\over\dot\rho^4} \dot\zeta \bigl[ 
\zeta \partial^k\dot\zeta - \dot\zeta \partial^k\zeta \bigr] \biggr\} 
+ \partial_k \biggl\{ 
e^\rho {1\over\dot\rho^2} \bigl[ 
\dot\zeta \partial_l\zeta - \zeta \partial_l\dot\zeta 
\bigr] \partial^k\partial^l\chi \biggr\} 
\nonumber \\
&& 
- \partial_k \biggl\{ 
e^\rho {1\over\dot\rho^2} \bigl[ 
\zeta \partial_l\zeta \partial^k\partial^l\dot\chi 
- \partial^k\zeta \partial_l \bigl( \zeta \partial^l\dot\chi \bigr)
+ \dot\chi \partial_l \bigl( \partial^k\zeta \partial^l\zeta \bigr) 
\bigr] \biggr\} 
\nonumber \\
&& 
+ \partial_k \biggl\{ 
e^\rho {1\over\dot\rho} \bigl[ 
7 \partial_l\zeta 
\bigl( \partial^k\zeta \partial^l\chi - \zeta \partial^k\partial^l\chi \bigr) 
- 2 \partial_l\zeta \partial^l\zeta \partial^k\chi 
- 3 \chi \partial_l \bigl( \partial^k\zeta \partial^l\zeta \bigr) 
\bigr] \biggr\} 
\nonumber \\
&& 
- \partial_k \biggl\{ 
e^\rho {\dot\phi^2\over\dot\rho^3} \bigl[ 
\zeta \partial_l \bigl( \partial^k\zeta \partial^l\chi \bigr) 
+ \zeta \partial^k\partial^l\zeta \partial_l\chi 
- \partial_l \bigl( \zeta \partial^l\zeta \bigr) \partial^k\chi 
\bigr] \biggr\} .
\nonumber  
\end{eqnarray}
\vskip18truept

\noindent{\it IV.C. The first $e^{3\rho}$ term}
\vskip6truept

There remain only the $e^{3\rho}$ terms.  This set will be broken in two, with one of the terms being treated by itself,   
$$
e^{3\rho} {\dot\phi^2\over\dot\rho^2} 
\biggl( 3 \zeta - {\dot\zeta\over\dot\rho} \biggr) \dot\zeta^2 . 
$$
It produces the leading contribution to the three-point function that does not contain any spatial derivatives.  Superficially it is proportional to $\epsilon$; but this appearance too is deceptive.

Start this time by integrating {\it one\/} of the $\dot\zeta$ factors in the second term by parts,
\begin{eqnarray}
e^{3\rho} {\dot\phi^2\over\dot\rho^2} 
\biggl( 3 \zeta - {\dot\zeta\over\dot\rho} \biggr) \dot\zeta^2 
&\!\!\!\!=\!\!\!\!& 
3 e^{3\rho} {\dot\phi^2\over\dot\rho^2} \dot\zeta^2 \zeta 
- e^{3\rho} {\dot\phi^2\over\dot\rho^3} \dot\zeta^2 {d\over dt} \zeta
\nonumber \\
&\!\!\!\!=\!\!\!\!& 
e^{3\rho} {\dot\phi^2\over\dot\rho^2} \biggl[ 
\ddot\zeta + {\ddot\phi\over\dot\phi} \dot\zeta 
+ {3\over 4} {\dot\phi^2\over\dot\rho} \dot\zeta 
+ 3 \dot\rho \dot\zeta 
\biggr] 
{2\over\dot\rho} \dot\zeta \zeta 
\nonumber \\
&& 
- {d\over dt} \biggl\{ e^{3\rho} {\dot\phi^2\over\dot\rho^2} \dot\zeta^2\zeta 
\biggr\} 
- 3 e^{3\rho} {\dot\phi^2\over\dot\rho^4} 
\biggl[ \ddot\rho + {1\over 2} \dot\phi^2 \biggr] \dot\zeta^2\zeta . 
\nonumber 
\end{eqnarray}
The dynamical term looks a bit like the time-derivative piece of the $\zeta$ equation of motion, 
$$
e^{3\rho} {\dot\phi^2\over\dot\rho^2} \biggl[ 
\ddot\zeta + 2 {\ddot\phi\over\dot\phi} \dot\zeta 
+ {\dot\phi^2\over\dot\rho} \dot\zeta 
+ 3 \dot\rho \dot\zeta 
\biggr] 
= {d\over dt} \biggl\{ e^{3\rho} {\dot\phi^2\over\dot\rho^2} \dot\zeta \biggr\} 
+ 2 e^{3\rho} {\dot\phi^2\over\dot\rho^3} 
\biggl[ \ddot\rho + {1\over 2} \dot\phi^2 \biggr] \dot\zeta ,
$$
though with a few differences in some of the coefficients of the terms in the brackets.  Adding and subtracting the necessary pieces yields 
\begin{eqnarray}
e^{3\rho} {\dot\phi^2\over\dot\rho^2} 
\biggl( 3 \zeta - {\dot\zeta\over\dot\rho} \biggr) \dot\zeta^2 
&\!\!\!\!=\!\!\!\!& 
{d\over dt} \biggl\{ e^{3\rho} {\dot\phi^2\over\dot\rho^2} \dot\zeta \biggr\} 
{2\over\dot\rho} \dot\zeta \zeta 
- 2 e^{3\rho} {\dot\phi^2\over\dot\rho^2} \biggl[ 
{\ddot\phi\over\dot\phi\dot\rho} 
+ {1\over 2} {\dot\phi^2\over\dot\rho^2} \biggr] 
\dot\zeta^2 \zeta 
+ {1\over 2} e^{3\rho} {\dot\phi^4\over\dot\rho^4} \dot\zeta^2 \zeta 
\nonumber \\
&& 
- {d\over dt} \biggl\{ e^{3\rho} {\dot\phi^2\over\dot\rho^2} \dot\zeta^2\zeta 
\biggr\} 
+ e^{3\rho} {\dot\phi^2\over\dot\rho^4} 
\biggl[ \ddot\rho + {1\over 2} \dot\phi^2 \biggr] \dot\zeta^2\zeta . 
\nonumber 
\end{eqnarray}
It might not be immediately obvious why the last term on the first line was explicitly extracted, since it is already proportional to one of the pieces in the second term, but it will produce a term that matches with one already present on the second line of the definition for ${\cal F}_B$, which contains the spatial-derivative part of the $\zeta$ equation of motion.

Notice that the second of the dynamical terms looks as though it is second order in the slow-roll parameters, although it is actually of even higher order.  To make this fact apparent, integrate a factor $\dot\zeta\zeta$ by parts, so that 
\begin{eqnarray}
- 2 e^{3\rho} {\dot\phi^2\over\dot\rho^2} \biggl[ 
{\ddot\phi\over\dot\phi\dot\rho} 
+ {1\over 2} {\dot\phi^2\over\dot\rho^2} \biggr] 
\dot\zeta^2 \zeta 
&\!\!\!\!=\!\!\!\!& 
- e^{3\rho} {\dot\phi^2\over\dot\rho^2} \dot\zeta \biggl[ 
{\ddot\phi\over\dot\phi\dot\rho} 
+ {1\over 2} {\dot\phi^2\over\dot\rho^2} \biggr] 
{d\over dt} \zeta^2
\nonumber \\
&\!\!\!\!=\!\!\!\!& 
e^{3\rho} {\dot\phi^2\over\dot\rho^2} \dot\zeta \zeta^2
{d\over dt} \biggl[ {\ddot\phi\over\dot\phi\dot\rho} 
+ {1\over 2} {\dot\phi^2\over\dot\rho^2} \biggr] 
\nonumber \\
&& 
+ {d\over dt} \biggl\{ e^{3\rho} {\dot\phi^2\over\dot\rho^2} \dot\zeta \biggr\} 
\biggl[ {\ddot\phi\over\dot\phi\dot\rho} \zeta^2
+ {1\over 2} {\dot\phi^2\over\dot\rho^2} \zeta^2 \biggr] 
\nonumber \\
&& 
- {d\over dt} \biggl\{ 
e^{3\rho} {\dot\phi^2\over\dot\rho^2} 
\biggl[ {\ddot\phi\over\dot\phi\dot\rho} 
+ {1\over 2} {\dot\phi^2\over\dot\rho^2} \biggr] \dot\zeta \zeta^2
\biggr\} . 
\nonumber 
\end{eqnarray}

What has thus been learned is that the first term in the cubic action can be rewritten as 
$$
e^{3\rho} {\dot\phi^2\over\dot\rho^2} 
\biggl( 3 \zeta - {\dot\zeta\over\dot\rho} \biggr) \dot\zeta^2
= {1\over 2} e^{3\rho} {\dot\phi^4\over\dot\rho^4} \dot\zeta^2 \zeta 
+ e^{3\rho} {\dot\phi^2\over\dot\rho^2} \dot\zeta \zeta^2 
{d\over dt} \biggl[ {\ddot\phi\over\dot\phi\dot\rho} 
+ {1\over 2} {\dot\phi^2\over\dot\rho^2} \biggr] 
+ {\cal F}_C + {\cal D}_C , 
$$
where 
$$
{\cal D}_C 
= - {d\over dt} \biggl\{ e^{3\rho} {\dot\phi^2\over\dot\rho^2}  
\biggl[ {\ddot\phi\over\dot\phi\dot\rho} 
+ {1\over 2} {\dot\phi^2\over\dot\rho^2} \biggr] \dot\zeta \zeta^2
+ e^{3\rho} {\dot\phi^2\over\dot\rho^3} \dot\zeta^2 \zeta
\biggr\} 
+ e^{3\rho} {\dot\phi^2\over\dot\rho^4} 
\biggl[ \ddot\rho + {1\over 2} \dot\phi^2 \biggr] \dot\zeta^2 \zeta
$$
is something that has no dynamical effect in an inflationary background, and where 
$$
{\cal F}_C = 
{d\over dt} \biggl\{ e^{3\rho} {\dot\phi^2\over\dot\rho^2} \dot\zeta \biggr\} 
\biggl[ {\ddot\phi\over\dot\phi\dot\rho} \zeta^2 
+ {1\over 2} {\dot\phi^2\over\dot\rho^2} \zeta^2 
+ {2\over\dot\rho} \dot\zeta \zeta
\biggr] 
$$
will contribute yet another few terms proportional to the $\zeta$ equation of motion.
\vskip18truept

\noindent{\it IV.D. The remaining $e^{3\rho}$ terms}
\vskip6truept

There is left just one final set to examine,
\begin{eqnarray}
2{\cal L}^{(3)}\bigr|_{e^{3\rho}} 
&\!\!\!=\!\!\!& 
3 e^{3\rho} \zeta \partial_k\partial_l\chi \partial^k\partial^l\chi 
- e^{3\rho} {1\over\dot\rho} 
\dot\zeta \partial_k\partial_l\chi \partial^k\partial^l\chi 
- {3\over 4} e^{3\rho} {\dot\phi^4\over\dot\rho^4} \dot\zeta^2 \zeta
+ {1\over 4} e^{3\rho} {\dot\phi^4\over\dot\rho^5} \dot\zeta^3
\nonumber \\
&& 
- 2 e^{3\rho} {\dot\phi^2\over\dot\rho^2} \dot\zeta \partial_k\zeta \partial^k\chi , 
\nonumber 
\end{eqnarray}
those proportional to $e^{3\rho}$, aside from the one that was just evaluated.  By now, the strategy for rearranging the terms should more or less be clear, and this same strategy will be applied to this set too.

Begin by integrating the second term by parts to remove the time derivative from the $\dot\zeta$ and then integrate further some of the spatial derivatives to remove them from the term that contains a $\dot\chi$ factor,
\begin{eqnarray}
- e^{3\rho} {1\over\dot\rho} 
\dot\zeta \partial_k\partial_l\chi \partial^k\partial^l\chi 
&\!\!\!\!=\!\!\!\!&
{1\over 2} e^{3\rho} {\dot\phi^2\over\dot\rho^2} 
\zeta \partial_k\partial_l\chi \partial^k\partial^l\chi 
+ 3 e^{3\rho} \zeta \partial_k\partial_l\chi \partial^k\partial^l\chi 
\nonumber  \\
&& 
- 2 e^{3\rho} {1\over\dot\rho} \dot\chi 
\partial_k\partial_l \bigl( \partial^k\zeta \partial^l\chi \bigr)
- 2 e^{3\rho} {1\over\dot\rho} \zeta 
\partial_k\chi \partial^k \partial_l\partial^l\dot\chi 
\nonumber \\ 
&& 
- {d\over dt} \biggl\{ 
e^{3\rho} {1\over\dot\rho} 
\zeta \partial_k\partial_l\chi \partial^k\partial^l\chi \biggr\}
- {e^{3\rho}\over\dot\rho^2} 
\biggl[ \ddot\rho + {1\over 2} \dot\phi^2 \biggr]
\zeta \partial_k\partial_l\chi \partial^k\partial^l\chi 
\nonumber \\
&& 
+ \partial_k \biggl\{
2 e^{3\rho} {1\over\dot\rho} \bigl[ 
\zeta \partial_l\chi \partial^k\partial^l\dot\chi
- \partial^k\zeta \partial_l\chi \partial^l\dot\chi
+ \dot\chi \partial_l \bigl( \partial^l\zeta \partial^k\chi \bigr) \bigr]
\biggr\} . 
\nonumber 
\end{eqnarray}
The second term in this expression is the same as the first term in ${\cal L}^{(3)}\bigr|_{e^{3\rho}}$.  Consider the first two terms of the order $e^{3\rho}$ Lagrangian together, and integrate the spatial derivatives on one of the $\chi$ fields by parts, to produce 
\begin{eqnarray}
&&\!\!\!\!\!\!\!\!\!\!\!\!\!\!\!\!\!\!\!\!\!\!\!\!\!\!\!
3 e^{3\rho} 
\zeta \partial_k\partial_l\chi \partial^k\partial^l\chi 
- e^{3\rho} {1\over\dot\rho} 
\dot\zeta \partial_k\partial_l\chi \partial^k\partial^l\chi 
\nonumber \\
&\!\!\!=\!\!\!&
{1\over 2} e^{3\rho} {\dot\phi^2\over\dot\rho^2} 
\zeta \partial_k\partial_l\chi \partial^k\partial^l\chi 
+ 2 e^{3\rho} {1\over\dot\rho} \zeta 
\partial_k\partial^k\chi \partial_l\partial^l
\bigl[ \dot\chi + 3 \dot\rho \chi \bigr] 
\nonumber \\
&& 
+ 2 e^{3\rho} {1\over\dot\rho} \partial_k \zeta 
\partial^k\chi \partial_l\partial^l
\bigl[ \dot\chi + 3 \dot\rho \chi \bigr] 
- 2 e^{3\rho} {1\over\dot\rho} 
\bigl[ \dot\chi + 3 \dot\rho \chi \bigr] 
\partial_k\partial_l \bigl( \partial^k\zeta \partial^l\chi \bigr)
\nonumber \\ 
&& 
- {d\over dt} \biggl\{ 
e^{3\rho} {1\over\dot\rho} 
\zeta \partial_k\partial_l\chi \partial^k\partial^l\chi \biggr\}
- e^{3\rho} {1\over\dot\rho^2} 
\biggl[ \ddot\rho + {1\over 2} \dot\phi^2 \biggr]
\zeta \partial_k\partial_l\chi \partial^k\partial^l\chi 
\nonumber \\
&& 
+ \partial_k \biggl\{
2 e^{3\rho} {1\over\dot\rho} \bigl[ 
\zeta \partial_l\chi \partial^k\partial^l\dot\chi
- \partial^k\zeta \partial_l\chi \partial^l\dot\chi
+ \dot\chi \partial_l \bigl( \partial^l\zeta \partial^k\chi \bigr) \bigr]
\biggr\}
\nonumber \\
&& 
+ \partial_k \biggl\{
6 e^{3\rho} \bigl[ 
\zeta \partial_l\chi \partial^k\partial^l\chi
- \partial^k\zeta \partial_l\chi \partial^l\chi
+ \chi \partial_l \bigl( \partial^l\zeta \partial^k\chi \bigr) \bigr]
\biggr\}
\nonumber \\
&& 
- \partial_k \biggl\{ 
2 e^{3\rho} {1\over\dot\rho} \zeta 
\partial^k\chi \partial_l\partial^l
\bigl[ \dot\chi + 3 \dot\rho \chi \bigr] \biggr\} . 
\nonumber
\end{eqnarray}
Now apply the same trick that was used for the order $e^\rho$ terms, where some of the $\chi$ fields were replaced with
$$
\chi = {1\over 2} {\dot\phi^2\over\dot\rho^2} \partial^{-2}\dot\zeta 
$$
and then integrate the $\partial^{-2}$ operators by parts to remove them from the time derivative of the $\zeta$ field, to arrive at
\begin{eqnarray}
&&\!\!\!\!\!\!\!\!\!\!\!\!\!\!\!\!\!\!\!\!\!\!\!\!\!\!\!
3 e^{3\rho} 
\zeta \partial_k\partial_l\chi \partial^k\partial^l\chi 
- e^{3\rho} {1\over\dot\rho} 
\dot\zeta \partial_k\partial_l\chi \partial^k\partial^l\chi 
\nonumber \\
&\!\!\!=\!\!\!&
{1\over 2} e^{3\rho} {\dot\phi^2\over\dot\rho^2} 
\zeta \partial_k\partial_l\chi \partial^k\partial^l\chi 
+ {1\over 2} {\dot\phi^2\over\dot\rho^3} \dot\zeta \zeta 
{d\over dt} \biggl\{ e^{3\rho} {\dot\phi^2\over\dot\rho^2} \dot\zeta \biggr\}
\nonumber \\
&& 
+ {1\over\dot\rho} 
{d\over dt} \biggl\{ e^{3\rho} {\dot\phi^2\over\dot\rho^2} \dot\zeta \biggr\}
\bigl[ \partial_k \zeta \partial_k\chi 
- \partial^{-2} \partial_k\partial_l \bigl( \partial^k\zeta \partial^l\chi \bigr) \bigr] 
+ \cdots . 
\nonumber 
\end{eqnarray}

If the last two terms on the first line of ${\cal L}^{(3)}\bigr|_{e^{3\rho}}$ are combined with the second term of this last expression, together they yield 
\begin{eqnarray}
&&\!\!\!\!\!\!\!\!\!\!\!\!\!\!\!\!\!\!\!\!\!\!\!\!\!\!\!
- {3\over 4} e^{3\rho} {\dot\phi^4\over\dot\rho^4} \dot\zeta^2 \zeta
+ {1\over 4} e^{3\rho} {\dot\phi^4\over\dot\rho^5} \dot\zeta^3
+ {1\over 2} {\dot\phi^2\over\dot\rho^3} \dot\zeta \zeta 
{d\over dt} \biggl\{ e^{3\rho} {\dot\phi^2\over\dot\rho^2} \dot\zeta \biggr\}
\nonumber \\
&\!\!\!=\!\!\!&
- {1\over 8} e^{3\rho} {\dot\phi^6\over\dot\rho^6} 
\dot\zeta^2 \zeta
+ {d\over dt} \biggl\{ 
{1\over 4} e^{3\rho} {\dot\phi^4\over\dot\rho^5} \dot\zeta^2 \zeta \biggr\}
+ {1\over 4} e^{3\rho} {\dot\phi^4\over\dot\rho^6} 
\biggl[ \ddot\rho + {1\over 2} \dot\phi^2 \biggr] \dot\zeta^2 \zeta . 
\nonumber 
\end{eqnarray}
Putting all the terms together, one obtains 
$$
2{\cal L}^{(3)}\bigr|_{e^{3\rho}}
= - 2 e^{3\rho} {\dot\phi^2\over\dot\rho^2} \dot\zeta \partial_k\zeta \partial^k\chi
+ {1\over 2} e^{3\rho} {\dot\phi^2\over\dot\rho^2} 
\zeta \partial_k\partial_l\chi \partial^k\partial^l\chi 
- {1\over 8} e^{3\rho} {\dot\phi^6\over\dot\rho^6} 
\dot\zeta^2 \zeta
+ {\cal F}_D + {\cal D}_D , 
$$
with
$$
{\cal F}_D 
= {1\over\dot\rho} 
{d\over dt} \biggl\{ e^{3\rho} {\dot\phi^2\over\dot\rho^2} \dot\zeta \biggr\}
\bigl[ \partial_k \zeta \partial^k\chi 
- \partial^{-2} \partial_k\partial_l \bigl( \partial^k\zeta \partial^l\chi \bigr) \bigr] 
$$
and 
\begin{eqnarray}
{\cal D}_D
&\!\!\!=\!\!\!&
{d\over dt} \biggl\{ 
{1\over 4} e^{3\rho} {\dot\phi^4\over\dot\rho^5} \dot\zeta^2 \zeta \biggr\}
+ {1\over 4} e^{3\rho} {\dot\phi^4\over\dot\rho^6} 
\biggl[ \ddot\rho + {1\over 2} \dot\phi^2 \biggr] \dot\zeta^2 \zeta 
\nonumber \\ 
&& 
- {d\over dt} \biggl\{ 
e^{3\rho} {1\over\dot\rho} 
\zeta \partial_k\partial_l\chi \partial^k\partial^l\chi \biggr\}
- e^{3\rho} {1\over\dot\rho^2} 
\biggl[ \ddot\rho + {1\over 2} \dot\phi^2 \biggr]
\zeta \partial_k\partial_l\chi \partial^k\partial^l\chi 
\nonumber \\
&& 
+ \partial_k \biggl\{
2 e^{3\rho} {1\over\dot\rho} \bigl[ 
\zeta \partial_l\chi \partial^k\partial^l\dot\chi
- \partial^k\zeta \partial_l\chi \partial^l\dot\chi
+ \dot\chi \partial_l \bigl( \partial^l\zeta \partial^k\chi \bigr) \bigr]
\biggr\}
\nonumber \\
&& 
+ \partial_k \biggl\{
6 e^{3\rho} \bigl[ 
\zeta \partial_l\chi \partial^k\partial^l\chi
- \partial^k\zeta \partial_l\chi \partial^l\chi
+ \chi \partial_l \bigl( \partial^l\zeta \partial^k\chi \bigr) \bigr]
\biggr\}
\nonumber \\
&& 
- \partial_k \biggl\{ 
2 e^{3\rho} {1\over\dot\rho} \zeta 
\partial^k\chi \partial_l\partial^l
\bigl[ \dot\chi + 3 \dot\rho \chi \bigr] \biggr\} . 
\nonumber 
\end{eqnarray}
Note that the second cubic term was not altered at all from its form in the original interaction.
\vskip18truept

\noindent{\it IV.E. Reassembling the cubic action}
\vskip6truept

Finally, the results for each of these separate calculations for the different pieces of the cubic action must be reassembled to see what is its true size in terms of the slow-roll parameters.  Most importantly, it is still necessary to show that when all of the ${\cal F}$-terms are combined, they really are proportional to the variation of the quadratic action.  Adding together the results of the four groupings of the cubic terms yields the following set of interactions
\begin{eqnarray}
S^{(3)} &\!\!\!=\!\!\!& 
{1\over 2} \int d^4x\, \biggl\{
{1\over 2} e^{3\rho} {\dot\phi^4\over\dot\rho^4} \dot\zeta^2 \zeta 
+ {1\over 2} e^\rho {\dot\phi^4\over\dot\rho^4} 
\zeta \partial_k\zeta \partial^k\zeta 
- 2 e^{3\rho} {\dot\phi^2\over\dot\rho^2} \dot\zeta \partial_k\zeta \partial^k\chi
\nonumber \\
&&\qquad\quad
+ e^{3\rho} {\dot\phi^2\over\dot\rho^2} \dot\zeta \zeta^2 
{d\over dt} \biggl[ {\ddot\phi\over\dot\phi\dot\rho} 
+ {1\over 2} {\dot\phi^2\over\dot\rho^2} \biggr] 
- {1\over 8} e^{3\rho} {\dot\phi^6\over\dot\rho^6} 
\dot\zeta^2 \zeta
+ {1\over 2} e^{3\rho} {\dot\phi^2\over\dot\rho^2} 
\zeta \partial_k\partial_l\chi \partial^k\partial^l\chi 
\nonumber \\
&&\qquad\quad
+ {\cal F} + {\cal D} \biggl\} , 
\nonumber 
\end{eqnarray}
where ${\cal D}$ represents the accumulated total derivative terms, 
$$
{\cal D} = {\cal D}_0 + {\cal D}_A + {\cal D}_B + {\cal D}_C + {\cal D}_D , 
$$
together with the terms the vanish when $\ddot\rho = - {1\over 2}\dot\phi^2$.

Analogously, ${\cal F}$ represents the sum terms that are proportional to  
$$
{d\over dt} \biggl[ e^{3\rho} {\dot\phi^2\over\dot\rho^2} \dot\zeta \biggr] 
- e^\rho {\dot\phi^2\over\dot\rho^2} \partial_k\partial^k\zeta . 
$$
Together, these terms are 
\begin{eqnarray}
{\cal F} 
&\!\!\!=\!\!\!& {\cal F}_A + {\cal F}_B + {\cal F}_C + {\cal F}_D 
\nonumber \\
&\!\!\!=\!\!\!& 
\biggl\{ 
{d\over dt} \biggl[ e^{3\rho} {\dot\phi^2\over\dot\rho^2} \dot\zeta \biggr] 
- e^\rho {\dot\phi^2\over\dot\rho^2} \partial_k\partial^k\zeta 
\biggr\} 
\biggl[ {\ddot\phi\over\dot\phi\dot\rho} \zeta^2 
+ {1\over 2} {\dot\phi^2\over\dot\rho^2} \zeta^2 
+ 2 {1\over\dot\rho} \dot\zeta \zeta
\biggr] 
\nonumber \\
&&
+ {d\over dt} \biggl[ e^{3\rho} {\dot\phi^2\over\dot\rho^2} \dot\zeta \biggr]
{1\over\dot\rho} \bigl[ \partial_k \zeta \partial^k\chi 
- \partial^{-2} \partial_k\partial_l \bigl( \partial^k\zeta \partial^l\chi \bigr) \bigr] 
\nonumber \\
&&
- {d\over dt} \biggl[ e^{3\rho} {\dot\phi^2\over\dot\rho^2} \dot\zeta \biggr] 
{1\over 2} {e^{-2\rho}\over\dot\rho^2} \bigl[ \partial_k\zeta \partial^k\zeta 
- \partial^{-2} \partial_k\partial_l \bigl( \partial^k\zeta \partial^l\zeta \bigr) \bigr]
\nonumber \\
&&
- e^\rho {\dot\phi^2\over\dot\rho^3} \bigl[ 
\partial_k\zeta \partial^k\chi \partial_l\partial^l\zeta 
- \zeta \partial_k\partial_l \bigl( \partial^k\zeta \partial^l\chi \bigr)
\bigr] 
\nonumber \\
&&
+ {1\over 2} e^{-\rho}{\dot\phi^2\over\dot\rho^4} \bigl[ 
\partial_k\zeta \partial^k\zeta \partial_l\partial^l\zeta 
- \zeta \partial_k\partial_l \bigl( \partial^k\zeta \partial^l\zeta \bigr)
\bigr] . 
\nonumber 
\end{eqnarray}
The first line is already proportional to the $\zeta$ equation of motion, but the last two lines do not yet quite match with the two preceding them; but they can be converted by taking two of the $\zeta$'s in them and inserting a factor of the identity operator in the form, $\partial^{-2}\partial_k\partial^k\zeta$, and then integrating by parts one last time so that these lines become, 
\begin{eqnarray}
&\!\!\!=\!\!\!& \cdots 
\nonumber \\
&&
+ \biggl\{ - e^\rho {\dot\phi^2\over\dot\rho^2} \partial_k\partial^k\zeta \biggr\} 
{1\over\dot\rho} \bigl[ \partial_l\zeta \partial^l\chi 
- \partial^{-2}\partial_j\partial_l \bigl( \partial^j\zeta \partial^l\chi \bigr)
\bigr] 
\nonumber \\
&&
- \biggl\{ - e^\rho {\dot\phi^2\over\dot\rho^2} \partial_k\partial^k\zeta 
\biggr\} 
{1\over 2} {e^{-2\rho}\over\dot\rho^2} 
\bigl[ \partial_l\zeta \partial^l\zeta 
- \partial^{-2}\partial_j\partial_l \bigl( \partial^j\zeta \partial^l\zeta \bigr)
\bigr] . 
\nonumber 
\end{eqnarray}
The expression for ${\cal F}$ then becomes, 
\begin{eqnarray}
{\cal F} 
&\!\!\!=\!\!\!& 
\biggl\{ 
{d\over dt} \biggl[ e^{3\rho} {\dot\phi^2\over\dot\rho^2} \dot\zeta \biggr] 
- e^\rho {\dot\phi^2\over\dot\rho^2} \partial_j\partial^j\zeta 
\biggr\}
\biggl[ {\ddot\phi\over\dot\phi\dot\rho} \zeta^2 
+ {1\over 2} {\dot\phi^2\over\dot\rho^2} \zeta^2 
+ 2 {1\over\dot\rho} \dot\zeta \zeta \biggr] 
\nonumber \\
&& 
+ \biggl\{ 
{d\over dt} \biggl[ e^{3\rho} {\dot\phi^2\over\dot\rho^2} \dot\zeta \biggr] 
- e^\rho {\dot\phi^2\over\dot\rho^2} \partial_j\partial^j\zeta 
\biggr\}
{1\over\dot\rho} \bigl[ \partial_k \zeta \partial^k\chi 
- \partial^{-2} \partial_k\partial_l \bigl( \partial^k\zeta \partial^l\chi \bigr) \bigr] 
\nonumber \\
&& 
- \biggl\{ 
{d\over dt} \biggl[ e^{3\rho} {\dot\phi^2\over\dot\rho^2} \dot\zeta \biggr] 
- e^\rho {\dot\phi^2\over\dot\rho^2} \partial_j\partial^j\zeta 
\biggr\}
{1\over 2} {e^{-2\rho}\over\dot\rho^2} \bigl[ \partial_k\zeta \partial^k\zeta 
- \partial^{-2} \partial_k\partial_l \bigl( \partial^k\zeta \partial^l\zeta \bigr) \bigr] . 
\nonumber 
\end{eqnarray}
Collecting all the terms yields
\begin{eqnarray}
{\cal F} 
&\!\!\!=\!\!\!& 
\biggl\{ 
{d\over dt} \biggl[ e^{3\rho} {\dot\phi^2\over\dot\rho^2} \dot\zeta \biggr] 
- e^\rho {\dot\phi^2\over\dot\rho^2} \partial_j\partial^j\zeta 
\biggr\}
\biggl\{ 
{\ddot\phi\over\dot\phi\dot\rho} \zeta^2 
+ {1\over 2} {\dot\phi^2\over\dot\rho^2} \zeta^2 
+ 2 {1\over\dot\rho} \dot\zeta \zeta 
\nonumber \\
&&\qquad\qquad\qquad\qquad\qquad\qquad 
-\,\, {1\over 2} {e^{-2\rho}\over\dot\rho^2} \bigl[ \partial_k\zeta \partial^k\zeta 
- \partial^{-2} \partial_k\partial_l \bigl( \partial^k\zeta \partial^l\zeta \bigr) \bigr] 
\nonumber \\
&&\qquad\qquad\qquad\qquad\qquad\qquad
+\,\, 
{1\over\dot\rho} \bigl[ \partial_k \zeta \partial^k\chi 
- \partial^{-2} \partial_k\partial_l \bigl( \partial^k\zeta \partial^l\chi \bigr) \bigr] 
\biggr\} . 
\nonumber
\end{eqnarray}

So, up to total derivatives, the cubic part of the action for an inflationary theory with a single inflaton field, written in terms of the coordinates where the inflaton has no fluctuations and the independent scalar field $\zeta(t,\vec x)$ corresponds to the fluctuation in the scale factor that multiplies a flat spatial metric, is
\begin{eqnarray}
S^{(3)} &\!\!\!\!\!\!=\!\!\!\!\!\!& 
\int d^4x\, \biggl\{
{1\over 4} {e^{3\rho}\over M_{\rm pl}^2} 
{\dot\phi^4\over\dot\rho^4} \dot\zeta^2 \zeta 
+ {1\over 4} {e^\rho\over M_{\rm pl}^2} {\dot\phi^4\over\dot\rho^4} 
\zeta \partial_k\zeta \partial^k\zeta 
- {1\over 2} {e^{3\rho}\over M_{\rm pl}^2} {\dot\phi^4\over\dot\rho^4} \dot\zeta \partial_k\zeta \partial^k(\partial^{-2}\dot\zeta)
\nonumber \\
&&\qquad\ 
+\,\, {1\over 2} e^{3\rho} {\dot\phi^2\over\dot\rho^2} \dot\zeta \zeta^2 
{d\over dt} \biggl[ {\ddot\phi\over\dot\phi\dot\rho} 
+ {1\over 2} {1\over M_{\rm pl}^2} {\dot\phi^2\over\dot\rho^2} \biggr] 
- {1\over 16} {e^{3\rho}\over M_{\rm pl}^4} {\dot\phi^6\over\dot\rho^6} 
\bigl[ \dot\zeta^2 \zeta
- \zeta \partial_k\partial_l( \partial^{-2}\dot\zeta) \partial^k\partial^l ( \partial^{-2}\dot\zeta)
\bigr]
\nonumber \\
&&\qquad\ 
+\,\, \biggl\{ 
{d\over dt} \biggl[ e^{3\rho} {\dot\phi^2\over\dot\rho^2} \dot\zeta \biggr] 
- e^\rho {\dot\phi^2\over\dot\rho^2} \partial_j\partial^j\zeta 
\biggr\}
\biggl\{ 
{1\over 2} \biggl[{\ddot\phi\over\dot\phi\dot\rho} 
+ {1\over 2} {1\over M_{\rm pl}^2} {\dot\phi^2\over\dot\rho^2} \biggr] \zeta^2 
+ {1\over\dot\rho} \dot\zeta \zeta 
\nonumber \\
&&\qquad\qquad\qquad\qquad\qquad\qquad\qquad\quad\ \ \,
-\,\, {1\over 4} {e^{-2\rho}\over\dot\rho^2} \bigl[ \partial_k\zeta \partial^k\zeta 
- \partial^{-2} \partial_k\partial_l \bigl( \partial^k\zeta \partial^l\zeta \bigr) \bigr] 
\nonumber \\
&&\qquad\qquad\qquad\qquad\qquad\qquad\qquad\quad\ \ \,
+\,\, {1\over 4} {1\over M_{\rm pl}^2} {\dot\phi^2\over\dot\rho^3} 
\bigl[ \partial_k \zeta \partial^k( \partial^{-2}\dot\zeta) 
- \partial^{-2} \partial_k\partial_l \bigl( \partial^k\zeta \partial^l( \partial^{-2}\dot\zeta) \bigr) \bigr] 
\biggr\} . 
\nonumber 
\end{eqnarray}
Here, all of the proper factors of $M_{\rm pl}^2$ have once again been restored.  To see more clearly how each of these interactions scales in the slow-roll parameters, this set of cubic interactions can be expressed equivalently as 
\begin{eqnarray}
S^{(3)} &\!\!\!\!\!\!=\!\!\!\!\!\!& 
M_{\rm pl}^2 \int d^4x\, \biggl\{
\epsilon^2 e^{3\rho} \dot\zeta^2 \zeta 
+ \epsilon^2 e^\rho \zeta \partial_k\zeta \partial^k\zeta 
-  2\epsilon^2 e^{3\rho}\dot\zeta \partial_k\zeta \partial^k(\partial^{-2}\dot\zeta)
\nonumber \\
&&\qquad\quad\ \ \ 
+\,\, \epsilon (\dot\delta + \dot\epsilon) e^{3\rho} \dot\zeta \zeta^2  
- {1\over 2} e^{3\rho} \epsilon^3 
\bigl[ \dot\zeta^2 \zeta
- \zeta \partial_k\partial_l( \partial^{-2}\dot\zeta) \partial^k\partial^l ( \partial^{-2}\dot\zeta)
\bigr]
\nonumber \\
&&\qquad\quad\ \ \ 
+\,\, \biggl\{ 
{d\over dt} \bigl[ \epsilon e^{3\rho} \dot\zeta \bigr] 
- \epsilon e^\rho \partial_j\partial^j\zeta 
\biggr\}
\biggl\{ 
(\delta+\epsilon) \zeta^2 
+ {2\over\dot\rho} \dot\zeta \zeta 
\nonumber \\
&&\qquad\qquad\qquad\qquad\qquad\qquad\qquad\quad\ \  
-\,\, {1\over 2} {e^{-2\rho}\over\dot\rho^2} \bigl[ \partial_k\zeta \partial^k\zeta 
- \partial^{-2} \partial_k\partial_l \bigl( \partial^k\zeta \partial^l\zeta \bigr) \bigr] 
\nonumber \\
&&\qquad\qquad\qquad\qquad\qquad\qquad\qquad\quad\ \ 
+\,\, {1\over\dot\rho} \epsilon 
\bigl[ \partial_k \zeta \partial^k( \partial^{-2}\dot\zeta) 
- \partial^{-2} \partial_k\partial_l \bigl( \partial^k\zeta \partial^l( \partial^{-2}\dot\zeta) \bigr) \bigr] 
\biggr\} . 
\nonumber 
\end{eqnarray}
The operators on the first line are now obviously suppressed by $\epsilon^2$ while those on the next two lines are even more suppressed.  However, several of the terms amongst the ones proportional to the linear equation of motion for $\zeta(t,\vec x)$ are accompanied only by a single factor of $\epsilon$.  They are not as strongly suppressed by the slow-roll parameters as the rest of the terms; they turn out to be unimportant for the three-point function when it is evaluated in the observationally relevant limit, as will be explained in the following section.

Amongst the part proportional to the linear equation of motion for $\zeta$, only the first term---the one containing the factor $(\delta+\epsilon) \zeta^2 $---survives in this limit.  With a few more suitable integrations by parts, this particular term can be combined with some of those on the first two lines,
\begin{eqnarray}
S^{(3)} &\!\!\!\!\!\!=\!\!\!\!\!\!& 
M_{\rm pl}^2 \int d^4x\, \biggl\{
- \epsilon (2\delta+\epsilon) e^{3\rho} \dot\zeta^2 \zeta
+ \epsilon (2\delta+3\epsilon) e^\rho \zeta\partial_k\zeta\partial^k\zeta 
-  2\epsilon^2 e^{3\rho}\dot\zeta \partial_k\zeta \partial^k(\partial^{-2}\dot\zeta)
\nonumber \\
&&\qquad\quad\ \ \ 
-\,\, {1\over 2} e^{3\rho} \epsilon^3 
\bigl[ \dot\zeta^2 \zeta
- \zeta \partial_k\partial_l( \partial^{-2}\dot\zeta) \partial^k\partial^l ( \partial^{-2}\dot\zeta)
\bigr]
\nonumber \\
&&\qquad\quad\ \ \ 
+\,\, \biggl\{ 
{d\over dt} \bigl[ \epsilon e^{3\rho} \dot\zeta \bigr] 
- \epsilon e^\rho \partial_j\partial^j\zeta 
\biggr\}
\biggl\{ 
{2\over\dot\rho} \dot\zeta \zeta 
- {1\over 2} {e^{-2\rho}\over\dot\rho^2} \bigl[ \partial_k\zeta \partial^k\zeta 
- \partial^{-2} \partial_k\partial_l \bigl( \partial^k\zeta \partial^l\zeta \bigr) \bigr] 
\nonumber \\
&&\qquad\qquad\qquad\qquad\qquad\qquad\qquad\quad\ \,
+\,\, {1\over\dot\rho} \epsilon 
\bigl[ \partial_k \zeta \partial^k( \partial^{-2}\dot\zeta) 
- \partial^{-2} \partial_k\partial_l \bigl( \partial^k\zeta \partial^l( \partial^{-2}\dot\zeta) \bigr) \bigr] 
\biggr\} . 
\nonumber 
\end{eqnarray}
Essentially only the first three operators are responsible for the leading non-Gaussian part of the primordial fluctuations.
\vskip24truept

\noindent{\bf\large V. THE THREE-POINT FUNCTION}
\vskip9truept

\noindent 
The cubic action that was just derived determines the standard inflationary prediction for the three-point function of the primordial fluctuations.  A theory that contains any information beyond what is already included in the two-point function is not a Gaussian theory, and the three-point function is usually the best place to look for these non-Gaussian structures.  For the simple model being analysed here, the non-Gaussian parts of the correlation functions are predicted to be quite small.  Most of the terms in the cubic action are manifestly suppressed by a factor of at least $\epsilon^2$.  The only terms where this is not the case occurs among the terms proportional to the equation of motion---what were collectively written as ${\cal F}$.  A few of these terms are accompanied by only one factor of $\epsilon$.  However, it will be seen that these too are suppressed:  the only effects that survive at the end of inflation from the ${\cal F}$ term also contribute at second order in the slow-roll parameters---just like the rest of the action.

As was done for the two-point function, it will be helpful to express the three-point function in terms of a standard Fourier amplitude with a few factors removed as a matter of convention.  For the two-point function, this standard form of the amplitude was the power spectrum, $P_k(t)$, defined by 
$$
\langle 0(t)|\zeta(t,\vec x)\zeta(t,\vec y)|0(t)\rangle 
= \int {d^3\vec k\over (2\pi)^3}\, e^{i\vec k\cdot(\vec x-\vec y)}
{2\pi^2\over k^3} P_k(t) .
$$
Because there is only one way to construct a translationally invariant vector from two vectors---their difference, $\vec x-\vec y$---the power spectrum was defined with just one momentum, rather than with one accompanying each spatial coordinate.  For the three-point function, it is more convenient to keep a separate momentum for each of the coordinates by defining a momentum-dependent amplitude ${\cal A}_{k_1,k_2,k_3}(t)$ through
\begin{eqnarray}
&&\!\!\!\!\!\!\!\!\!\!\!\!\!\!\!\!\!\!\!\!\!\!\!
\langle 0(t)|\zeta(t,\vec x)\zeta(t,\vec y)\zeta(t,\vec z)|0(t)\rangle 
\nonumber \\
&\!\!\!\!\!\!=\!\!\!\!\!\!& 
\int {d^3\vec k_1\over (2\pi)^3} {d^3\vec k_2\over (2\pi)^3} 
{d^3\vec k_3\over (2\pi)^3}\, e^{i\vec k_1\cdot\vec x}
e^{i\vec k_2\cdot\vec y} e^{i\vec k_3\cdot\vec z}\,
(2\pi)^3\, \delta^3(\vec k_1+\vec k_2+\vec k_3)
\nonumber \\
&&\quad\times
{1\over 32\epsilon^2} {H^4\over M_{\rm pl}^4} {1\over k_1^3k_2^3k_3^3} 
{\cal A}_{k_1,k_2,k_3}(t) . 
\nonumber 
\end{eqnarray}
Of course, the $\delta$-function means that one of these three spatial momenta could be eliminated.  But since doing so would spoil the manifest symmetry of the amplitude, this $\delta$-function is usually left intact.

While it is possible to evaluate the three-point function of the fluctuations $\zeta(t,\vec x)$ directly, using the cubic action derived in the last section the fact that several of its terms are proportional to the linear equation of motion for $\zeta(t,\vec x)$ permits a trick which simplifies the calculation---which is why they were gathered together in the first place.  Writing the cubic action once again in its penultimate form, it is
\begin{eqnarray}
S^{(3)} &\!\!\!\!\!\!=\!\!\!\!\!\!& 
\int d^4x\, \biggl\{
{1\over 4} {e^{3\rho}\over M_{\rm pl}^2} 
{\dot\phi^4\over\dot\rho^4} \dot\zeta^2 \zeta 
+ {1\over 4} {e^\rho\over M_{\rm pl}^2} {\dot\phi^4\over\dot\rho^4} 
\zeta \partial_k\zeta \partial^k\zeta 
- {1\over 2} {e^{3\rho}\over M_{\rm pl}^2} {\dot\phi^4\over\dot\rho^4} \dot\zeta \partial_k\zeta \partial^k(\partial^{-2}\dot\zeta)
\nonumber \\
&&\qquad\ 
+\,\, {1\over 2} e^{3\rho} {\dot\phi^2\over\dot\rho^2} \dot\zeta \zeta^2 
{d\over dt} \biggl[ {\ddot\phi\over\dot\phi\dot\rho} 
+ {1\over 2} {1\over M_{\rm pl}^2} {\dot\phi^2\over\dot\rho^2} \biggr] 
- {1\over 16} {e^{3\rho}\over M_{\rm pl}^4} {\dot\phi^6\over\dot\rho^6} 
\bigl[ \dot\zeta^2 \zeta
- \zeta \partial_k\partial_l( \partial^{-2}\dot\zeta) \partial^k\partial^l ( \partial^{-2}\dot\zeta)
\bigr]
\nonumber \\
&&\qquad\ 
+\,\, f(\zeta) \biggl\{ 
{d\over dt} \biggl[ e^{3\rho} {\dot\phi^2\over\dot\rho^2} \dot\zeta \biggr] 
- e^\rho {\dot\phi^2\over\dot\rho^2} \partial_k\partial^k\zeta 
\biggr\} , 
\nonumber 
\end{eqnarray}
up to the total derivatives.  $f(\zeta)$ is a quadratic function of the field.  Comparing the form of the $f$-term with the expression for ${1\over 2}{\cal F}$ from before, this function is seen to be
\begin{eqnarray}
f(\zeta) &\!\!\!\!\!\!=\!\!\!\!\!\!& 
{1\over 2} \biggl[{\ddot\phi\over\dot\phi\dot\rho} 
+ {1\over 2} {1\over M_{\rm pl}^2} {\dot\phi^2\over\dot\rho^2} \biggr] \zeta^2 
+ {1\over\dot\rho} \dot\zeta \zeta 
- {1\over 4} {e^{-2\rho}\over\dot\rho^2} \bigl[ \partial_k\zeta \partial^k\zeta 
- \partial^{-2} \partial_k\partial_l \bigl( \partial^k\zeta \partial^l\zeta \bigr) \bigr] 
\nonumber \\
&&
+\,\, {1\over 4} {1\over M_{\rm pl}^2} {\dot\phi^2\over\dot\rho^3} 
\bigl[ \partial_k \zeta \partial^k( \partial^{-2}\dot\zeta) 
- \partial^{-2} \partial_k\partial_l \bigl( \partial^k\zeta \partial^l( \partial^{-2}\dot\zeta) \bigr) \bigr] . 
\nonumber 
\end{eqnarray}
The part of cubic action proportional to the linear equation of motion can then be removed by defining a new field $\zeta_n(t,\vec x)$ which is related to $\zeta(t,\vec x)$ through a nonlinear shift in the original field, 
$$
\zeta = \zeta_n + f(\zeta_n) . 
$$
This $f(\zeta_n)$ is exactly the same quadratic function that appears in the cubic action. 

If all of the terms generated by this shift contributed more or less equally, then not much would be gained by performing it.  However most of these terms turn out to be strongly suppressed.  As was shown during the discussion of the two-point function, the fluctuations that are relevant for the subsequent inhomogeneities of the universe are found by taking the late-time limit.  In this limit many of the terms generated by the shift grow essentially negligible.

To understand in a little more detail how it helps to make this shift from $\zeta$ to $\zeta_n$, notice first that the shift generates cubic (and quartic) interactions for $\zeta_n$ from the quadratic part of the action for $\zeta$, 
$$
S^{(2)}(\zeta) = S^{(2)}(\zeta_n) + \tilde S^{(3)}(\zeta_n) + {\cal O}(\zeta_n^4).
$$
Since there are no linear terms in $\zeta(t,\vec x)$ appearing in the action, the quadratic action for $\zeta_n(t,\vec x)$ has exactly the same form as it did before for $\zeta(t,\vec x)$, 
$$
S^{(2)}(\zeta_n) = \int d^4x\, e^{3\rho} {1\over 2}{\dot\phi^2\over\dot\rho^2} 
\Bigl\{ \dot\zeta_n^2 - e^{-2\rho}\partial_k\zeta_n\partial^k\zeta_n \Bigr\} . 
$$
This means that the free evolution of the field $\zeta_n(t,\vec x)$ will also be exactly the same as it was for $\zeta(t,\vec x)$.  Looking at the next order, the cubic terms generated from the quadratic part of the action for $\zeta(t,\vec x)$ are 
$$
\tilde S^{(3)} = \int d^4x\, e^{3\rho} {\dot\phi^2\over\dot\rho^2} 
\Bigl\{ \dot\zeta_n\dot f(\zeta_n) - e^{-2\rho}\partial_k\zeta_n \partial^kf(\zeta_n) \Bigr\} .
$$
Integrating these two terms by parts in the appropriate variable removes the derivatives from $f(\zeta_n)$.  $\tilde S^{(3)}$ can then be written as
$$
\tilde S^{(3)} = - \int d^4x\, f(\zeta_n) 
\biggl\{ {d\over dt} \biggl[ e^{3\rho} {\dot\phi^2\over\dot\rho^2} \dot\zeta_n \biggr] 
- e^\rho {\dot\phi^2\over\dot\rho^2} \partial_k\partial^k\zeta_n \biggr\} 
= - \int d^4x\, \bigl\{ {\textstyle{1\over 2}} {\cal F} \bigr\} , 
$$
up to total derivatives.  This term precisely cancels the corresponding part of $S^{(3)}(\zeta_n)$ that is proportional to the linear equation of motion,
$$
S^{(3)}(\zeta) = S^{(3)}(\zeta_n) + {\cal O}(\zeta_n^4) . 
$$
The leading, order $\epsilon^2$, terms amongst the cubic terms in $\zeta_n(t,\vec x)$ correspond then to just the following three operators, 
\begin{eqnarray}
S_n^{(3)}(\zeta_n) &\!\!\!\!\!\!\equiv\!\!\!\!\!\!& 
S^{(3)}(\zeta_n) + \tilde S^{(3)}(\zeta_n) 
\nonumber \\
&\!\!\!\!\!\!=\!\!\!\!\!\!& 
\int d^4x\, {1\over 4} {1\over M_{\rm pl}^2} {\dot\phi^4\over\dot\rho^4} 
\Bigl\{
e^{3\rho} \dot\zeta_n^2 \zeta_n 
+ e^\rho \zeta_n \partial_k\zeta_n \partial^k\zeta_n 
- 2 e^{3\rho} \dot\zeta_n \partial_k\zeta_n \partial^k(\partial^{-2}\dot\zeta_n)
+ \cdots \Bigr\} . 
\nonumber 
\end{eqnarray}

In the interaction picture of quantum field theory that will be used to compute the leading form of the three-point function, this action is responsible for the evolution of $\zeta_n$-states in the theory.  Defining the interaction Hamiltonian associated with this action to be\footnote{In writing the interaction thus, there is an implicit assumption that these interactions, written in a Lagrangian form as $S^{(3)}_n=\int dt\, L_I(t)$, can be immediately converted into the corresponding part of the Hamiltonian, $H_I=-L_I$.  This simple relation holds for the cubic terms; but at quartic order and beyond one needs to be more careful.  The reason is that the canonical momentum $\pi_n=\partial L/\partial\dot\zeta_n$ is modified by the time-derivatives that appear amongst the cubic terms.  The Hamiltonian is given by $H=\pi_n\dot\zeta_n-L$.  It is not always necessarily the case {\it in general\/} that $H_I=-L_I$ when there are derivative interactions. }
$$
H_I(t) = - \int d^3\vec x\, {1\over 4} {1\over M_{\rm pl}^2} 
{\dot\phi^4\over\dot\rho^4} \Bigl\{
e^{3\rho} \dot\zeta_n^2 \zeta_n 
+ e^\rho \zeta_n \partial_k\zeta_n \partial^k\zeta_n 
- 2e^{3\rho} \dot\zeta_n \partial_k\zeta_n \partial^k(\partial^{-2}\dot\zeta_n)
+ \cdots \Bigr\} , 
$$
then the time evolution of the state is generated by 
$$
|0(t)\rangle = T e^{-i\int_{t_0}^t dt'\, H_I(t')}|0(t_0)\rangle .
$$
In the simplest inflationary picture $t_0$ is usually taken to be infinitely far in the past, $t_0\to -\infty$; and since the fluctuations that are relevant for the inhomogeneities of the universe are those that---by the end of inflation---have been stretched so that their wavelength is much larger than the size of the horizon, the final time can more or less be taken to be a very large but finite value,
$$
|0(t)\rangle = T e^{-i\int_{-\infty}^t dt'\, H_I(t')}|0\rangle .
$$
$|0\rangle$ will always mean the initial state, which here is $|0\rangle = |0(-\infty)\rangle$, when the vacuum is written without any explicit argument.  

The three-point function is meant to be evaluated in the vacuum state of the full theory.  But when the correlation functions are being evaluated perturbatively, it is the vacuum state of the free field theory that is actually used in practice.  Although the vacua for the free and interacting theories are not the same, the time integrals associated with the evolution of the state will always implicitly be applying an $i\epsilon$ prescription.  In the limit where the initial time is taken into the infinite past, $t_0\to -\infty$, any admixture of the excited states in an expansion of the free vacuum in the states of the full theory will be projected away by this prescription, leaving only the true vacuum contribution.

Upon making this shift in the field, the calculation of the three-point function naturally divides into two sets of terms which can each be evaluated separately, 
\begin{eqnarray}
\langle 0(t)| \zeta(t,\vec x)\zeta(t,\vec y)\zeta(t,\vec z) |0(t)\rangle 
&\!\!\!\!\!\!=\!\!\!\!\!\!& 
\langle 0(t)| \zeta_n(t,\vec x)\zeta_n(t,\vec y)\zeta_n(t,\vec z) |0(t)\rangle 
\nonumber \\
&&
+\,\, \langle 0(t)| \zeta_n(t,\vec x)\zeta_n(t,\vec y)f\bigl(\zeta_n(t,\vec z)\bigr) |0(t)\rangle 
\nonumber \\
&&
+\,\, \langle 0(t)| \zeta_n(t,\vec x)f\bigl(\zeta_n(t,\vec y)\bigr)\zeta_n(t,\vec z) |0(t)\rangle 
\nonumber \\
&&
+\,\, \langle 0(t)| f\bigl(\zeta_n(t,\vec x)\bigr)\zeta_n(t,\vec y)\zeta_n(t,\vec z) |0(t)\rangle 
\nonumber \\
&&
+\,\, \hbox{higher order terms} . 
\nonumber 
\end{eqnarray}
The `higher order terms' are those that include at least five factors of the shifted fluctuations, $\zeta_n(t,\vec x)$.  They are suppressed in part because they contain further factors of the slow-roll parameters, beyond the terms that have been explicitly written, and also because some parts effectively vanish in the late-time limit.

The three-point function of $\zeta_n(t,\vec x)$ would of course vanish entirely if evaluated in the initial state.  In the interaction picture, its value is generated entirely by the evolution of the state, 
\begin{eqnarray}
&&\!\!\!\!\!\!\!\!\!\!\!\!\!\!\!\!\!\!\!\!\!\!\!\!\!\!\!\!\!\!\!\!\!\!\!\!
\langle 0(t)| \zeta_n(t,\vec x)\zeta_n(t,\vec y)\zeta_n(t,\vec z) |0(t)\rangle
\nonumber \\
&\!\!\!\!\!\!=\!\!\!\!\!\!& 
\bigl\langle 0\big| 
\bigl( T e^{-i\int_{-\infty}^t dt'\, H_I(t')}\bigr)^\dagger 
\zeta_n(t,\vec x)\zeta_n(t,\vec y)\zeta_n(t,\vec z) 
\bigl( T e^{-i\int_{-\infty}^t dt'\, H_I(t')}\bigr) \big|0\bigr\rangle . 
\nonumber
\end{eqnarray}
The method for evaluating a matrix element of this form is slightly different from the more familiar techniques used to analyse scattering processes and which appear in any textbook on quantum field theory.  Since the interaction Hamiltonian is small, being explicitly of order $\epsilon^2$, the time-ordered exponentials can be expanded in a Taylor series.  The leading contribution to the three-point function of $\zeta_n$ is produced by the linear terms in the interaction Hamiltonian
$$
\langle 0(t)| \zeta_n(t,\vec x)\zeta_n(t,\vec y)\zeta_n(t,\vec z) |0(t)\rangle
= - i\int_{-\infty}^t dt'\, \bigl\langle 0\big| 
\bigl[ \zeta_n(t,\vec x)\zeta_n(t,\vec y)\zeta_n(t,\vec z), H_I(t') \bigr] \big|0\bigr\rangle + \cdots . 
$$
One way to evaluate the right side is to expand all of the fields directly as sums of creation and annihilation operators, 
$$
\zeta_n(t,\vec x) 
= \int {d^3\vec k\over (2\pi)^3}\, \Bigl\{ 
\zeta_{n,k}(t)e^{i\vec k\cdot x} a_{n,\vec k} 
+ \zeta_{n,k}^*(t)e^{-i\vec k\cdot x} a_{n,\vec k}^\dagger \Bigr\} , 
$$
where $a_{n,\vec k}\,|0\rangle = 0$ annihilates the initial---the free---vacuum state.  While this approach is straightforward enough, it is a little tedious.  Therefore this matrix element will be instead treated using the Schwinger-Keldysh formalism.  This formalism will be briefly reviewed here once the other---the `quartic'---contribution to the three-point function of $\zeta$ has been evaluated.

These `quartic' contributions correspond to the next three terms, each of which contains an appearance of $f(\zeta_n)$.  They only differ from one another by a cyclic swapping of the spatial coordinates.  So it is only necessary to evaluate one of them to determine all of them.  Since they are quartic in the field $\zeta_n$, they do contribute in the initial state $|0\rangle$; in their case the time-evolution of the state only contributes at a higher order in the slow-roll parameter $\epsilon$.  The leading piece from this type of matrix element is found by calculating,
$$
\langle 0(t)| \zeta_n(t,\vec x)\zeta_n(t,\vec y)f\bigl(\zeta_n(t,\vec z)\bigr) |0(t)\rangle
= \langle 0| \zeta_n(t,\vec x)\zeta_n(t,\vec y)f\bigl(\zeta_n(t,\vec z)\bigr) |0\rangle 
+ \cdots . 
$$
Thus, the original calculation of the three-point function for the fluctuations $\zeta$ reduces to the calculation of the three-point function of the shifted field $\zeta_n$ (using a simpler cubic action) and the expectation value of a certain quartic operator in $\zeta_n$.  The latter is the easier of the two calculations, so it will be done first.
\vskip18truept

\noindent{\it V.A. The late-time behaviour of the $\zeta_n$ Wightman functions}
\vskip6truept

Before plunging into the calculation of these expectation values, it is useful first to review what was found earlier for the two-point function of $\zeta(t,\vec x)$ to extract some of the limiting behaviour.  The quadratic part of the action for $\zeta_n(t,\vec x)$ is exactly the same as it was for $\zeta(t,\vec x)$.  Correspondingly, the structure of its two-point function will also be exactly the same.  The two-point function is the basic element for calculating the expectation values of arbitrary operators in perturbation theory.

The expectation value of a pair of quantum fields evaluated at different space-time points is called a {\it Wightman function\/}.  Here the state in which they are evaluated is effectively the asymptotic, initial vacuum state of the free theory of $\zeta_n(t,\vec x)$, $|0\rangle = |0(-\infty)\rangle$.  In principle there are two Wightman functions, one for each of the two orderings of the fields, 
\begin{eqnarray}
G^>(t,\vec x;t',\vec y) &\!\!\!\!\!\!=\!\!\!\!\!\!& 
\langle 0| \zeta_n(t,\vec x)\zeta_n(t',\vec y) |0\rangle 
= \int {d^3\vec k\over (2\pi)^3}\, e^{i\vec k\cdot(\vec x-\vec y)} 
G_k^>(t,t')
\nonumber \\
G^<(t,\vec x;t',\vec y) &\!\!\!\!\!\!=\!\!\!\!\!\!& 
\langle 0| \zeta_n(t',\vec y)\zeta_n(t,\vec x) |0\rangle 
= \int {d^3\vec k\over (2\pi)^3}\, e^{i\vec k\cdot(\vec x-\vec y)} 
G_k^<(t,t') ,
\nonumber 
\end{eqnarray}
although it should be clear that they really are the same.  Under an exchange of their arguments,
$$
G^<(t,\vec x;t',\vec y) = G^>(t',\vec y;t,\vec x) . 
$$
The Fourier modes inherit this symmetry too:  since $\zeta_n(t,\vec x)$ is a real scalar field,
$$
G_k^<(t,t') = G_k^>(t',t) ,
$$
and 
$$
G_k^<(t,t') = \bigl( G_k^>(t,t')\bigr) ^*
$$
as well.  The specific forms of these Wightman functions were already derived much earlier in the course of investigating the power spectrum for inflation.  In terms of the canonically normalized field,
$$
\zeta_n(t,\vec x) = e^{\rho(t)} {\dot\phi(t)\over\rho(t)} \varphi(t,\vec x) ,
$$
the momentum representations of the Wightman functions are
\begin{eqnarray}
G_k^>(t,t') &\!\!\!\!\!\!=\!\!\!\!\!\!&
e^{-\rho(t)} e^{-\rho(t')} 
{\dot\rho(t)\over\dot\phi(t)} {\dot\rho(t')\over\dot\phi(t')} 
\varphi_k(t)\varphi_k^*(t')
\nonumber \\
G_k^<(t,t') &\!\!\!\!\!\!=\!\!\!\!\!\!&
e^{-\rho(t)} e^{-\rho(t')} 
{\dot\rho(t)\over\dot\phi(t)} {\dot\rho(t')\over\dot\phi(t')} 
\varphi_k^*(t)\varphi_k(t') , 
\nonumber 
\end{eqnarray}
where $\varphi_k(t)$ is the mode function that accompanies the annihilation operator of the free $\varphi(t,\vec x)$ vacuum.  

In the case of the power spectrum, it made sense to work beyond the leading dependence in the slow-roll parameters because observations of the universe are already sufficiently precise to be able to constrain not only the overall amplitude of the power spectrum (a leading effect in $\epsilon$, $\delta$), but its dependence on $k$ as well (a sub-leading effect).  In contrast, even the amplitude of the three-point function has not been seen yet; so it is a bit premature to work beyond the leading behaviour in the slow-roll parameters.  This allows the calculation to be simplified.  The Wightman functions will be treated in the $\epsilon,\delta\to 0$ limit, at least as far as is possible.  The only exception occurs in the $1/\epsilon$ due to the factors
$$
{\dot\rho(t)\over\dot\phi(t)} {\dot\rho(t')\over\dot\phi(t')} \to
{\dot\rho^2\over\dot\phi^2} = {1\over 2\epsilon} {1\over M_{\rm pl}^2}
$$
that appear in the Wightman functions.  Obviously this occurrence of $\epsilon$ must be kept finite, for otherwise the factor would diverge.  But for the rest of what makes up the Wightman functions---the scale factor and the modes of the canonically normalized field---they can be put into their de Sitter forms,
$$
e^{\rho(t)} \to - {1\over H\eta} 
\qquad\hbox{and}\qquad
\varphi_k(t) = {i\over\sqrt{2k^3}} {1+ik\eta\over -\eta} e^{-ik\eta} . 
$$
The expression for $\varphi_k(t)$ was derived as follows.  Earlier, when treating the two-point function, $\varphi_k(t)$ was presented in the more general form, 
$$
\varphi_k(\eta) = -{\sqrt{\pi}\over 2} \sqrt{-\eta} H_\nu^{(1)}(-k\eta) , 
\qquad\hbox{where $\nu={3\over 2}\sqrt{1+{4\over 3}(2\epsilon+\delta)}$.}
$$
The de Sitter limit corresponds to setting $\nu={3\over 2}$, for which value the Hankel function becomes 
$$
H_{3/2}^{(1)}(-z) = - \sqrt{{2\over\pi}} {1\over\sqrt{-z}} 
\Bigl( 1 - {i\over z} \Bigr) e^{-iz} . 
$$

Notice that $e^{\rho(t)}$ and $\varphi_k(t)$ are now being written in terms of the conformal time $\eta(t)$.  The dependence of the Wightman functions on $\eta$ is simpler in appearance than it would have been had they been written as function of the `cosmological time' coordinate $t$.  The relation between the two is quite straightforward in the de Sitter limit, 
$$
d\eta = e^{-Ht}\, dt .
$$
In terms of the conformal time, the leading behaviour of the Wightman functions in the slow-roll limit is thus
\begin{eqnarray}
G_k^>(t,t') &\!\!\!\!\!\!=\!\!\!\!\!\!&
{1\over 4\epsilon} {H^2\over M_{\rm pl}^2} {1\over k^3} 
(1+ik\eta)(1-ik\eta') e^{-ik(\eta-\eta')} + \cdots
\nonumber \\
G_k^<(t,t') &\!\!\!\!\!\!=\!\!\!\!\!\!&
{1\over 4\epsilon} {H^2\over M_{\rm pl}^2} {1\over k^3} 
(1-ik\eta)(1+ik\eta') e^{ik(\eta-\eta')} + \cdots . 
\nonumber 
\end{eqnarray}

Many of the operators that appear in the cubic interaction contain time derivatives.  Wick contractions involving these operators will then naturally produce time derivatives of the Wightman functions too.  As a convention which will be sufficient here, let the time derivatives always be implicitly acting on the {\it second\/} of the arguments of the Wightman functions; so,
$$
\dot G_k^>(t,t') \equiv {d\over dt'} G_k^>(t,t') 
= {d\eta'\over dt'}{d\over d\eta'} G_k^>(t,t') . 
$$
Using the relation 
$$
{d\eta'\over dt'} = -H\eta'
$$
which is valid in the de Sitter limit, the leading part of the derivative of a Wightman function has even a simpler dependence on the conformal times, 
$$
\dot G_k^>(t,t') = - {1\over 4\epsilon} {H^3\over M_{\rm pl}^2} 
{\eta^{\prime 2}\over k} (1+ik\eta) e^{-ik(\eta-\eta')} + \cdots . 
$$
Similarly 
$$
\dot G_k^<(t,t') = - {1\over 4\epsilon} {H^3\over M_{\rm pl}^2} 
{\eta^{\prime 2}\over k} (1-ik\eta) e^{ik(\eta-\eta')} + \cdots . 
$$

Finally, the fluctuations that are important for the subsequent inhomogeneities of the universe are those that have been stretched beyond the Hubble horizon by the end of inflation.  As was explained during the derivation of the two-point function, this really corresponds to the $-k\eta\to 0$ limit, though usually it is sufficient to take just the $\eta\to 0$ limit.  In evaluating the three-point function, $\langle 0(t)|\zeta(t,\vec x)\zeta(t,\vec y)\zeta(t,\vec z)|0(t) \rangle$, the Wightman functions will always be arranged here so that their first argument (and sometimes the second argument too) will be this time $t$.  Taking $t\to\infty$ or $\eta\to 0$, the Wightman functions and their derivatives become 
\begin{eqnarray}
G_k^>(0,\eta') &\!\!\!\!\!\!\approx\!\!\!\!\!\!&
{1\over 4\epsilon} {H^2\over M_{\rm pl}^2} 
{1\over k^3} (1-ik\eta') e^{ik\eta'}, 
\qquad\quad\ \ \,
\dot G_k^>(0,\eta') \approx - {1\over 4\epsilon} {H^3\over M_{\rm pl}^2} 
{\eta^{\prime 2}\over k} e^{ik\eta'} , 
\nonumber \\
G_k^<(0,\eta') &\!\!\!\!\!\!\approx\!\!\!\!\!\!&
{1\over 4\epsilon} {H^2\over M_{\rm pl}^2} 
{1\over k^3} (1+ik\eta') e^{-ik\eta'}, 
\qquad\quad
\dot G_k^<(0,\eta') \approx - {1\over 4\epsilon} {H^3\over M_{\rm pl}^2} 
{\eta^{\prime 2}\over k} e^{-ik\eta'} .
\nonumber 
\end{eqnarray}
These expressions will be very useful when evaluating the expectation values of the operators needed for the three-point function.  Notice that when the second argument is also taken to zero, the time-derivatives of the Wightman functions {\it vanish\/}.  Operators that contain time-derivatives will generally be suppressed in the late-time limit, unless there other sources for compensating factors of $1/\eta^{\prime 2}$ about.
\vskip18truept

\noindent{\it V.B. The `quartic' contributions}
\vskip6truept

Of the two general types of contributions to the three-point function, those that are quartic in the shifted field $\zeta_n$ are the more straightforward to evaluate.  They differ from one another only according to their spatial arguments; so it is sufficient to calculate one of them in detail, 
$$
 \langle 0| \zeta_n(t,\vec x)\zeta_n(t,\vec y)f\bigl(\zeta_n(t,\vec z)\bigr) |0\rangle , 
$$
for example.  Substituting the explicit form of $f(\zeta_n(t,\vec z))$ into this matrix element produces a fair number of terms, 
\begin{eqnarray}
&&\!\!\!\!\!\!\!\!\!\!\!\!\!\!\!\!\!\!\!\!\!\!\!\!\!\!\!
\langle 0| \zeta_n(t,\vec x)\zeta_n(t,\vec y)f\bigl(\zeta_n(t,\vec z)\bigr) |0\rangle 
\nonumber \\
&\!\!\!\!\!\!=\!\!\!\!\!\!& 
\biggl[ {1\over 2} {\ddot\phi\over\dot\phi\dot\rho} 
+ {1\over 4} {1\over M_{\rm pl}^2} {\dot\phi^2\over\dot\rho^2}\biggr] 
\langle 0| \zeta_n(t,\vec x)\zeta_n(t,\vec y)\zeta_n(t,\vec z)\zeta_n(t,\vec z) |0\rangle 
\nonumber \\
&& 
+\,\, {1\over\dot\rho} 
\langle 0| \zeta_n(t,\vec x)\zeta_n(t,\vec y)\dot\zeta_n(t,\vec z)\zeta_n(t,\vec z) |0\rangle 
\nonumber \\
&& 
-\,\, {1\over 4}{e^{-2\rho}\over\dot\rho^2} 
\langle 0| \zeta_n(t,\vec x)\zeta_n(t,\vec y)\partial_k\zeta_n(t,\vec z) \partial^k\zeta_n(t,\vec z) |0\rangle 
\nonumber \\
&& 
+\,\, {1\over 4}{e^{-2\rho}\over\dot\rho^2} 
\langle 0| \zeta_n(t,\vec x)\zeta_n(t,\vec y)\partial^{-2}\partial_k\partial_l \bigl( \partial^k\zeta_n(t,\vec z) \partial^l\zeta_n(t,\vec z) \bigr) |0\rangle 
\nonumber \\
&& 
+\,\, {1\over 4}{1\over M_{\rm pl}^2}{\dot\phi^2\over\dot\rho^3} 
\langle 0| \zeta_n(t,\vec x)\zeta_n(t,\vec y) \partial_k\zeta_n(t,\vec z) \partial^k\dot\zeta_n(t,\vec z) |0\rangle 
\nonumber \\
&& 
-\,\, {1\over 4}{1\over M_{\rm pl}^2}{\dot\phi^2\over\dot\rho^3} 
\langle 0| \zeta_n(t,\vec x)\zeta_n(t,\vec y)\partial^{-2}\partial_k\partial_l \bigl( \partial^k\zeta_n(t,\vec z) \partial^l\dot\zeta_n(t,\vec z) \bigr) |0\rangle . 
\nonumber 
\end{eqnarray}
As it will turn out, only the first line contributes appreciably in the late-time limit.  The reason is two-fold.  Many of the terms contain time derivatives of the field; these lead to time derivatives of the Wightman functions which, as was just shown, vanish when $t\to\infty$.  Two of the terms only have spatial derivatives.  These terms, however, are both accompanied by a factor $e^{-2\rho(t)}$.  In the de Sitter limit, $e^{-2\rho(t)} = H^2(-\eta)^2$, which also vanishes as $t\to\infty$, or $\eta\to 0$.

The matrix elements are converted into Wightman functions by taking Wick contractions of pairs of fields.  To produce a connected diagram, both the field at $\vec x$ and the field at $\vec y$ must each be contracted with one of the fields at $\vec z$.  A sketch of the sort of connected diagram that is generated by these contractions looks like 
$$
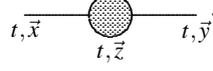
\beginpicture
\setcoordinatesystem units <1.00truept,1.00truept>
\setplotarea x from -36 to 36, y from -8 to 8
\circulararc 360 degrees from 8 0 center at 0 0
\plot  8 0   32 0 /
\plot -8 0  -32 0 /
\put {\footnotesize $t,\vec x$} [c] at -32  -5.8
\put {\footnotesize $t,\vec y$} [c] at  32  -6
\put {\footnotesize $t,\vec z$} [c] at  0  -13.5
\setshadesymbol ({\tmrms .})
\setshadegrid span <0.9pt>
\setquadratic
\hshade -5.65 -5.65 -5.65 <,z,,>  0 -8.00 -5.65   5.65 -5.65 -5.65 /
\vshade -5.65 -5.65  5.65 <z,z,,> 0 -8.00  8.00   5.65 -5.65  5.65 /
\hshade -5.65  5.65  5.65 <z,,,>  0  5.65  8.00   5.65  5.65  5.65 /
\endpicture .
$$
The blob represents the various derivative operators acting on the fields at $(t,\vec z)$.  Since all of the fields are evaluated at the same time $t$, and since $\zeta_n$ is a scalar field, there is no need to worry about the time-ordering when taking a contraction, 
$$
\beginpicture
\setcoordinatesystem units <1.00truept,1.00truept>
\setplotarea x from 0 to 175, y from 0 to 16
\put {$\zeta_n(t,\vec x)\zeta_n(t,\vec z) = G^>(t,\vec x;t,\vec z) 
= G^<(t,\vec x;t,\vec z).$} [l] at 0 6
\putrule from 3 12.5 to 3 16 
\putrule from 3 16 to 35.5 16
\putrule from 35.5 16 to 35.5 12.5
\endpicture
$$
Taking these contractions and expanding the resulting Wightman functions in Fourier modes, the expectation value of the quartic operator $\zeta_n(t,\vec x)\zeta_n(t,\vec y)f(\zeta_n(t,\vec x))$ in the vacuum state becomes
\begin{eqnarray}
&&\!\!\!\!\!\!\!\!\!\!\!\!\!\!
\langle 0| \zeta_n(t,\vec x)\zeta_n(t,\vec y)f\bigl(\zeta_n(t,\vec z)\bigr) |0\rangle 
\nonumber \\
&\!\!\!\!\!\!=\!\!\!\!\!\!& 
\int {d^3\vec k_1\over (2\pi)^3} {d^3\vec k_2\over (2\pi)^3}\, 
e^{i\vec k_1\cdot (\vec x-\vec z)} e^{i\vec k_2\cdot (\vec y-\vec z)}
\nonumber \\
&&\times 
\biggl\{ 
\biggl[ {\ddot\phi\over\dot\phi\dot\rho} 
+ {1\over 2} {1\over M_{\rm pl}^2} {\dot\phi^2\over\dot\rho^2}\biggr] G_{k_1}^>(t,t)G_{k_2}^>(t,t)
+ {1\over\dot\rho} \bigl[ 
\dot G_{k_1}^>(t,t)G_{k_2}^>(t,t) + G_{k_1}^>(t,t)\dot G_{k_2}^>(t,t) \bigr]
\nonumber \\
&&\quad
+\,\, {1\over 2}{e^{-2\rho}\over\dot\rho^2} (\vec k_1\cdot\vec k_2) 
G_{k_1}^>(t,t)G_{k_2}^>(t,t) 
- {1\over 2}{e^{-2\rho}\over\dot\rho^2} 
{(k_1^2 + \vec k_1\cdot\vec k_2)(k_2^2 + \vec k_1\cdot\vec k_2) \over 
|\vec k_1 + \vec k_2|^2} G_{k_1}^>(t,t)G_{k_2}^>(t,t) 
\nonumber \\
&&\quad 
+\,\, {1\over 4}{1\over M_{\rm pl}^2}{\dot\phi^2\over\dot\rho^3} \biggl[ 
{\vec k_1\cdot\vec k_2\over k_2^2} G_{k_1}^>(t,t)\dot G_{k_2}^>(t,t)
+ {\vec k_1\cdot\vec k_2\over k_1^2} \dot G_{k_1}^>(t,t) G_{k_2}^>(t,t) \biggr]
\nonumber \\
&&\quad 
-\,\, {1\over 4}{1\over M_{\rm pl}^2}{\dot\phi^2\over\dot\rho^3} {(k_1^2 + \vec k_1\cdot\vec k_2)(k_2^2 + \vec k_1\cdot\vec k_2) \over |\vec k_1 + \vec k_2|^2} 
\biggl[ {1\over k_2^2} G_{k_1}^>(t,t)\dot G_{k_2}^>(t,t)
+ {1\over k_1^2} \dot G_{k_1}^>(t,t) G_{k_2}^>(t,t) \biggr]
\biggr\} . 
\nonumber 
\end{eqnarray}
After the expansions in Fourier modes, the spatial derivatives have become factors of the appropriate spatial momenta.  In this form it is clear that there are no factors that are growing in the conformal time to compensate for the time derivatives or the inverse scale factors, both of which are vanishing quadratically as $\eta\to 0$,
$$
\dot G_k^>(t,t) \to - {1\over 4\epsilon} {H^3\over M_{\rm pl}^2} 
{\eta^2\over k} 
\qquad\hbox{and}\qquad
e^{-2\rho(t)}\to H^2\eta^2 .
$$
Thus, as promised, only the first term contributes in this limit,
\begin{eqnarray}
&&\!\!\!\!\!\!\!\!\!\!\!\!\!\!\!\!\!\!\!\!\!\!\!\!\!\!\!
\langle 0| \zeta_n(t,\vec x)\zeta_n(t,\vec y)f\bigl(\zeta_n(t,\vec z)\bigr) |0\rangle 
\nonumber \\
&\!\!\!\!\!\!=\!\!\!\!\!\!& 
\int {d^3\vec k_1\over (2\pi)^3} {d^3\vec k_2\over (2\pi)^3}\, 
e^{i\vec k_1\cdot (\vec x-\vec z)} e^{i\vec k_2\cdot (\vec y-\vec z)}
\biggl\{ 
\biggl[ {\ddot\phi\over\dot\phi\dot\rho} 
+ {1\over 2} {1\over M_{\rm pl}^2} {\dot\phi^2\over\dot\rho^2}\biggr] G_{k_1}^>(t,t)G_{k_2}^>(t,t)
+ \cdots \biggr\} .
\nonumber 
\end{eqnarray}
In the late-time limit, the Wightman function becomes a constant, 
$$
\lim_{t\to\infty} G_k^>(t,t) = {1\over 4\epsilon} {H^2\over M_{\rm pl}^2} {1\over k^3} , 
$$
so that 
\begin{eqnarray}
&&\!\!\!\!\!\!\!\!\!\!\!\!\!\!\!\!\!\!\!\!\!\!\!\!\!\!\!
\langle 0| \zeta_n(t,\vec x)\zeta_n(t,\vec y)f\bigl(\zeta_n(t,\vec z)\bigr) |0\rangle 
\nonumber \\
&\!\!\!\!\!\!=\!\!\!\!\!\!& 
\int {d^3\vec k_1\over (2\pi)^3} {d^3\vec k_2\over (2\pi)^3}\, 
e^{i\vec k_1\cdot\vec x} e^{i\vec k_2\cdot\vec y} 
e^{-i(\vec k_1+\vec k_2)\cdot\vec z} 
{1\over 16\epsilon^2} {H^4\over M_{\rm pl}^4} {1\over k_1^3} {1\over k_2^3}
\biggl[ {\ddot\phi\over\dot\phi\dot\rho} 
+ {1\over 2} {1\over M_{\rm pl}^2} {\dot\phi^2\over\dot\rho^2}\biggr] 
+ \cdots . 
\nonumber 
\end{eqnarray}
The three quartic contributions to the three-point function all have the same structure up to a cyclic permutation of the spatial positions of the fields.  The form that was just derived for $\langle 0| \zeta_n(t,\vec x)\zeta_n(t,\vec y)f\bigl(\zeta_n(t,\vec z)\bigr) |0\rangle$ does not treat the three coordinates very symmetrically; but it can easily be converted into a more symmetric form, with a different momentum associated with each of the spatial coordinates, by introducing a $\delta$-function,
\begin{eqnarray}
&&\!\!\!\!\!\!\!\!\!\!\!\!\!\!\!\!\!\!\!\!\!\!\!\!\!\!\!
\langle 0| \zeta_n(t,\vec x)\zeta_n(t,\vec y)f\bigl(\zeta_n(t,\vec z)\bigr) |0\rangle 
\nonumber \\
&\!\!\!\!\!\!=\!\!\!\!\!\!& 
\int {d^3\vec k_1\over (2\pi)^3} {d^3\vec k_2\over (2\pi)^3} 
{d^3\vec k_3\over (2\pi)^3}\, 
e^{i\vec k_1\cdot\vec x} e^{i\vec k_2\cdot\vec y} e^{i\vec k_3\cdot\vec z}\,
(2\pi)^3\, \delta^3(\vec k_1+\vec k_2+\vec k_3) 
\nonumber \\
&&\quad\times
{1\over 16\epsilon^2} {H^4\over M_{\rm pl}^4} {1\over k_1^3k_2^3k_3^3} 
\biggl\{ 
\biggl[ {\ddot\phi\over\dot\phi\dot\rho} 
+ {1\over 2} {1\over M_{\rm pl}^2} {\dot\phi^2\over\dot\rho^2}\biggr] k_3^3 + \cdots \biggr\}
\nonumber 
\end{eqnarray}
Integrating over $d^3\vec k_3$ reverts this expression back to its previous form.  Now everything outside the braces is completely symmetric in the three momenta and three coordinates.  It is then the work of a moment to symmetrize over all three of the positions by replacing the $k_3^3$ inside the braces with $k_1^3+k_2^3+k_3^3$, 
\begin{eqnarray}
&&\!\!\!\!\!\!\!\!\!\!\!\!\!\!\!\!\!\!\!\!\!\!\!\!\!\!\!
\langle 0| \zeta_n(t,\vec x)\zeta_n(t,\vec y)f\bigl(\zeta_n(t,\vec z)\bigr) |0\rangle 
+ \hbox{two more}
\nonumber \\
&\!\!\!\!\!\!=\!\!\!\!\!\!& 
\int {d^3\vec k_1\over (2\pi)^3} {d^3\vec k_2\over (2\pi)^3} 
{d^3\vec k_3\over (2\pi)^3}\, 
e^{i\vec k_1\cdot\vec x} e^{i\vec k_2\cdot\vec y} e^{i\vec k_3\cdot\vec z}\,
(2\pi)^3\, \delta^3(\vec k_1+\vec k_2+\vec k_3) 
\nonumber \\
&&\quad\times
{1\over 32\epsilon^2} {H^4\over M_{\rm pl}^4} {1\over k_1^3k_2^3k_3^3} 
\biggl\{ 
2 \biggl[ {\ddot\phi\over\dot\phi\dot\rho} 
+ {1\over 2} {1\over M_{\rm pl}^2} {\dot\phi^2\over\dot\rho^2}\biggr] 
\bigl( k_1^3 + k_2^3 + k_3^3 \bigr) + \cdots \biggr\} . 
\nonumber 
\end{eqnarray}
Notice that the prefactors have been rearranged slightly to match with the general expression for the three-point function that was introduced at the beginning of this section.
\vskip18truept

\noindent{\it V.C. The Schwinger-Keldysh formalism}
\vskip6truept

The three-point function of the shifted fluctuations,
$$
\langle 0(t)|\zeta_n(t,\vec x)\zeta_n(t,\vec y)\zeta_n(t,\vec z)|0(t)\rangle ,
$$
is just a particular instance of a general expectation value in quantum field theory, 
$$
\langle 0(t)|{\cal O}(t)|0(t)\rangle ,
$$
where ${\cal O}(t)$ is an arbitrary operator.  For scattering processes, one starts with a initial state, evolves it from $t=-\infty$ to $t=\infty$ and then evaluates its overlap with a final state.  What is to be computed here is not quite the same.  Here, {\it both\/} of the states in the expectation value are evolving in time.  Although the methods for treating these objects have existed for many decades and though these methods do not differ very much from the standard diagrammatic techniques used to solve scattering processes, nonetheless they do not seem to be as widely known amongst quantum field theorists.  This subsection will present a short review of the techniques that are needed to evaluate expectation values with the general structure $\langle 0(t)|{\cal O}(t)|0(t)\rangle$.

To begin, consider a theory whose Hamiltonian is 
$$
H = H_0 + H_I .
$$
The $H_0$ corresponds to the `free' part of the theory---terms that are quadratic in the field and that contain no more than two derivatives.  $H_I$ contains everything else:  all of the higher order self-interactions of the field and the interactions amongst different fields, if there are any.  In the interaction picture of quantum field theory, $H_0$ determines the time dependence of the fields and $H_I$ determines how the quantum states evolve over time.

The detailed derivation of the time-dependence of the fluctuations $\zeta(t,\vec x)$ presented earlier---and that of the canonically normalized field $\varphi(t,\vec x)$ too---was already following precisely this picture, based on the free Hamiltonian
$$
H_0(t) = \int d^3\vec x\, \biggl\{ 
{1\over 2}e^{3\rho}{\dot\phi^2\over\dot\rho^2} \dot\zeta^2
+ {1\over 2} e^\rho{\dot\phi^2\over\dot\rho^2} \vec\nabla\zeta\cdot\vec\nabla\zeta \biggr\} .
$$ 
The free Hamiltonian for $\zeta_n(t,\vec x)$ is the same.  The leading part of $H_I$ is the set of cubic interactions calculated in the last section; this cubic action is really only the leading part since the theory has an infinite set of higher-order corrections counted according to the numbers of fields appearing in each set of operators---quartic, quintic, {\it etc\/}.

The evolution of a state is found by solving the equation
$$
i {d\over dt}|0(t)\rangle = H_I(t)\, |0(t)\rangle ,
$$
using just the interacting part of the theory.  Introducing a time-evolution operator $U_I(t,t')$ that evolves any state from $t'$ to $t$, for example
$$
|0(t)\rangle = U_I(t,t')\, |0(t')\rangle ,
$$
this operator is found by solving the equation 
$$
{d\over dt}U_I(t,t') = -i H_I(t)\, U_I(t,t') . 
$$
Its solution is given by the operator that was already met a little earlier,
$$
U_I(t,t') = T e^{-i \int_{t'}^t dt^{\prime\prime}\, H_I(t^{\prime\prime})} . 
$$
$T$ orders the fields so that those occurring at the later times are placed to the left and those occurring at the earlier times are placed to the right.  For a pair of fields, for example, this time-ordering can be written as
$$
T\bigl( \zeta_n(t,\vec x)\zeta_n(t',\vec y) \bigr) 
= \Theta(t-t')\, \zeta_n(t,\vec x)\zeta_n(t',\vec y)
+ \Theta(t'-t)\, \zeta_n(t',\vec y)\zeta_n(t,\vec x) . 
$$
Note that while the formalism is being developed here using the vacuum state as an example---because that is the state used in the inflationary calculation---this same time-evolution operator works more generally.  $U_I(t,t')$ generates the time evolution of any quantum state.

Suppose that the system starts at some time $t_0$ in some state $|0(t_0)\rangle$.  An expectation value at an arbitrary time later is given by evolving the states forward from $t_0$ to $t$,
$$
\langle 0(t)|{\cal O}(t)|0(t)\rangle 
= \langle 0(t_0)|U_I^\dagger(t,t_0){\cal O}(t)U_I(t,t_0)|0(t_0)\rangle . 
$$
In this form, it is a bit difficult to apply the diagrammatic techniques that are used to evaluate matrix elements of scattering processes perturbatively.  However, a trick can be invoked to recast this expectation value in a form that is more reminiscent of what is encountered in a scattering calculation.

Reading the expectation value from right to left, 
$$
\langle 0(t_0)|U_I^\dagger(t,t_0){\cal O}(t)U_I(t,t_0)|0(t_0)\rangle , 
$$
the system starts in a state at $t_0$, evolves forward until $t$ where the operator occurs, and then evolves back again to the original state at $t_0$.  The reason for the backwards evolution is that when $U_I^\dagger(t,t_0)$ is conjugated it becomes $U_I(t_0,t)$, which generates a backwards evolution from $t$ to $t_0$,
$$
U_I^\dagger(t,t_0) = U_I(t_0,t) . 
$$
If desired, the time-evolution can also be extended into the infinite future by including the identity in the form
$$
{\mathbb I} = U_I^\dagger(\infty,t) U_I(\infty,t) . 
$$
Then, 
\begin{eqnarray}
\langle 0(t)|{\cal O}(t)|0(t)\rangle 
&\!\!\!\!\!\!=\!\!\!\!\!\!& 
\langle 0(t_0)|U_I^\dagger(t,t_0){\cal O}(t)U_I(t,t_0)|0(t_0)\rangle 
\nonumber \\
&\!\!\!\!\!\!=\!\!\!\!\!\!& 
\langle 0(t_0)|U_I^\dagger(t,t_0){\mathbb I}{\cal O}(t)U_I(t,t_0)|0(t_0)\rangle 
\nonumber \\
&\!\!\!\!\!\!=\!\!\!\!\!\!& 
\langle 0(t_0)|U_I^\dagger(t,t_0)U_I^\dagger(\infty,t) U_I(\infty,t) {\cal O}(t)U_I(t,t_0)|0(t_0)\rangle 
\nonumber \\
&\!\!\!\!\!\!=\!\!\!\!\!\!& 
\langle 0(t_0)|U_I^\dagger(\infty,t_0) U_I(\infty,t) {\cal O}(t)U_I(t,t_0)|0(t_0)\rangle . 
\nonumber 
\end{eqnarray}
Now the little tale has been extended a bit further:  reading from right to left, the system starts in the initial state $|0(t_0)\rangle$, evolves to $t$ where the operator occurs, then continues to evolve out to the infinite future.  Once there, the evolution goes all the way back again to $t_0$ and the original initial state.  Seen thus, the evolution can be viewed as though it flows along a single contour of time as shown in this sketch:
$$
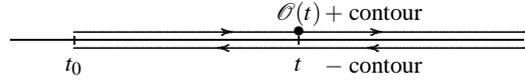
\beginpicture
\setcoordinatesystem units <1.00truept,1.00truept>
\setplotarea x from -48 to 144, y from -8 to 8
\putrule from -48 0 to 144 0
\putrule from -24 1.5 to -24 -1.5
\putrule from  60 1.5 to  60 -1.5
\circulararc 180 degrees from 144 -3 center at 144 0
\plot  -24 3   144  3 /
\plot -24 -3   144 -3 /
\arrow <4pt> [0.2,0.67] from 32  3 to 35  3
\arrow <4pt> [0.2,0.67] from 32 -3 to 30 -3
\arrow <4pt> [0.2,0.67] from 88  3 to 91  3
\arrow <4pt> [0.2,0.67] from 88 -3 to 86 -3
\put {\footnotesize $\bullet$} [c] at 60  3
\put {\footnotesize ${\cal O}(t)$} [c] at 60 10
\put {\footnotesize $t_0$} [c] at -24  -10
\put {\footnotesize $t$} [c] at 60  -9
\put {\footnotesize $+$ contour} [c] at  88  10
\put {\footnotesize $-$ contour} [c] at  88 -10
\endpicture 
$$
The forward and backward parts are considered to be distinct parts of this single continuous contour, rather than a simple retreading of the same path in two directions.  This might appear to be no more than a heuristic trick, but regarding the fields as time-ordered according to the points along {\it this contour\/} makes it possible to write the expectation value in terms of a single time-ordered expression.  To do so, a way must be found to indicate when a field is on one part of the contour or the other.  Let the fields on the forward-running part of the contour, which will be called the `$+$' part, be labelled with a $+$ too: 
$$
\zeta_n^+(t,\vec x) \quad\to\quad\hbox{is on the $+$ contour}.
$$
A field on the backward-running part of the contour, which will be called the `$-$' part, is labelled with a $-$, 
$$
\zeta_n^-(t,\vec x) \quad\to\quad\hbox{is on the $-$ contour}.
$$
Because the `$-$' fields always occur later along the contour than the `$+$' fields, the time-ordering can be extended to a time-ordering along the whole contour.  Two `$+$' fields are ordered in the usual way 
$$
T\bigl( \zeta_n^+(t,\vec x)\zeta_n^+(t',\vec y) \bigr) 
= \Theta(t-t')\, \zeta_n^+(t,\vec x)\zeta_n^+(t',\vec y)
+ \Theta(t'-t)\, \zeta_n^+(t',\vec y)\zeta_n^+(t,\vec x) . 
$$
And because {\it everything\/} on the `$-$' contour occurs after anything on the `$+$' contour, 
\begin{eqnarray}
T\bigl( \zeta_n^+(t,\vec x)\zeta_n^-(t',\vec y) \bigr) 
&\!\!\!\!\!\!=\!\!\!\!\!\!& 
\zeta_n^-(t',\vec y)\zeta_n^+(t,\vec x) . 
\nonumber \\
T\bigl( \zeta_n^-(t,\vec x)\zeta_n^+(t',\vec y) \bigr) 
&\!\!\!\!\!\!=\!\!\!\!\!\!& 
\zeta_n^-(t,\vec x)\zeta_n^+(t',\vec y). 
\nonumber 
\end{eqnarray}
The orientation of the contour on its `$-$' part with respect to the usual sense of the flow of time is the reverse of that on its `$+$' part.  For two fields that are both on the `$-$' part of the contour, the time ordering is in the contour sense, which is the opposite of the coordinate sense,
$$
T\bigl( \zeta_n^-(t,\vec x)\zeta_n^-(t',\vec y) \bigr) 
= \Theta(t'-t)\, \zeta_n^-(t,\vec x)\zeta_n^-(t',\vec y)
+ \Theta(t-t')\, \zeta_n^-(t',\vec y)\zeta_n^-(t,\vec x) . 
$$

Since all of the operators in $U_I(\infty,t){\cal O}U_I(t,t_0)$ are associated with the $+$ contour, they should all be written as $+$ fields,
\begin{eqnarray}
U_I(\infty,t){\cal O}(t)U_I(t,t_0) 
&\!\!\!\!\!\!=\!\!\!\!\!\!& 
T \Bigl( e^{-i \int_t^\infty dt'\, H_I^+(t')} \Bigr)
{\cal O}^+(t)
T \Bigl( e^{-i \int_{t_0}^t dt'\, H_I^+(t')} \Bigr)
\nonumber \\
&\!\!\!\!\!\!=\!\!\!\!\!\!& 
T \Bigl( {\cal O}^+(t) e^{-i \int_{t_0}^\infty dt'\, H_I^+(t')} \Bigr) . 
\nonumber 
\end{eqnarray}
A $+$ or a $-$ superscript just means that the fields in the operator ${\cal O}$ or in the interaction Hamiltonian are `$+$' fields or `$-$' fields; {\it e.g.\/}
$$
H_I^\pm(t) \equiv H_I[\zeta_n^\pm(t,\vec x)] . 
$$
Similarly, the operator $U_I^\dagger(\infty,t_0)=U_I(t_0,\infty)$ is on the $-$ contour; so it should be written in terms of `$-$' fields
$$
U_I^\dagger(\infty,t_0) 
= T \Bigl( e^{-i \int_\infty^{t_0} dt'\, H_I^-(t')} \Bigr)
= T \Bigl( e^{i \int_{t_0}^\infty dt'\, H_I^-(t')} \Bigr) , 
$$
remembering that the time-ordering is the one appropriate for the $\zeta_n^-$ fields.  Since $T$ is now the time-ordering along the entire contour, everything can be combined into a single time-ordered expression, 
\begin{eqnarray}
U_I^\dagger(\infty,t_0)U_I(\infty,t){\cal O}(t)U_I(t,t_0) 
&\!\!\!\!\!\!=\!\!\!\!\!\!& 
T \Bigl( e^{i \int_{t_0}^\infty dt'\, H_I^-(t')} \Bigr)T \Bigl( {\cal O}^+(t) e^{-i \int_{t_0}^\infty dt'\, H_I^+(t')} \Bigr) 
\nonumber \\
&\!\!\!\!\!\!=\!\!\!\!\!\!& 
T \Bigl( {\cal O}^+(t) e^{-i \int_{t_0}^\infty dt'\, [H_I^+(t')-H_I^-(t')]} \Bigr) . 
\nonumber 
\end{eqnarray}
Taking the expectation value of both sides in the initial state $|0(t_0)\rangle$ produces a general expression for how to compute a time-evolving expectation value,
$$
\langle 0(t)|{\cal O}(t)|0(t)\rangle  
= \bigl\langle 0(t)\big|T \bigl( {\cal O}^+(t) e^{-i \int_{t_0}^\infty dt'\, [H_I^+(t')-H_I^-(t')]} \bigr) \big|0(t)\bigr\rangle . 
$$

When the interactions are small, this expectation value can be evaluated perturbatively.  The exponential is expanded in a Taylor series, 
\begin{eqnarray}
&&\!\!\!\!\!\!\!\!\!\!\!\!\!\!\!\!\!\!\!\!\!\!\!\!\!
\langle 0(t)|{\cal O}(t)|0(t)\rangle 
\nonumber \\
&\!\!\!\!\!\!=\!\!\!\!\!\!& 
\langle 0(t_0)| {\cal O}^+(t) |0(t_0)\rangle 
\nonumber \\
&& 
-\,\, i \int_{t_0}^\infty dt'\, 
\langle 0(t_0)|T \bigl( {\cal O}^+(t)[H_I^+(t')-H_I^-(t')] \bigr) |0(t_0)\rangle 
\nonumber \\
&& 
-\,\, {1\over 2} \int_{t_0}^\infty dt'\, \int_{t_0}^\infty dt^{\prime\prime}\, 
\langle 0(t)\big|T \bigl( {\cal O}^+(t) [H_I^+(t')-H_I^-(t')] 
[H_I^+(t^{\prime\prime})-H_I^-(t^{\prime\prime})] \bigr) \big|0(t)\bigr\rangle 
\nonumber \\
&& 
+\,\, \cdots , 
\nonumber 
\end{eqnarray}
and the individual terms in the series are evaluated in more or less the usual way.  For each term, one takes the Wick contractions of the fields to form all the possible connected diagrams.  The only difference from the analogous step in calculating a scattering matrix element is that there are now two types of fields, $\zeta_n^+(t,\vec x)$ and $\zeta_n^-(t,\vec x)$, and correspondingly four possible Wick contractions, 
$$
\beginpicture
\setcoordinatesystem units <1.00truept,1.00truept>
\setplotarea x from 0 to 283, y from 0 to 15
\putrule from  4 14 to  4 18
\putrule from  4 18 to 39 18 
\putrule from 39 18 to 39 14 
\put {$\zeta_n^\pm(t,\vec x)\zeta_n^\pm(t',\vec y) 
= \langle 0(t_0)|T\bigl( \zeta_n^\pm(t,\vec x)\zeta_n^\pm(t',\vec y) \bigr) |0(t_0)\rangle = G^{\pm\pm}(t,\vec x;t',\vec y).$} [l] at 0  6
\endpicture 
$$
The functions $G^{\pm\pm}(t,\vec x;t',\vec y)$ are the four propagators of the theory.  Taking the expectation values of the time-ordered pairs of fields that has already been introduced, these propagators can be written in terms of the two Wightman functions, 
\begin{eqnarray}
G^{++}(t,\vec x;t',\vec y) 
&\!\!\!\!\!\!=\!\!\!\!\!\!& 
\Theta(t-t')\, G^>(t,\vec x;t',\vec y) + \Theta(t'-t)\, G^<(t,\vec x;t',\vec y) 
\nonumber \\
G^{+-}(t,\vec x;t',\vec y) 
&\!\!\!\!\!\!=\!\!\!\!\!\!& 
G^<(t,\vec x;t',\vec y) 
\nonumber \\
G^{-+}(t,\vec x;t',\vec y) 
&\!\!\!\!\!\!=\!\!\!\!\!\!& 
G^>(t,\vec x;t',\vec y) 
\nonumber \\
G^{--}(t,\vec x;t',\vec y) 
&\!\!\!\!\!\!=\!\!\!\!\!\!& 
\Theta(t'-t)\, G^>(t,\vec x;t',\vec y) + \Theta(t-t')\, G^<(t,\vec x;t',\vec y) 
\nonumber 
\end{eqnarray}
where
\begin{eqnarray}
G^>(t,\vec x;t',\vec y) &\!\!\!\!\!\!=\!\!\!\!\!\!& 
\langle 0(t_0)| \zeta_n(t,\vec x)\zeta_n(t',\vec y) |0(t_0)\rangle
\nonumber \\
G^<(t,\vec x;t',\vec y) &\!\!\!\!\!\!=\!\!\!\!\!\!& 
\langle 0(t_0)| \zeta_n(t',\vec y)\zeta_n(t,\vec x) |0(t_0)\rangle . 
\nonumber 
\end{eqnarray}
For the calculation to be done here, $t_0\to -\infty$, and $|0(-\infty)\rangle\equiv |0\rangle$ is the Bunch-Davies state described during the analysis of the two-point function.

This formalism for evaluating the time-evolution of expectation values was introduced by Julian Schwinger and then further developed by Kalyana Mahanthappa and Pradip Bakshi and also by Leonid Keldysh [4].
\vskip18truept

\noindent{\it V.D. The three-point function of the shifted field $\zeta_n$}
\vskip6truept

The three-point function for $\zeta_n$ is generated entirely by the evolution of the state.  This is because odd $n$-point functions vanish when evaluated in the vacuum of the free theory.  Using the general formula for how to evaluate the expectation value in the Schwinger-Keldysh formalism for the case where the operator is the product of three fields produces 
\begin{eqnarray}
&&\!\!\!\!\!\!\!\!\!\!\!\!\!\!\!\!\!\!\!\!\!\!\!\!\!\!\!
\langle 0(t)| \zeta_n(t,\vec x)\zeta_n(t,\vec y)\zeta_n(t,\vec z) |0(t)\rangle 
\nonumber \\
&\!\!\!\!\!\!=\!\!\!\!\!\!& 
\bigl\langle 0\big| 
T \bigl( 
\zeta_n^+(t,\vec x)\zeta_n^+(t,\vec y)\zeta_n^+(t,\vec z) 
e^{-i\int_{-\infty}^t dt'\, [H_I^+(t')-H_I^-(t')]} \bigr) \big|0\bigr\rangle . 
\nonumber 
\end{eqnarray}
To be able to calculate the right side of this equation, at least in a perturbative sense, the interactions must be {\it small\/}.  This is certainly true here since $H_I(t)$ is proportional to $\epsilon^2$.  At leading order in the slow-roll parameters, it suffices to evaluate just the linear term that is produced by the Taylor-series expansion of the exponentials, 
\begin{eqnarray}
&&\!\!\!\!\!\!\!\!\!\!\!\!\!\!\!\!\!\!\!\!\!\!\!\!\!\!\!
\langle 0(t)| \zeta_n(t,\vec x)\zeta_n(t,\vec y)\zeta_n(t,\vec z) |0(t)\rangle 
\nonumber \\
&\!\!\!\!\!\!=\!\!\!\!\!\!& 
-i \int_{-\infty}^t dt'\, \langle 0| 
T \bigl( \zeta_n^+(t,\vec x)\zeta_n^+(t,\vec y)\zeta_n^+(t,\vec z) 
[H_I^+(t')-H_I^-(t')] \bigr) |0\rangle 
+ \cdots .
\nonumber 
\end{eqnarray}
This is the same as the commutator form that was written earlier:  the time-ordering would move the $H_I^-(t')$ to the left of the three $\zeta_n$ fields while the $H_I^+(t')$ would stay where it is, since $t'\le t$.  Of course, here it will be instead evaluated using the Schwinger-Keldysh picture.

The interacting part of the Hamiltonian contains three operators that are of order $\epsilon^2$, 
$$
H_I(t) = - \int d^3\vec x\, \Bigl\{
{\cal O}_1(t,\vec x) + {\cal O}_2(t,\vec x) + {\cal O}_3(t,\vec x) 
+ \cdots \Bigr\} , 
$$
where
\begin{eqnarray}
{\cal O}_1
&\!\!\!\!\!\!=\!\!\!\!\!\!& 
{1\over 4} e^{3\rho} {1\over M_{\rm pl}^2} {\dot\phi^4\over\dot\rho^4} \dot\zeta_n^2 \zeta_n 
\nonumber \\
{\cal O}_2
&\!\!\!\!\!\!=\!\!\!\!\!\!& 
{1\over 4} e^\rho {1\over M_{\rm pl}^2}{\dot\phi^4\over\dot\rho^4} 
\zeta_n \partial_k\zeta_n \partial^k\zeta_n 
\nonumber \\
{\cal O}_3
&\!\!\!\!\!\!=\!\!\!\!\!\!& 
- e^{3\rho} {\dot\phi^2\over\dot\rho^2} \dot\zeta_n \partial_k\zeta_n \partial^k\chi_n 
= - {1\over 2} e^{3\rho} {1\over M_{\rm pl}^2}{\dot\phi^4\over\dot\rho^4} \dot\zeta_n \partial_k\zeta_n \partial^k\bigl( \partial^{-2} \dot\zeta_n \bigr) . 
\nonumber 
\end{eqnarray}
To make the calculation a bit easier to follow, it will be broken into three steps:  the contributions from each of these operators will be considered separately, using the notation
$$
\langle{\cal O}_i(t)\rangle \equiv 
i \int_{-\infty}^t dt'\, \int d^3\vec w\, \bigl\langle 0\big| 
T \Bigl( \zeta_n^+(t,\vec x)\zeta_n^+(t,\vec y)\zeta_n^+(t,\vec z) 
\bigl[ {\cal O}_i^+(t',\vec w) - {\cal O}_i^-(t',\vec w) \bigr] \Bigr) 
\big|0\bigr\rangle .
$$
In each case, the connected diagrams have the same overall structure,
$$
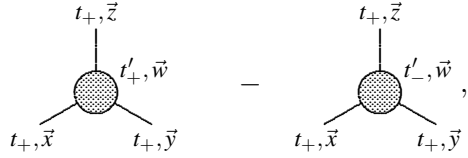
\beginpicture
\setcoordinatesystem units <1.00truept,1.00truept>
\setplotarea x from -30 to 30, y from -20 to 30
\circulararc 360 degrees from 8 0 center at 0 0
\plot  6.93 -4   20.78 -12 /
\plot -6.93 -4  -20.78 -12 /
\plot 0 8  0 24 /
\put {\footnotesize $t_+,\vec x$}  [c] at -24 -18
\put {\footnotesize $t_+,\vec y$}  [c] at  22 -18
\put {\footnotesize $t_+,\vec z$}  [c] at   0  30
\put {\footnotesize $t'_+,\vec w$} [c] at  18   8
\setshadesymbol ({\tmrms .})
\setshadegrid span <0.9pt>
\setquadratic
\hshade -5.65 -5.65 -5.65 <,z,,>  0 -8.00 -5.65   5.65 -5.65 -5.65 /
\vshade -5.65 -5.65  5.65 <z,z,,> 0 -8.00  8.00   5.65 -5.65  5.65 /
\hshade -5.65  5.65  5.65 <z,,,>  0  5.65  8.00   5.65  5.65  5.65 /
\endpicture
\qquad - \quad
\beginpicture
\setcoordinatesystem units <1.00truept,1.00truept>
\setplotarea x from -30 to 30, y from -20 to 30
\circulararc 360 degrees from 8 0 center at 0 0
\plot  6.93 -4   20.78 -12 /
\plot -6.93 -4  -20.78 -12 /
\plot 0 8  0 24 /
\put {\footnotesize $t_+,\vec x$}  [c] at -24 -18
\put {\footnotesize $t_+,\vec y$}  [c] at  22 -18
\put {\footnotesize $t_+,\vec z$}  [c] at   0  30
\put {\footnotesize $t'_-,\vec w$} [c] at  18   8
\setshadesymbol ({\tmrms .})
\setshadegrid span <0.9pt>
\setquadratic
\hshade -5.65 -5.65 -5.65 <,z,,>  0 -8.00 -5.65   5.65 -5.65 -5.65 /
\vshade -5.65 -5.65  5.65 <z,z,,> 0 -8.00  8.00   5.65 -5.65  5.65 /
\hshade -5.65  5.65  5.65 <z,,,>  0  5.65  8.00   5.65  5.65  5.65 /
\endpicture , 
$$
where each of the three `external' fields is contracted with one of the fields in the cubic operator.  Unlike the quartic contributions before, there are now two times appearing here:  the time at which the three-point operator is being evaluated, $t$, and the time $t'$ associated with the cubic interaction.  The $\pm$ subscripts on the times refer to whether the fields are $\zeta_n^+$ or $\zeta_n^-$ fields, or, equivalently, on the `$\pm$' parts of the contour.

Taking all of the possible contractions that produce such a connected graph, and then expanding in Fourier modes, the first of these interactions contributes
\begin{eqnarray}
\langle{\cal O}_1(t)\rangle 
&\!\!\!\!\!\!=\!\!\!\!\!\!& 
{i\over 4} {1\over M_{\rm pl}^2} 
\int_{-\infty}^t dt'\, e^{3\rho(t')} {\dot\phi^4\over\dot\rho^4}
\int d^3\vec w\, \Bigl\{ 
\bigl\langle 0\big| 
T \Bigl( \zeta_n^+(t,\vec x)\zeta_n^+(t,\vec y)\zeta_n^+(t,\vec z) 
\bigl[ \dot\zeta_n^+(t',\vec w)\bigr]^2 \zeta_n^+(t',\vec w) \Bigr) 
\big|0\bigr\rangle 
\nonumber \\
&&\qquad\qquad\qquad\qquad\qquad\quad\ \  
-\,\, \bigl\langle 0\big| 
T \Bigl( \zeta_n^+(t,\vec x)\zeta_n^+(t,\vec y)\zeta_n^+(t,\vec z) 
\bigl[ \dot\zeta_n^-(t',\vec w)\bigr]^2 \zeta_n^-(t',\vec w) \Bigr) 
\big|0\bigr\rangle \Bigr\}
\nonumber \\
&\!\!\!\!\!\!=\!\!\!\!\!\!& 
{i\over 2} {1\over M_{\rm pl}^2} 
\int_{-\infty}^t dt'\, e^{3\rho(t')} {\dot\phi^4\over\dot\rho^4} 
\int {d^3\vec k_1\over (2\pi)^3} {d^3\vec k_2\over (2\pi)^3} 
{d^3\vec k_3\over (2\pi)^3}\, e^{i\vec k_1\cdot\vec x}
e^{i\vec k_2\cdot\vec y} e^{i\vec k_3\cdot\vec z}\,
(2\pi)^3\, \delta^3(\vec k_1+\vec k_2+\vec k_3)
\nonumber \\
&&\qquad\qquad\qquad\qquad\qquad
\Bigl\{ 
\dot G_{k_1}^>(t,t')\dot G_{k_2}^>(t,t') G_{k_3}^>(t,t')
- \dot G_{k_1}^<(t,t')\dot G_{k_2}^<(t,t') G_{k_3}^<(t,t')
\nonumber \\
&&\qquad\qquad\qquad\qquad\qquad
+\,\, \dot G_{k_1}^>(t,t') G_{k_2}^>(t,t')\dot G_{k_3}^>(t,t')
- \dot G_{k_1}^<(t,t') G_{k_2}^<(t,t')\dot G_{k_3}^<(t,t')
\nonumber \\
&&\qquad\qquad\qquad\qquad\qquad
+\,\, G_{k_1}^>(t,t')\dot G_{k_2}^>(t,t')\dot G_{k_3}^>(t,t')
- G_{k_1}^<(t,t')\dot G_{k_2}^<(t,t')\dot G_{k_3}^<(t,t')
\Bigr\} . 
\nonumber 
\end{eqnarray}
As before, the convention is that the time derivatives are with respect to the {\it second\/} of the arguments of the Wightman functions.

Time derivatives of Wightman functions are usually suppressed in the late-time limit; but here there are growing factors present to compensate.  For one, there is the scale factor, $e^{3\rho(t')}$.  And there is another secret factor too, hidden in the measure of the time integration.  In the de Sitter limit,
$$
dt'\, e^{3\rho(t')} = d\eta'\, {dt'\over d\eta'}\, e^{3\rho(\eta')} 
= d\eta'\, e^{\rho(\eta')}\, e^{3\rho(\eta')} 
= {1\over H^4}\, {d\eta'\over\eta^{\prime 4}} . 
$$
The four powers of $1/\eta'$ are just what are needed to cancel the two factors of $\eta^{\prime 2}$ coming from the two differentiated Wightman functions that appear in every term.  Taking the arguments of the `external' fields, $t$ or $\eta$, to late times, and using the appropriate expressions for the Wightman functions that were derived earlier, yields 
\begin{eqnarray}
&&\!\!\!\!\!\!\!\!\!\!\!\!\!\!\!\!
\lim_{t\to\infty} \bigl\{ 
\dot G_{k_1}^>(t,t')\dot G_{k_2}^>(t,t') G_{k_3}^>(t,t')
- \dot G_{k_1}^<(t,t')\dot G_{k_2}^<(t,t') G_{k_3}^<(t,t')\bigr\}
\nonumber \\
&\!\!\!\!\!\!=\!\!\!\!\!\!& 
\biggl( {-1\over 4\epsilon} {H^3\over M_{\rm pl}^2} 
{\eta^{\prime 2}\over k_1} e^{ik_1\eta'} \biggr)
\biggl( {-1\over 4\epsilon} {H^3\over M_{\rm pl}^2} 
{\eta^{\prime 2}\over k_2} e^{ik_2\eta'} \biggr)
\biggl( {1\over 4\epsilon} {H^2\over M_{\rm pl}^2} 
{1\over k_3^3} (1-ik_3\eta') e^{ik_3\eta'} \biggr)
\nonumber \\
&&
-\,\, \biggl( {-1\over 4\epsilon} {H^3\over M_{\rm pl}^2} 
{\eta^{\prime 2}\over k_1} e^{-ik_1\eta'} \biggr)
\biggl( {-1\over 4\epsilon} {H^3\over M_{\rm pl}^2} 
{\eta^{\prime 2}\over k_2} e^{-ik_2\eta'} \biggr)
\biggl( {1\over 4\epsilon} {H^2\over M_{\rm pl}^2} 
{1\over k_3^3} (1+ik_3\eta') e^{-ik_3\eta'} \biggr)
\nonumber \\
&\!\!\!\!\!\!=\!\!\!\!\!\!& 
{1\over 64\epsilon^3} {H^8\over M_{\rm pl}^6} 
{\eta^{\prime 4}\over k_1k_2k_3^3} 
\Bigl[ (1-ik_3\eta') e^{i(k_1+k_2+k_3)\eta'} 
- (1+ik_3\eta') e^{-i(k_1+k_2+k_3)\eta'} \Bigr]
\nonumber \\
&\!\!\!\!\!\!=\!\!\!\!\!\!& 
{i\over 32\epsilon^3} {H^8\over M_{\rm pl}^6} 
{\eta^{\prime 4}\over k_1k_2k_3^3} 
\Bigl[ \sin\bigl[ (k_1+k_2+k_3)\eta'\bigr] 
- k_3\eta' \cos\bigl[(k_1+k_2+k_3)\eta'\bigr]
\Bigr] ,
\nonumber 
\end{eqnarray}
for example.  The expectation value of ${\cal O}_1$ then becomes in the limit $\eta\to 0$
\begin{eqnarray}
\langle{\cal O}_1\rangle 
&\!\!\!\!\!\!=\!\!\!\!\!\!& 
\int {d^3\vec k_1\over (2\pi)^3} {d^3\vec k_2\over (2\pi)^3} 
{d^3\vec k_3\over (2\pi)^3}\, e^{i\vec k_1\cdot\vec x}
e^{i\vec k_2\cdot\vec y} e^{i\vec k_3\cdot\vec z}\,
(2\pi)^3\, \delta^3(\vec k_1+\vec k_2+\vec k_3)
\nonumber \\
&&\times
{-1\over 64\epsilon^3} {H^4\over M_{\rm pl}^8} {\dot\phi^4\over\dot\rho^4}
{1\over k_1^3k_2^3k_3^3}  \int_{-\infty}^0 d\eta'\, 
\Bigl\{ 
k_1^2k_2^2 \bigl[ \sin K\eta' - k_3\eta'\cos K\eta' \bigr] 
\nonumber \\
&&\qquad\qquad\qquad\qquad\qquad\qquad\!\!
+\,\, k_1^2k_3^2 \bigl[ \sin K\eta' - k_2\eta'\cos K\eta' \bigr] 
\nonumber \\
&&\qquad\qquad\qquad\qquad\qquad\qquad\!\!
+\,\, k_2^2k_3^2 \bigl[ \sin K\eta' - k_1\eta'\cos K\eta' \bigr] 
\Bigr\} . 
\nonumber 
\end{eqnarray}
Here, 
$$
K\equiv k_1+k_2+k_3 
$$
is the sum of the {\it magnitudes\/} of the spatial momenta.  Performing the $\eta'$ integrations then produces
\begin{eqnarray}
\langle{\cal O}_1\rangle 
&\!\!\!\!\!\!=\!\!\!\!\!\!& 
\int {d^3\vec k_1\over (2\pi)^3} {d^3\vec k_2\over (2\pi)^3} 
{d^3\vec k_3\over (2\pi)^3}\, e^{i\vec k_1\cdot\vec x}
e^{i\vec k_2\cdot\vec y} e^{i\vec k_3\cdot\vec z}\,
(2\pi)^3\, \delta^3(\vec k_1+\vec k_2+\vec k_3)
\nonumber \\
&&
{1\over 32\epsilon^2} {H^4\over M_{\rm pl}^4} 
{1\over k_1^3k_2^3k_3^3} 
\biggl\{ 
{1\over 2\epsilon} {1\over M_{\rm pl}^4}{\dot\phi^4\over\dot\rho^4} \biggl[ 
{1\over K} \bigl[ k_1^2k_2^2 + k_1^2k_3^2 + k_2^2k_3^2 \bigr] 
+ {k_1k_2k_3\over K^2} \bigl[ k_1k_2 + k_1k_3 + k_2k_3 \bigr] 
\biggr] \biggr\} . 
\nonumber 
\end{eqnarray}
Notice that in order to put this expectation value into the standard form introduced at the beginning of this section, one factor of $1/\epsilon$ has been separated from the others and included in the expression within the braces.  Using the fact that $\epsilon = {1\over 2}{1\over M_{\rm pl}^2}{\dot\phi^2\over\dot\rho^2}$, this expectation value can be also written as 
\begin{eqnarray}
\langle{\cal O}_1\rangle 
&\!\!\!\!\!\!=\!\!\!\!\!\!& 
\int {d^3\vec k_1\over (2\pi)^3} {d^3\vec k_2\over (2\pi)^3} 
{d^3\vec k_3\over (2\pi)^3}\, e^{i\vec k_1\cdot\vec x}
e^{i\vec k_2\cdot\vec y} e^{i\vec k_3\cdot\vec z}\,
(2\pi)^3\, \delta^3(\vec k_1+\vec k_2+\vec k_3)
\nonumber \\
&&
{1\over 32\epsilon^2} {H^4\over M_{\rm pl}^4} 
{1\over k_1^3k_2^3k_3^3} 
\biggl\{ 
{1\over M_{\rm pl}^2} {\dot\phi^2\over\dot\rho^2} \biggl[ 
{1\over K} \bigl[ k_1^2k_2^2 + k_1^2k_3^2 + k_2^2k_3^2 \bigr] 
+ {k_1k_2k_3\over K^2} \bigl[ k_1k_2 + k_1k_3 + k_2k_3 \bigr] 
\biggr] \biggr\} . 
\nonumber 
\end{eqnarray}

The contribution from the second of the operators, ${\cal O}_2$, is found by following exactly the same procedure.  First, the explicit form of ${\cal O}_2$ is substituted into the general formula, 
\begin{eqnarray}
\langle{\cal O}_2\rangle 
&\!\!\!\!\!\!=\!\!\!\!\!\!& 
{i\over 4} \int_{-\infty}^t dt'\, e^{\rho(t')} 
{1\over M_{\rm pl}^2} {\dot\phi^4\over\dot\rho^4}
\nonumber \\
&&
\int d^3\vec w\, \Bigl\{ 
\bigl\langle 0\big| 
T \bigl( \zeta_n^+(t,\vec x)\zeta_n^+(t,\vec y)\zeta_n^+(t,\vec z) 
\zeta_n^+(t',\vec w) \partial_k\zeta_n^+(t',\vec w) 
\partial^k\zeta_n^+(t',\vec w) \bigr) 
\big|0\bigr\rangle 
\nonumber \\
&&\qquad\ \,\,
-\,\, \bigl\langle 0\big| 
T \bigl( \zeta_n^+(t,\vec x)\zeta_n^+(t,\vec y)\zeta_n^+(t,\vec z) 
\zeta_n^-(t',\vec w) \partial_k\zeta_n^-(t',\vec w) 
\partial^k\zeta_n^-(t',\vec w) \bigr) 
\big|0\bigr\rangle \Bigr\} . 
\nonumber 
\end{eqnarray}
Taking all the Wick contractions that lead to connected diagrams, and then expanding the resulting Wightman functions in their Fourier modes, yields
\begin{eqnarray}
\langle{\cal O}_2\rangle 
&\!\!\!\!\!\!=\!\!\!\!\!\!& 
- {i\over 2} \int_{-\infty}^t dt'\, e^{\rho(t')} 
{1\over M_{\rm pl}^2} {\dot\phi^4\over\dot\rho^4} 
\int {d^3\vec k_1\over (2\pi)^3} {d^3\vec k_2\over (2\pi)^3} 
{d^3\vec k_3\over (2\pi)^3}\, e^{i\vec k_1\cdot\vec x}
e^{i\vec k_2\cdot\vec y} e^{i\vec k_3\cdot\vec z}\,
(2\pi)^3\, \delta^3(\vec k_1+\vec k_2+\vec k_3)
\nonumber \\
&&\quad\times
\bigl[ \vec k_1\cdot\vec k_2 + \vec k_1\cdot\vec k_3 + \vec k_2\cdot\vec k_3 
\bigr] \bigl\{ 
G_{k_1}^>(t,t') G_{k_2}^>(t,t') G_{k_3}^>(t,t')
- G_{k_1}^<(t,t') G_{k_2}^<(t,t') G_{k_3}^<(t,t')
\bigr\} . 
\nonumber 
\end{eqnarray}
The prefactor on the second line depends on the scalar products of the various momenta.  It can also be expressed solely in terms of the magnitudes of these momenta.  The sum of all of the momenta must vanish.  So squaring their sum, 
$$
\bigl( \vec k_1+\vec k_2+\vec k_3 \bigr)^2 
= k_1^2 + k_2^2 + k_3^2 
+ 2\vec k_1\cdot\vec k_2 + 2\vec k_1\cdot\vec k_3 + 2\vec k_2\cdot\vec k_3 = 0 ,
$$
generates the relation
$$\textstyle 
\vec k_1\cdot\vec k_2 + \vec k_1\cdot\vec k_3 + \vec k_2\cdot\vec k_3 
= - {1\over 2} \bigl[ k_1^2 + k_2^2 + k_3^2 \bigr] . 
$$

Since there are no time derivatives in this operator, the inverse powers of the conformal time from the measure and the scale factor, 
$$
\int_{-\infty}^t dt'\, e^{\rho(t')} \cdots 
= \int_{-\infty}^\eta d\eta'\, {dt'\over\eta'}\, e^{\rho(\eta')} \cdots 
= \int_{-\infty}^\eta d\eta'\, {1\over H^2\eta^{\prime 2}} \cdots , 
$$
are not cancelled by a overall power of $\eta^{\prime\, 2}$ from the products of Wightman functions.  Of course, the contribution from this operator is still finite in the late-time limit, $\eta\to 0$; but this happens because the divergences cancel amongst the individual terms once the $d\eta'$ integral has been performed.  The product of the Wightman functions that appears in the integrand is 
\begin{eqnarray}
&&\!\!\!\!\!\!\!\!\!\!\!\!\!\!\!\!\!\! 
\int_{-\infty}^\eta {d\eta'\over H^2\eta^{\prime 2}} 
\bigl[ G_{k_1}^>(t,t') G_{k_2}^>(t,t') G_{k_3}^>(t,t')
- G_{k_1}^<(t,t') G_{k_2}^<(t,t') G_{k_3}^<(t,t') \bigr] 
\nonumber \\
&\!\!\!\!\!\!=\!\!\!\!\!\!&
{i\over 32\epsilon^3} {H^4\over M_{\rm pl}^6} {1\over k_1^3k_2^3k_3^3} 
\int_{-\infty}^\eta d\eta'  
\biggl\{ 
{\sin\bigl[K(\eta'-\eta)\bigr] - K(\eta'-\eta) \cos\bigl[K(\eta'-\eta)\bigr] 
\over\eta^{\prime 2}}
\nonumber \\
&&\qquad\qquad\qquad\qquad
-\,\, [k_1k_2+k_1k_3+k_2k_3] \sin\bigl[K(\eta'-\eta)\bigr] 
+ k_1k_2k_3\eta' \cos\bigl[K(\eta'-\eta)\bigr] 
+ \cdots 
\biggr\} . 
\nonumber \\
&\!\!\!\!\!\!=\!\!\!\!\!\!&
{i\over 32\epsilon^3} {H^4\over M_{\rm pl}^6} {1\over k_1^3k_2^3k_3^3} 
\biggl\{ 
- K + {1\over K} [k_1k_2+k_1k_3+k_2k_3] + {1\over K^2} k_1k_2k_3 
+ \cdots \biggr\} ,
\nonumber 
\end{eqnarray}
where the terms that have not been written explicitly do not contribute in the limit where $\eta\to 0$.  Putting this result into the expression for the contribution from the second operator yields  
\begin{eqnarray}
\langle{\cal O}_2\rangle 
&\!\!\!\!\!\!=\!\!\!\!\!\!&  
\int {d^3\vec k_1\over (2\pi)^3} {d^3\vec k_2\over (2\pi)^3} 
{d^3\vec k_3\over (2\pi)^3}\, e^{i\vec k_1\cdot\vec x}
e^{i\vec k_2\cdot\vec y} e^{i\vec k_3\cdot\vec z}\,
(2\pi)^3\, \delta^3(\vec k_1+\vec k_2+\vec k_3)
\nonumber \\
&&\times
{1\over 32\epsilon^2} {H^4\over M_{\rm pl}^4} {1\over k_1^3k_2^3k_3^3}
\biggl\{ {1\over 4\epsilon} {1\over M_{\rm pl}^4}{\dot\phi^4\over\dot\rho^4} 
\bigl[ k_1^2 + k_2^2 + k_3^2 \bigr] \biggl[ 
K - {1\over K} \bigl[ k_1k_2 + k_1k_3 + k_2k_3 \bigr] 
- {1\over K^2} k_1k_2k_3
\biggr] \biggr\} . 
\nonumber 
\end{eqnarray}
The $\epsilon$ inside the braces can be converted back into ${1\over 2}{1\over M_{\rm pl}^2}{\dot\phi^2\over\dot\rho^2}$ to give 
\begin{eqnarray}
\langle{\cal O}_2\rangle 
&\!\!\!\!\!\!=\!\!\!\!\!\!&  
\int {d^3\vec k_1\over (2\pi)^3} {d^3\vec k_2\over (2\pi)^3} 
{d^3\vec k_3\over (2\pi)^3}\, e^{i\vec k_1\cdot\vec x}
e^{i\vec k_2\cdot\vec y} e^{i\vec k_3\cdot\vec z}\,
(2\pi)^3\, \delta^3(\vec k_1+\vec k_2+\vec k_3)
\nonumber \\
&&\times
{1\over 32\epsilon^2} {H^4\over M_{\rm pl}^4} {1\over k_1^3k_2^3k_3^3}
\biggl\{ {1\over 2} {1\over M_{\rm pl}^2} {\dot\phi^2\over\dot\rho^2} 
\bigl[ k_1^2 + k_2^2 + k_3^2 \bigr] \biggl[ 
K - {1\over K} \bigl[ k_1k_2 + k_1k_3 + k_2k_3 \bigr] 
- {1\over K^2} k_1k_2k_3
\biggr] \biggr\} . 
\nonumber 
\end{eqnarray}

Before going on to the third operator, and as a preparation for combining all of the different contributions together it will be useful to rearrange the momentum-dependent quantity within the braces.  Multiplying the rest of the terms by the $k_1^2+k_2^2+k_3^2$ factor produces
\begin{eqnarray}
&&\!\!\!\!\!\!\!\!\!\!\!\!\!\!\!
\bigl[ k_1^2 + k_2^2 + k_3^2 \bigr] \biggl[ 
K - {1\over K} \bigl[ k_1k_2 + k_1k_3 + k_2k_3 \bigr] 
- {1\over K^2} k_1k_2k_3 \biggr] 
\nonumber \\
&\!\!\!\!\!\!=\!\!\!\!\!\!& 
(k_1^3 + k_2^3 + k_3^3) 
+ (k_1^2k_2 + k_1k_2^2 + k_1^2k_3 + k_1k_3^2 + k_2^2k_3 + k_2k_3^2) 
\nonumber \\
&&
-\,\, {1\over K} (k_1^3k_2 + k_1k_2^3 + k_1^3k_3 + k_1k_3^3 + k_2^3k_3 + k_2k_3^3) 
- k_1k_2k_3 - {1\over K^2} k_1k_2k_3 \bigl[ k_1^2 + k_2^2 + k_3^2 \bigr]
\nonumber .
\end{eqnarray}
The first term on the last line can be rewritten by using the relation
$$
k_1^3k_2+k_1k_2^3+k_1^3k_3+k_1k_3^3+k_2^3k_3+k_2k_3^3 
= K \bigl( k_1^3+k_2^3+k_3^3 \bigr) - \bigl( k_1^4 + k_2^4 + k_3^4 \bigr) . 
$$
Applying this relation, the contribution from the second operator becomes 
\begin{eqnarray}
\langle{\cal O}_2\rangle 
&\!\!\!\!\!\!=\!\!\!\!\!\!&  
\int {d^3\vec k_1\over (2\pi)^3} {d^3\vec k_2\over (2\pi)^3} 
{d^3\vec k_3\over (2\pi)^3}\, e^{i\vec k_1\cdot\vec x}
e^{i\vec k_2\cdot\vec y} e^{i\vec k_3\cdot\vec z}\,
(2\pi)^3\, \delta^3(\vec k_1+\vec k_2+\vec k_3)
\nonumber \\
&&\times
{1\over 32\epsilon^2} {H^4\over M_{\rm pl}^4} {1\over k_1^3k_2^3k_3^3}
\biggl\{ {1\over 2} {1\over M_{\rm pl}^2} {\dot\phi^2\over\dot\rho^2} \biggr\} 
\biggl\{ 
(k_1^2k_2 + k_1k_2^2 + k_1^2k_3 + k_1k_3^2 + k_2^2k_3 + k_2k_3^2) 
\nonumber \\
&&\qquad\qquad\qquad\qquad
+ {1\over K} \bigl( k_1^4 + k_2^4 + k_3^4 \bigr)
- k_1k_2k_3 - {1\over K^2} k_1k_2k_3 \bigl[ k_1^2 + k_2^2 + k_3^2 \bigr]
\biggr\} . 
\nonumber 
\end{eqnarray}

This leaves only one more operator to evaluate,
$$
{\cal O}_3 = - {1\over 2} e^{3\rho} {1\over M_{\rm pl}^2} 
{\dot\phi^4\over\dot\rho^4} \dot\zeta_n \partial_k\zeta_n \partial^k\bigl( \partial^{-2} \dot\zeta_n \bigr) . 
$$
Because each of the fields in this operator has a different set of derivatives acting on it---the first has a time derivative, the second has a spatial derivative, and the third has a bit of everything---its analysis is slightly more complicated than the previous two operators.  Taking the Wick contractions that produce connected graphs, and expanding in Fourier modes, the leading contribution from the third operator to the three-point function is 
\begin{eqnarray}
\langle{\cal O}_3\rangle 
&\!\!\!\!\!\!=\!\!\!\!\!\!& 
- {i\over 2} \int_{-\infty}^t dt'\, e^{3\rho(t')} 
{1\over M_{\rm pl}^2} {\dot\phi^4\over\dot\rho^4}
\nonumber \\
&&
\int d^3\vec w\, \Bigl\{ 
\bigl\langle 0\big| 
T \bigl( \zeta_n^+(t,\vec x)\zeta_n^+(t,\vec y)\zeta_n^+(t,\vec z) 
\dot\zeta_n^+(t',\vec w) \partial_k\zeta_n^+(t',\vec w) \partial^k\bigl( \partial^{-2} \dot\zeta_n^+(t',\vec w) \bigr) \bigr) 
\big|0\bigr\rangle 
\nonumber \\
&&\qquad\ \,\,
-\,\, \bigl\langle 0\big| 
T \bigl( \zeta_n^+(t,\vec x)\zeta_n^+(t,\vec y)\zeta_n^+(t,\vec z) 
\dot\zeta_n^-(t',\vec w) \partial_k\zeta_n^-(t',\vec w) \partial^k\bigl( \partial^{-2} \dot\zeta_n^-(t',\vec w) \bigr) \bigr) 
\big|0\bigr\rangle \Bigr\} 
\nonumber \\
&\!\!\!\!\!\!=\!\!\!\!\!\!& 
- {i\over 2} {1\over M_{\rm pl}^2} 
\int_{-\infty}^t dt'\, e^{3\rho(t')} {\dot\phi^4\over\dot\rho^4} 
\int {d^3\vec k_1\over (2\pi)^3} {d^3\vec k_2\over (2\pi)^3} 
{d^3\vec k_3\over (2\pi)^3}\, e^{i\vec k_1\cdot\vec x}
e^{i\vec k_2\cdot\vec y} e^{i\vec k_3\cdot\vec z}\,
(2\pi)^3\, \delta^3(\vec k_1+\vec k_2+\vec k_3)
\nonumber \\
&&\quad\times
\biggl\{ 
\biggl[ {\vec k_1\cdot\vec k_3\over k_1^2} + {\vec k_2\cdot\vec k_3\over k_2^2} \biggr]
\Bigl[ \dot G_{k_1}^>(t,t') \dot G_{k_2}^>(t,t') G_{k_3}^>(t,t')
- \dot G_{k_1}^<(t,t') \dot G_{k_2}^<(t,t') G_{k_3}^<(t,t') \Bigr]
\nonumber \\
&&\quad\ \ \,
+\,\,  
\biggl[ {\vec k_1\cdot\vec k_2\over k_1^2} + {\vec k_3\cdot\vec k_2\over k_3^2} \biggr]
\Bigl[ \dot G_{k_1}^>(t,t') G_{k_2}^>(t,t') \dot G_{k_3}^>(t,t')
- \dot G_{k_1}^<(t,t') G_{k_2}^<(t,t') \dot G_{k_3}^<(t,t') \Bigr]
\nonumber \\
&&\quad\ \ \,
+\,\, 
\biggl[ {\vec k_2\cdot\vec k_1\over k_2^2} + {\vec k_3\cdot\vec k_1\over k_3^2} \biggr]
\Bigl[ G_{k_1}^>(t,t') \dot G_{k_2}^>(t,t') \dot G_{k_3}^>(t,t')
- G_{k_1}^<(t,t') \dot G_{k_2}^<(t,t') \dot G_{k_3}^<(t,t') \Bigr]
\biggr\} . 
\nonumber 
\end{eqnarray}
The presence of so many time derivatives again simplifies the late-time limit of the combinations of Wightman functions.  Although the spatial momentum structure is noticeably different, the same combinations of Wightman functions appear here as appeared in the expectation value of the first operator.  And once again there are four inverse powers of $\eta'$ to cancel the $\eta^{\prime 4}$ factor from these time derivatives of the Wightman functions, 
$$
\int_{-\infty}^\infty dt'\, e^{3\rho(t')} \cdots 
= \int_{-\infty}^0 d\eta'\, {dt'\over\eta'}\, e^{3\rho(\eta')} \cdots 
= \int_{-\infty}^0 d\eta'\, {1\over H^4\eta^{\prime 4}} \cdots .
$$
Integrating the combination of Wightman functions then yields
\begin{eqnarray}
&&\!\!\!\!\!\!\!\!\!\!\!\!\!\!\!\!\!\!\!\!\!\!\!\!\!
\int_{-\infty}^\infty dt'\, e^{3\rho(t')} 
\Bigl[ \dot G_{k_1}^>(\infty,t') \dot G_{k_2}^>(\infty,t') G_{k_3}^>(\infty,t')
- \dot G_{k_1}^<(\infty,t') \dot G_{k_2}^<(\infty,t') G_{k_3}^<(\infty,t') \Bigr]
\nonumber \\
&\!\!\!\!\!\!=\!\!\!\!\!\!& 
{i\over 32\epsilon^3} {H^4\over M_{\rm pl}^6} {1\over k_1k_2k_3^3} \int_{-\infty}^0 d\eta'\, \bigl[ \sin K\eta' - k_3\eta'\cos K\eta' \bigr] 
\nonumber \\
&\!\!\!\!\!\!=\!\!\!\!\!\!& 
- {i\over 32\epsilon^3} {H^4\over M_{\rm pl}^6} {1\over k_1k_2k_3^3} 
\biggl\{ {1\over K} + {k_3\over K^2} \biggr\}
= - {i\over 32\epsilon^3} {H^4\over M_{\rm pl}^6} 
{k_1^2k_2^2\over k_1^3k_2^3k_3^3} 
\biggl\{ {1\over K} + {k_3\over K^2} \biggr\} . 
\nonumber 
\end{eqnarray}

The other two lines in the earlier expression for $\langle{\cal O}_3\rangle$ are just cyclic permutations of this result; thus
\begin{eqnarray}
\langle{\cal O}_3\rangle 
&\!\!\!\!\!\!=\!\!\!\!\!\!& 
\int_{-\infty}^t dt'\, e^{3\rho(t')} 
\int {d^3\vec k_1\over (2\pi)^3} {d^3\vec k_2\over (2\pi)^3} 
{d^3\vec k_3\over (2\pi)^3}\, e^{i\vec k_1\cdot\vec x}
e^{i\vec k_2\cdot\vec y} e^{i\vec k_3\cdot\vec z}\,
(2\pi)^3\, \delta^3(\vec k_1+\vec k_2+\vec k_3)
\nonumber \\
&&\quad\times
{1\over 32\epsilon^2} {H^4\over M_{\rm pl}^6} {1\over k_1^3k_2^3k_3^3} 
\biggl\{ - {1\over 2\epsilon} {1\over M_{\rm pl}^2} 
{\dot\phi^4\over\dot\rho^4} \biggr\} 
\biggl\{ 
\Bigl[ k_1^2 (\vec k_2\cdot\vec k_3) + k_2^2 (\vec k_1\cdot\vec k_3) \Bigr]
\biggl\{ {1\over K} + {k_3\over K^2} \biggr\} 
\nonumber \\
&&\qquad\qquad\qquad\qquad\qquad\qquad\qquad\ \
+\,\,  
\Bigl[ k_1^2 (\vec k_2\cdot\vec k_3) + k_3^2 (\vec k_1\cdot\vec k_2) \Bigr]
\biggl\{ {1\over K} + {k_2\over K^2} \biggr\} 
\nonumber \\
&&\qquad\qquad\qquad\qquad\qquad\qquad\qquad\ \ 
+\,\, 
\Bigl[ k_2^2 (\vec k_1\cdot\vec k_3) + k_3^2 (\vec k_1\cdot\vec k_2) \Bigr]
\biggl\{ {1\over K} + {k_1\over K^2} \biggr\} 
\biggr\} . 
\nonumber 
\end{eqnarray}
The scalar products of momenta will again be converted into expressions that only depend on the magnitudes of the vectors, $k_1$, $k_2$, and $k_3$.  For example, the scalar product $\vec k_1\cdot\vec k_2$, can be written as
$$\textstyle 
\vec k_1\cdot\vec k_2 
= {1\over 2}\bigl[ k_1^2 + 2\vec k_1\cdot\vec k_2 + k_2^2 \bigr] 
- {1\over 2} k_1^2 - {1\over 2} k_2^2 
= {1\over 2}\bigl( \vec k_1+\vec k_2 \bigr)^2 
- {1\over 2} k_1^2 - {1\over 2} k_2^2 .
$$
Using the conservation of momentum, the sum of the two can be replaced by the third, $\vec k_1+\vec k_2= -\vec k_3$.  Thus 
$$\textstyle 
\vec k_1\cdot\vec k_2 
= {1\over 2} \bigl[ k_3^2 - k_1^2 - k_2^2 \bigr] ;
$$
and similarly 
$$\textstyle 
\vec k_1\cdot\vec k_3 = {1\over 2} \bigl[ k_2^2 - k_1^2 - k_3^2 \bigr] 
\qquad\hbox{and}\qquad
\vec k_2\cdot\vec k_3 = {1\over 2} \bigl[ k_1^2 - k_2^2 - k_3^2 \bigr] .
$$
Substituting these expressions into the contribution from the third operator, and collecting all of the terms with a common structure, gives 
\begin{eqnarray}
\langle{\cal O}_3\rangle 
&\!\!\!\!\!\!=\!\!\!\!\!\!& 
\int {d^3\vec k_1\over (2\pi)^3} {d^3\vec k_2\over (2\pi)^3} 
{d^3\vec k_3\over (2\pi)^3}\, e^{i\vec k_1\cdot\vec x}
e^{i\vec k_2\cdot\vec y} e^{i\vec k_3\cdot\vec z}\,
(2\pi)^3\, \delta^3(\vec k_1+\vec k_2+\vec k_3)
\nonumber \\
&&\times
{1\over 32\epsilon^2} {H^4\over M_{\rm pl}^4} {1\over k_1^3k_2^3k_3^3} 
\biggl\{ {1\over M_{\rm pl}^2} {\dot\phi^2\over\dot\rho^2} \biggr\} 
\biggl\{ 
- {1\over K} \bigl( k_1^4+k_2^4+k_3^4 \bigr)
+ {2\over K} \bigl( k_1^2k_2^2+k_1^2k_3^2+k_2^2k_3^2 \bigr)
\nonumber \\
&&\qquad\qquad\qquad\qquad\qquad\qquad\ \ 
-\,\, {1\over 2} {1\over K^2}
\bigl( k_1^4k_2+k_1k_2^4+k_1^4k_3+k_1k_3^4+k_2^4k_3+k_2k_3^4 \bigr)
\nonumber \\
&&\qquad\qquad\qquad\qquad\qquad\qquad\ \ 
+\,\, {1\over 2} {1\over K^2}
\bigl( k_1^3k_2^2+k_1^2k_2^3+k_1^3k_3^2+k_1^2k_3^3+k_2^3k_3^2+k_2^2k_3^3 \bigr)
\nonumber \\
&&\qquad\qquad\qquad\qquad\qquad\qquad\ \ 
+\,\, {1\over K^2} k_1k_2k_3 \bigl( k_1k_2+k_1k_3+k_2k_3 \bigr)
\biggr\} .
\nonumber 
\end{eqnarray}

Here too it is useful first to reorganise the momentum structure a bit.  The goal will be to try to extract factors of $K=k_1+k_2+k_3$ from the more complicated terms that have $1/K^2$ prefactors until the only such terms that remain have a factor $k_1k_2k_3/K^2$ since these terms already appear in the contributions from the first two operators.

To begin, notice that one of the combinations of the momenta can be rewritten as 
\begin{eqnarray}
k_1^4k_2+k_1k_2^4+k_1^4k_3+k_1k_3^4+k_2^4k_3+k_2k_3^4
&\!\!\!\!\!\!=\!\!\!\!\!\!&
K \bigl( k_1^3k_2+k_1k_2^3+k_1^3k_3+k_1k_3^3+k_2^3k_3+k_2k_3^3 \bigr)
\nonumber \\
&&
-\,\, \bigl( k_1^3k_2^2+k_1^2k_2^3+k_1^3k_3^2+k_1^2k_3^3+k_2^3k_3^2+k_2^2k_3^3 \bigr)
\nonumber \\
&&
-\,\, 2k_1k_2k_3 \bigl( k_1^2 + k_2^2 + k_3^2 \bigr) .
\nonumber 
\end{eqnarray}
The factor in the first term on the right side was already encountered while rearranging the momentum-dependent factors in the expectation value of the second operator.  The second term on the right side, which also occurs on its own in the expectation value of ${\cal O}_3$, can be expanded as 
\begin{eqnarray}
k_1^3k_2^2+k_1^2k_2^3+k_1^3k_3^2+k_1^2k_3^3+k_2^3k_3^2+k_2^2k_3^3 
&\!\!\!\!\!\!=\!\!\!\!\!\!&
K \bigl( k_1^2k_2^2+k_1^2k_3^2+k_2^2k_3^2 \bigr)
\nonumber \\
&&
-\,\, k_1k_2k_3 \bigl( k_1k_2 + k_1k_3 + k_2k_3 \bigr) ,
\nonumber 
\end{eqnarray}
Thus, taken together,
\begin{eqnarray}
&&\!\!\!\!\!\!\!\!\!\!\!\!\!\!\!\!\!\!\!\!
k_1^4k_2+k_1k_2^4+k_1^4k_3+k_1k_3^4+k_2^4k_3+k_2k_3^4
-\,\, \bigl( k_1^3k_2^2+k_1^2k_2^3+k_1^3k_3^2+k_1^2k_3^3+k_2^3k_3^2+k_2^2k_3^3 
\nonumber \\
&\!\!\!\!\!\!=\!\!\!\!\!\!&
K^2 \bigl( k_1^3+k_2^3+k_3^3 \bigr)
- K \bigl( k_1^4 + k_2^4 + k_3^4 \bigr) 
- 2K \bigl( k_1^2k_2^2+k_1^2k_3^2+k_2^2k_3^2 \bigr) 
\nonumber \\
&&
+\,\, 2k_1k_2k_3 \bigl( k_1k_2 + k_1k_3 + k_2k_3 \bigr)
- 2k_1k_2k_3 \bigl( k_1^2 + k_2^2 + k_3^2 \bigr) .
\nonumber 
\end{eqnarray}
Substituting this relation into the contribution from ${\cal O}_3$ produces 
\begin{eqnarray}
\langle{\cal O}_3\rangle 
&\!\!\!\!\!\!=\!\!\!\!\!\!& 
\int {d^3\vec k_1\over (2\pi)^3} {d^3\vec k_2\over (2\pi)^3} 
{d^3\vec k_3\over (2\pi)^3}\, e^{i\vec k_1\cdot\vec x}
e^{i\vec k_2\cdot\vec y} e^{i\vec k_3\cdot\vec z}\,
(2\pi)^3\, \delta^3(\vec k_1+\vec k_2+\vec k_3)
\nonumber \\
&&\times
{1\over 32\epsilon^2} {H^4\over M_{\rm pl}^4} {1\over k_1^3k_2^3k_3^3} 
\biggl\{ {1\over 2} {1\over M_{\rm pl}^2} {\dot\phi^2\over\dot\rho^2} \biggr\} 
\biggl\{ 
- \bigl( k_1^3+k_2^3+k_3^3 \bigr)
+ {6\over K} \bigl( k_1^2k_2^2+k_1^2k_3^2+k_2^2k_3^2 \bigr)
\nonumber \\
&&\qquad\qquad\qquad\qquad\qquad\qquad\ 
- {1\over K} \bigl( k_1^4+k_2^4+k_3^4 \bigr)
+ {2k_1k_2k_3\over K^2} \bigl( k_1^2 + k_2^2 + k_3^2 \bigr) 
\biggr\} .
\nonumber 
\end{eqnarray}
\vskip18truept

\noindent{\it V.E. Combining and simplifying the three-point function}
\vskip6truept

The ingredients for the the three-point function have now been calculated; all that remains to assemble them and to simplify the result.  At the very beginning of this section, the following standard form for the momentum representation of the three-point function was introduced,
\begin{eqnarray}
&&\!\!\!\!\!\!\!\!\!\!\!\!\!\!\!\!\!\!\!\!\!\!\!
\langle 0(t)|\zeta(t,\vec x)\zeta(t,\vec y)\zeta(t,\vec z)|0(t)\rangle 
\nonumber \\
&\!\!\!\!\!\!=\!\!\!\!\!\!& 
\int {d^3\vec k_1\over (2\pi)^3} {d^3\vec k_2\over (2\pi)^3} 
{d^3\vec k_3\over (2\pi)^3}\, e^{i\vec k_1\cdot\vec x}
e^{i\vec k_2\cdot\vec y} e^{i\vec k_3\cdot\vec z}\,
(2\pi)^3\, \delta^3(\vec k_1+\vec k_2+\vec k_3)
\nonumber \\
&&\quad\times
{1\over 32\epsilon^2} {H^4\over M_{\rm pl}^4} {1\over k_1^3k_2^3k_3^3} 
{\cal A}_{k_1,k_2,k_3}(t) ,
\nonumber 
\end{eqnarray}
which is evaluated in the late-time limit.  Each of the contributions that have been calculated have already been put into this general form.  The complete amplitude, at least to leading order in $\epsilon$ and $\delta$, is obtained by putting everything together 
$$
{\cal A}_{k_1,k_2,k_3} = {\cal A}_1 + {\cal A}_2 + {\cal A}_3 + {\cal A}_f .
$$
Here, ${\cal A}_f$ is the contribution from the quartic operator, 
$$
{\cal A}_f 
= \biggl[ 2 {\ddot\phi\over\dot\phi\dot\rho} 
+ {1\over M_{\rm pl}^2} {\dot\phi^2\over\dot\rho^2}\biggr] 
\bigl( k_1^3 + k_2^3 + k_3^3 \bigr) + \cdots ,
$$
while the ${\cal A}_i$'s are the contributions from the operators ${\cal O}_i$ in the cubic action, 
\begin{eqnarray}
{\cal A}_1
&\!\!\!\!\!\!=\!\!\!\!\!\!&
{1\over M_{\rm pl}^2} {\dot\phi^2\over\dot\rho^2} \biggl\{ 
{1\over K} \bigl( k_1^2k_2^2 + k_1^2k_3^2 + k_2^2k_3^2 \bigr) 
+ {k_1k_2k_3\over K^2} \bigl( k_1k_2 + k_1k_3 + k_2k_3 \bigr) \biggr\} 
+ \cdots
\nonumber \\
{\cal A}_2
&\!\!\!\!\!\!=\!\!\!\!\!\!&
{1\over 2} {1\over M_{\rm pl}^2} {\dot\phi^2\over\dot\rho^2} 
\biggl\{ 
(k_1^2k_2 + k_1k_2^2 + k_1^2k_3 + k_1k_3^2 + k_2^2k_3 + k_2k_3^2) 
\nonumber \\
&&\qquad\qquad
+ {1\over K} \bigl( k_1^4 + k_2^4 + k_3^4 \bigr)
- k_1k_2k_3 - {1\over K^2} k_1k_2k_3 \bigl[ k_1^2 + k_2^2 + k_3^2 \bigr]
\biggr\} 
+ \cdots
\nonumber \\
{\cal A}_3
&\!\!\!\!\!\!=\!\!\!\!\!\!&
{1\over 2} {1\over M_{\rm pl}^2} {\dot\phi^2\over\dot\rho^2}  
\biggl\{ 
- \bigl( k_1^3+k_2^3+k_3^3 \bigr)
+ {6\over K} \bigl( k_1^2k_2^2+k_1^2k_3^2+k_2^2k_3^2 \bigr)
\nonumber \\
&&\qquad\qquad\ 
- {1\over K} \bigl( k_1^4+k_2^4+k_3^4 \bigr)
+ {2k_1k_2k_3\over K^2} \bigl( k_1^2 + k_2^2 + k_3^2 \bigr) 
\biggr\} 
+ \cdots . 
\nonumber 
\end{eqnarray}

Combining together the `cubic' contributions produces
\begin{eqnarray}
{\cal A}_1 + {\cal A}_2 + {\cal A}_3
&\!\!\!\!\!\!=\!\!\!\!\!\!&
{1\over 2} {1\over M_{\rm pl}^2} {\dot\phi^2\over\dot\rho^2} \biggl\{ 
- \bigl( k_1^3+k_2^3+k_3^3 \bigr)
+ (k_1^2k_2 + k_1k_2^2 + k_1^2k_3 + k_1k_3^2 + k_2^2k_3 + k_2k_3^2) 
\nonumber \\
&&\qquad\qquad\ 
+ {8\over K} \bigl( k_1^2k_2^2 + k_1^2k_3^2 + k_2^2k_3^2 \bigr) 
- k_1k_2k_3 
\nonumber \\
&&\qquad\qquad\ 
+ {k_1k_2k_3\over K^2} \bigl( k_1^2 + k_2^2 + k_3^2 \bigr) 
+ 2{k_1k_2k_3\over K^2} \bigl( k_1k_2 + k_1k_3 + k_2k_3 \bigr) 
\biggr\} + \cdots .
\nonumber 
\end{eqnarray}
The last three terms, those that are proportional to $k_1k_2k_3$, vanish when summed.  This leaves only 
\begin{eqnarray}
{\cal A}_1 + {\cal A}_2 + {\cal A}_3
&\!\!\!\!\!\!=\!\!\!\!\!\!&
-\,\, {1\over 2} {1\over M_{\rm pl}^2} {\dot\phi^2\over\dot\rho^2} \bigl( k_1^3+k_2^3+k_3^3 \bigr)
+ {1\over 2} {1\over M_{\rm pl}^2} {\dot\phi^2\over\dot\rho^2} 
(k_1^2k_2 + k_1k_2^2 + k_1^2k_3 + k_1k_3^2 + k_2^2k_3 + k_2k_3^2) 
\nonumber \\
&&
+\,\, 4 {1\over M_{\rm pl}^2} {\dot\phi^2\over\dot\rho^2} {k_1^2k_2^2 + k_1^2k_3^2 + k_2^2k_3^2\over K}
+ \cdots .
\nonumber 
\end{eqnarray}

Finally, when this result is combined with the `quartic' contribution the amplitude becomes
\begin{eqnarray}
{\cal A}_{k_1,k_2,k_3}
&\!\!\!\!\!\!=\!\!\!\!\!\!&
2 {\ddot\phi\over\dot\phi\dot\rho} \bigl( k_1^3 + k_2^3 + k_3^3 \bigr) 
+ {1\over 2} {1\over M_{\rm pl}^2} {\dot\phi^2\over\dot\rho^2} \bigl( k_1^3 + k_2^3 + k_3^3 \bigr) 
\nonumber \\
&&
+\,\, {1\over 2} {1\over M_{\rm pl}^2} {\dot\phi^2\over\dot\rho^2} 
\bigl( k_1^2k_2 + k_1k_2^2 + k_1^2k_3 + k_1k_3^2 + k_2^2k_3 + k_2k_3^2 \bigr) 
\nonumber \\
&&
+\,\, 4 {1\over M_{\rm pl}^2} {\dot\phi^2\over\dot\rho^2} 
{k_1^2k_2^2+k_1^2k_3^2+k_2^2k_3^2\over k_1+k_2+k_3}
+ \cdots .
\nonumber 
\end{eqnarray}
This is the result for the three-point function predicted by the simplest sort of inflationary model.  Such a model is one that
\vskip3truept

\noindent\hskip0.125truein 
{--} contains a single inflaton field,
\vskip3truept

\noindent\hskip0.125truein 
{--} whose potential satisfies the slow-roll conditions (where $\epsilon$ and $\delta$ are small), and
\vskip3truept

\noindent\hangindent=0.25truein\hskip0.125truein 
{--} where the quantum fluctuations $\zeta(t,\vec x)$ are in the vacuum state defined in the infinite past.
\vskip3truept

\noindent
Thus, the predicted form for the three-point function in this simple family of inflationary models is 
\begin{eqnarray}
&&\!\!\!\!\!\!\!\!\!\!\!\!\!\!\!\!\!\!\!\!\!\!\!
\langle 0(t)|\zeta(t,\vec x)\zeta(t,\vec y)\zeta(t,\vec z)|0(t)\rangle 
\nonumber \\
&\!\!\!\!\!\!=\!\!\!\!\!\!& 
\int {d^3\vec k_1\over (2\pi)^3} {d^3\vec k_2\over (2\pi)^3} 
{d^3\vec k_3\over (2\pi)^3}\, e^{i\vec k_1\cdot\vec x}
e^{i\vec k_2\cdot\vec y} e^{i\vec k_3\cdot\vec z}\,
(2\pi)^3\, \delta^3(\vec k_1+\vec k_2+\vec k_3)
\nonumber \\
&&\times
{1\over 32\epsilon^2} {H^4\over M_{\rm pl}^4} {1\over k_1^3k_2^3k_3^3} 
\biggl\{ 
2 {\ddot\phi\over\dot\phi\dot\rho} \bigl( k_1^3 + k_2^3 + k_3^3 \bigr) 
+ {1\over 2} {1\over M_{\rm pl}^2} {\dot\phi^2\over\dot\rho^2} \bigl( k_1^3 + k_2^3 + k_3^3 \bigr) 
\nonumber \\
&&\qquad\qquad\qquad\qquad\ 
+\,\, {1\over 2} {1\over M_{\rm pl}^2} {\dot\phi^2\over\dot\rho^2} 
\bigl( k_1^2k_2 + k_1k_2^2 + k_1^2k_3 + k_1k_3^2 + k_2^2k_3 + k_2k_3^2 \bigr) 
\nonumber \\
&&\qquad\qquad\qquad\qquad\
+\,\, 4 {1\over M_{\rm pl}^2} {\dot\phi^2\over\dot\rho^2} 
{k_1^2k_2^2+k_1^2k_3^2+k_2^2k_3^2\over k_1+k_2+k_3}
+ \cdots \biggr\} .
\nonumber 
\end{eqnarray}

Since the three-point function has a more complicated momentum dependence than the power spectrum, sometimes various representative or limiting cases are presented to provide a more intuitive sense for how the three-point function scales in the momenta.  For example, when all of the momenta are approximately equal in magnitude, $k_1\sim k_2\sim k_3\equiv k$, then the Fourier transform of the three-point function, 
\begin{eqnarray}
&&\!\!\!\!\!\!\!\!\!\!\!\!\!\!\!\!\!\!\!\!\!\!\!
\langle 0(t)|\zeta_{\vec k_1}(t)\zeta_{\vec k_2}(t)\zeta_{\vec k_3}(t)|0(t)\rangle\,\,  
(2\pi)^3\, \delta^3(\vec k_1+\vec k_2+\vec k_3)
\nonumber \\
&\!\!\!\!\!\!=\!\!\!\!\!\!& 
\int d^3\vec x\, d^3\vec y\, d^3\vec z\, e^{-i\vec k_1\cdot\vec x}
e^{-i\vec k_2\cdot\vec y} e^{-i\vec k_3\cdot\vec z}\,
\langle 0(t)|\zeta(t,\vec x)\zeta(t,\vec y)\zeta(t,\vec z)|0(t)\rangle ,
\nonumber 
\end{eqnarray}
scales as 
\begin{eqnarray}
\langle 0(t)|\zeta_{\vec k_1}(t)\zeta_{\vec k_2}(t)\zeta_{\vec k_3}(t)|0(t)\rangle_{\rm equil}
&\!\!\!\!\!\!=\!\!\!\!\!\!& 
{1\over 32\epsilon^2} {H^4\over M_{\rm pl}^4} {1\over k^6} 
\biggl\{ 
6 {\ddot\phi\over\dot\phi\dot\rho}  
+ {17\over 2} {1\over M_{\rm pl}^2} {\dot\phi^2\over\dot\rho^2}  
+ \cdots \biggr\} .
\nonumber \\
&\!\!\!\!\!\!=\!\!\!\!\!\!& 
{1\over 32\epsilon^2} {H^4\over M_{\rm pl}^4} {1\over k^6} 
\bigl\{ 17\epsilon + 6\delta + \cdots \bigr\} .
\nonumber 
\end{eqnarray}
Because the total momentum is conserved, as vectors $\vec k_1$, $\vec k_2$, and $\vec k_3$ together add up to zero; since they are all equal in length in this limit, they thus form the sides of an equilateral triangle.

Another, more important limit, occurs when one of the momenta is soft---the first one, $\vec k_1\to \vec 0$, for example.  This time momentum conservation requires that $\vec k_3\approx -\vec k_2$, or $k_3\approx k_2$.  In this case the three-point function simplifies to 
\begin{eqnarray}
\lim_{\vec k_1\to\vec 0}\,\, \langle 0(t)|\zeta_{\vec k_1}(t)\zeta_{\vec k_2}(t)\zeta_{\vec k_3}(t)|0(t)\rangle
&\!\!\!\!\!\!=\!\!\!\!\!\!& 
{1\over 8\epsilon^2} {H^4\over M_{\rm pl}^4} {1\over k_1^3k_2^3} 
\biggl\{ 
{\ddot\phi\over\dot\phi\dot\rho}  
+ {1\over M_{\rm pl}^2} {\dot\phi^2\over\dot\rho^2} \biggr\} + \cdots 
\nonumber \\
&\!\!\!\!\!\!=\!\!\!\!\!\!& 
{1\over 8\epsilon^2} {H^4\over M_{\rm pl}^4} {1\over k_1^3k_2^3} 
\bigl\{ 2\epsilon + \delta \bigr\} + \cdots . 
\nonumber 
\end{eqnarray}

The coordinates that have been chosen for describing the fluctuations about the homogeneous background do not quite exhaust the freedom to make small changes while remaining within this same class of coordinates.  Looking back at the metric, $\zeta(t,\vec x)$ appears in a conformal factor that multiplies a flat spatial metric, 
$$
ds^2 = (N^2-N_iN^i)\, dt^2 - 2N_i\, dtdx^i 
- e^{2\rho(t)+2\zeta(t,\vec x)}\, 
\delta_{ij}\, dx^idx^j .
$$
Therefore, any small {\it spatial\/} conformal transformation can be absorbed by a corresponding change in $\zeta(t,\vec x)$ without changing the basic structure of these coordinates.  As a result, the standard method for describing the quantum fluctuations during inflation has a residual conformal symmetry, which in turn implies relations between different orders of correlation functions in the limit where one of the momenta is `soft'.  The simplest example of such a `consistency relation' is that beween the two and three-point functions, 
\begin{eqnarray}
&&\!\!\!\!\!\!\!\!\!\!\!\!\!\!\!\!\!\!\!\!\!\!\!
\lim_{\vec k_1\to\vec 0}\,\, \langle 0(t)|\zeta_{\vec k_1}(t)\zeta_{\vec k_2}(t)\zeta_{-\vec k_2}(t)|0(t)\rangle
\nonumber \\
&\!\!\!\!\!\!=\!\!\!\!\!\!& 
- \langle 0(t)|\zeta_{\vec k_1}(t)\zeta_{-\vec k_1}(t)|0(t)\rangle
\biggl[ 3 + k_2{d\over dk_2} \biggr] 
\langle 0(t)|\zeta_{\vec k_2}(t)\zeta_{-\vec k_2}(t)|0(t)\rangle . 
\nonumber 
\end{eqnarray}
This particular relation is a consequence of the invariance of the $\zeta$ coordinates under a dilation of the spatial coordinates.  Restoring the $2\pi^2/k^3$ factor that was removed when defining the power spectrum, 
$$
\langle 0(t)|\zeta_{\vec k}(t)\zeta_{-\vec k}(t)|0(t)\rangle
= {2\pi^2\over k^3} P_k(t)  
= {1\over 4\epsilon} {H^2\over M_{\rm pl}^2} {1\over k^3} (-k\eta)^{-4\epsilon-2\delta} + \cdots ,
$$
it is easily seen that this relation is satisfied by the class of inflationary theories considered here, at least in the leading-order limit in the slow-roll parameters that has been used.  This relation was already noted by Maldacena [1] in his original treatment of the non-Gaussianities in inflation.
\vskip24truept

\noindent{\bf\large APPENDIX. CHANGING COORDINATES}
\vskip9truept

\noindent{\it A. Putting the fluctuations in the scalar field}
\vskip6truept

\noindent 
The derivation of the action, arranged and grouped according to the powers of the fluctuations about a perfectly homogeneous background, was performed for a particular choice of the coordinates.  The freedom to make small changes in the coordinates, $t\to t+\delta t$ and $\vec x\to\vec x+\delta\vec x$, was used to select the coordinates so that there would not be any quantum fluctuations in the scalar field, 
$$
\phi(t,\vec x) = \phi(t) ,
$$
and also to keep the spatial part of the metric conformally flat---at least as long as the vector and tensor fluctuations are being neglected.  The scalar fluctuations are then reduced to three fields, $\{ \zeta,\Phi, B\}$,
$$
N = 1 + 2\Phi(t,\vec x), \qquad
N^i = \delta^{ij}\partial_i B(t,\vec x), \qquad
h_{ij} = e^{2\rho(t)+2\zeta(t,\vec x)}\delta_{ij} .
$$
Imposing the two constraint equations associated with the $N$ and $N^i$ equations of motion fixed $\Phi(t,\vec x)$ and $B(t,\vec x)$ in terms of the single field $\zeta(t,\vec x)$.

These coordinates are especially useful since the scalar field $\zeta(t,\vec x)$ becomes constant when stretched to scales much larger than the horizon size during inflation.  Moreover, it has a close relation with the fluctuations of the scalar spatial curvature used after the inflationary expansion has ended.  One significant disadvantage of these coordinates is that they obscure the fact that the cubic action really is proportional to $\epsilon^2$.  This important property eventually emerged, but only after a long calculation.

Another way to choose the coordinates, which has its own advantages, is to eliminate $\zeta(t,\vec x)$ and $\xi(t,\vec x)$, leaving
$$
N = 1 + 2\Phi(t,\vec x), \qquad
N^i = \delta^{ij}\partial_i \chi(t,\vec x), \qquad
h_{ij} = e^{2\rho(t)}\delta_{ij} ,
$$
and instead to keep the fluctuations in the scalar field,
$$
\phi(t,\vec x) = \phi_0(t) + \delta\phi(t,\vec x) .
$$
Here, when writing an equation that applies to {\it any\/} coordinate system, $\phi$ will mean the full scalar field $\phi(t,\vec x)$.  But in order to match more uniformly with the rest of these notes (where there was no need to distinguish the background, $\phi_0$, from the fluctuations, $\delta\phi$, because there were not any), once the particular set of coordinates has been chosen $\phi$ will usually mean just the background, $\phi_0(t)\to \phi(t)$---that is, the subscript will be dropped when there is little danger of confusion.  Also the fluctuation in the scalar field will be written as $\delta\phi(t,\vec x)=\psi(t,\vec x)$; this change prevents derivatives of $\delta\phi(t,\vec x)$ from looking too awkward and it also prevents any confusion with $\delta$, the slow-roll parameter.

The same series of steps---the application of the two constraint equations followed by a careful expansion of the action in the remaining independent scalar field (in this case, $\psi(t,\vec x)$)---leads to an action whose cubic terms in $\psi(t,\vec x)$ are more self-evidently suppressed by $\epsilon^2$, without applying the seemingly infinite series of integrations by parts that was necessary for understanding the $\zeta(t,\vec x)$ action.

The two constraint equations, 
\begin{eqnarray}
&
\hat R - {1\over N^2}\bigl( E_{ij}E^{ij}-E^2 \bigr) 
- {1\over N^2}\bigl( \dot\phi - N^i\partial_i\phi \bigr)^2 
- h^{ij}\partial_i\phi\partial_j\phi - 2V = 0 
&\nonumber \\
&
\hat\nabla_j \Bigl[ {1\over N} \bigl( E_i^j - \delta_i^j E \bigr) \Bigr]
= {1\over N} \bigl( \dot\phi - N^j\partial_j\phi \bigr) \partial_i\phi , 
&\nonumber
\end{eqnarray}
hold in any coordinate system, and the background equations are still,
\begin{eqnarray}
3\dot\rho^2 &\!\!\!\!\!\!=\!\!\!\!\!\!&\textstyle
{1\over 2} \dot\phi^2 + V 
\nonumber \\
\ddot\rho &\!\!\!\!\!\!=\!\!\!\!\!\!&\textstyle
- {1\over 2} \dot\phi^2 
\nonumber \\
\ddot\phi + 3 \dot\rho\dot\phi  
&\!\!\!\!\!\!=\!\!\!\!\!\!& - {\delta V\over\delta\phi} .
\nonumber 
\end{eqnarray}
In the new coordinates, $h_{ij}$ has no spatial dependence, so $\hat R=0$.  $E_{ij}$, which is related to the extrinsic curvature, is this time also given by a much simpler expression 
$$
E_{ij} = e^{2\rho(t)}\bigl[ \dot\rho\delta_{ij}-\partial_i\partial_j\chi\bigr] .
$$
As a consequence, the two combinations of the extrinsic curvature that appear in the constraint equations are given {\it exactly\/} by
\begin{eqnarray}
E_{ij}E^{ij} - E^2 &\!\!\!\!\!\!=\!\!\!\!\!\!& 
-\,\, 6\dot\rho^2 + 4\dot\rho \partial_k\partial^k\chi 
+ \partial_i\partial_j\chi\partial^i\partial^j\chi 
- (\partial_k\partial^k\chi)^2 
\nonumber \\
E_i^j - \delta_i^jE &\!\!\!\!\!\!=\!\!\!\!\!\!& 
-\,\, 2\dot\rho\delta_i^j - \partial_i\partial^j\chi 
+ \delta_i^j \partial_k\partial^k\chi . 
\nonumber 
\end{eqnarray}

The second of the constraint equations in these coordinates becomes 
$$
\hat\nabla_j \biggl[ {1\over 1+2\Phi} \bigl( - 2\dot\rho\delta_i^j 
- \partial_i\partial^j\chi + \delta_i^j \partial_k\partial^k\chi \bigr) \biggr]
- {1\over 1+2\Phi} \bigl( \dot\phi + \psidot 
- \delta^{jk}\partial_k\chi\partial_j\psi \bigr) \partial_i\psi 
= 0 .
$$
Working only to leading order in the small fluctuations, this equation reduces to 
$$
\partial_i \bigl[ 4\dot\rho\Phi - \dot\phi \psi \bigr] = 0 .
$$
This constraint fixes one of the scalar fluctuations in the metric, $\Phi(t,\vec x)$, in terms of $\psi(t,\vec x)$,
$$
\Phi = {1\over 4} {\dot\phi\over\dot\rho} \psi .
$$
In principle, the solution to the differential equation could have had a constant of integration.  But here it should be zero, since the result must return to the original background solution---that is, when $\psi$ vanishes, $\Phi$ should vanish too.

The first of the constraints will similarly be used to determine $\chi(t,\vec x)$.  Expanding this constraint equation to first order in the fluctuations yields
$$
- (1-4\Phi)\bigl( - 6\dot\rho^2 + 4\dot\rho \partial_k\partial^k\chi \bigr) 
- (1-4\Phi) \bigl( \dot\phi^2 + 2 \dot\phi\!\!\psidot \bigr)
- 2V_0 - 2\psi {\delta V\over\delta\phi} = 0 . 
$$
Here, $V_0 = V(\phi)$ (or what is really $V_0 = V(\phi_0)$) corresponds to the potential evaluated with $\psi=0$; ${\delta V\over\delta\phi}$ is also implicitly evaluated with $\psi=0$.  Both of these two appearances of the potential can be rewritten through the background equations as 
$$
V_0 = 3\dot\rho^2 - {1\over 2} \dot\phi^2
\qquad\hbox{and}\qquad 
{\delta V\over\delta\phi} = - \ddot\phi - 3 \dot\rho\dot\phi . 
$$
Once these have been substituted into the constraint equation, all of the zeroth order terms cancel, leaving
$$
4\dot\rho \partial_k\partial^k\chi 
= - 24\dot\rho^2\Phi 
- 2 \dot\phi\!\!\psidot 
+ 4 \dot\phi^2\Phi 
+ 2 \ddot\phi\psi + 6 \dot\rho\dot\phi\psi . 
$$
Substituting in the value of $\Phi = {1\over 4}{\dot\phi\over\dot\rho}\psi$ produces
$$
4\dot\rho \partial_k\partial^k\chi 
= - 2 \dot\phi\!\!\psidot 
+ \dot\phi^2{\dot\phi\over\dot\rho}\psi
+ 2\psi\ddot\phi 
= - 2 {\dot\phi^2\over\dot\rho} \biggl\{ 
{\dot\rho\over\dot\phi}\!\!\psidot 
- {1\over 2} \dot\phi^2 {1\over\dot\phi}\psi
- {\dot\rho\ddot\phi\over\dot\phi^2} \psi \biggr\} , 
$$
or, upon using another of the background equations, $\ddot\rho=-{1\over 2} \dot\phi^2$, 
$$
\partial_k\partial^k\chi 
= - {1\over 2} {\dot\phi^2\over\dot\rho^2} \biggl\{ 
{\dot\rho\over\dot\phi}\!\!\psidot + {\ddot\rho\over\dot\phi}\psi
- {\dot\rho\ddot\phi\over\dot\phi^2} \psi \biggr\} .
$$
The expression within the braces can now be recognised as being a total time derivative,
$$
\partial_k\partial^k\chi 
= - {1\over 2} {\dot\phi^2\over\dot\rho^2} 
{d\over dt} \biggl\{ {\dot\rho\over\dot\phi}\psi \biggr\} .
$$
Thus, just as in the case of $\zeta$, the two constraint equations have reduced the set of three scalar fields, $\{ \psi,\Phi,\chi \}$ to just one $\psi(t,\vec x)$---the fluctuations in the inflaton.

The derivation of the cubic action is far, far simpler in terms of $\psi$ than it was for $\zeta$.  Starting from the full action once again, 
$$
S = {1\over 2} \int d^4x\, \sqrt{h}\, \Bigl\{ 
N\hat R + {1\over N} \bigl( E_{ij}E^{ij} - E^2 \bigr) 
+ {1\over N} \bigl( \dot\phi - N^i\partial_i\phi \bigr)^2 
- Nh^{ij}\partial_i\phi\partial_j\phi - 2NV(\phi) \Bigr\} , 
$$
when it has been expanded in the fluctuations $\psi(t,\vec x)$, the quadratic terms are 
\begin{eqnarray}
S^{(2)} &\!\!\!\!\!\!=\!\!\!\!\!\!& 
{1\over 2} \int d^4x\, e^{3\rho} \biggl\{ 
\psidot^2 - {\dot\phi^2\over\dot\rho} \psi\!\!\psidot 
- e^{-2\rho} \partial_k\psi\partial^k\psi 
- \Bigl[ {\delta^2 V\over\delta\phi^2} 
- {3\over 2}\dot\phi^2 
- {1\over 4} {\dot\phi^4\over\dot\rho^2} 
- {\dot\phi\ddot\phi\over\dot\rho} \Bigr] \psi^2  
\nonumber \\
&&\qquad\qquad\quad
-\,\, 2 \partial_k\bigl[ \dot\phi\psi \partial^k\chi \bigr]
+ \partial_k \bigl[ \partial_j\chi\partial^j\partial^k\chi 
- \partial^k\chi\partial_j\partial^j\chi \bigr]
\biggr\} , 
\nonumber 
\end{eqnarray}
while the cubic terms are 
\begin{eqnarray}
S^{(3)} &\!\!\!\!\!\!=\!\!\!\!\!\!& 
\int d^4x\, e^{3\rho} \biggl\{ 
- {1\over 4} {\dot\phi\over\dot\rho} \psi\!\!\psidot^2 
- {1\over 4} e^{-2\rho} {\dot\phi\over\dot\rho}\psi \partial_k\psi\partial^k\psi 
- \psidot\partial_k\psi\partial^k\chi 
\nonumber \\
&&\qquad\qquad\ \ 
+\,\, {3\over 8} {\dot\phi^3\over\dot\rho} \psi^3
- {1\over 16} {\dot\phi^5\over\dot\rho^3}\psi^3
- {1\over 4}{\dot\phi\over\dot\rho}{\delta^2 V\over\delta\phi^2} \psi^3 
- {1\over 6}{\delta^3 V\over\delta\phi^3} \psi^3
+ {1\over 4} {\dot\phi^3\over\dot\rho^2} \psi^2\!\!\psidot 
\nonumber \\
&&\qquad\qquad\ \  
+\,\, {1\over 4}{\dot\phi^2\over\dot\rho} \psi^2\partial_k\partial^k\chi  
- {1\over 4} {\dot\phi\over\dot\rho}\psi
   \bigl[ \partial_i\partial_j\chi\partial^i\partial^j\chi 
   - (\partial_k\partial^k\chi)^2 \bigr]
\nonumber \\
&&\qquad\qquad\ \ 
+\,\, \partial_k \biggl[ {1\over 4}{\dot\phi^2\over\dot\rho} \psi^2\partial^k\chi \biggr]
\biggr\} . 
\nonumber 
\end{eqnarray}
Other than having collected a few of the terms together into total spatial derivatives, nothing else has been done beyond the straightforward expansion of the action and the substitution of $\Phi={1\over 4}{\dot\phi\over\dot\rho} \psi$ to arrive at these expressions.

Before examining the cubic parts of the action, notice that the fluctuations in the scalar field $\psi(t,\vec x)$ are related to the canonically normalized field $\varphi(t,\vec x)$ through a simpler rescaling using just the scale factor itself, 
$$
\varphi = e^\rho\psi .
$$
The derivative of $\psi$ is given in terms of the canonically normalized field by 
$$
\psidot\,\, = e^{-\rho} \bigl[ \dot\varphi - \dot\rho\varphi \bigr] .
$$
Substituting these relations into the quadratic part of the action yields
\begin{eqnarray}
S^{(2)} &\!\!\!\!\!\!=\!\!\!\!\!\!& 
{1\over 2} \int d^4x\, e^\rho \biggl\{ 
\dot\varphi^2 - 2\dot\rho\varphi\dot\varphi  
- {\dot\phi^2\over\dot\rho} \varphi \dot\varphi 
- e^{-2\rho} \partial_k\varphi\partial^k\varphi 
\nonumber \\
&&\qquad\qquad
+ \Bigl[ \dot\rho^2 + {5\over 2}\dot\phi^2 
+ {1\over 4} {\dot\phi^4\over\dot\rho^2} 
+ {\dot\phi\ddot\phi\over\dot\rho} 
- {\delta^2 V\over\delta\phi^2}
 \Bigr] \varphi^2  
\biggr\} , 
\nonumber 
\end{eqnarray}
up to total derivatives.  The terms with a factor of $\varphi\dot\varphi = {1\over 2} {d\over dt}\varphi^2$ can be partially integrated to generate more contributions to the mass of the field $\varphi$,
\begin{eqnarray}
- 2e^\rho\dot\rho\varphi\dot\varphi
&\!\!\!\!\!\!=\!\!\!\!\!\!&
- e^\rho\dot\rho{d\over dt}\varphi^2
= -\,\, {d\over dt} \bigl[ e^\rho\dot\rho\varphi^2 \bigr]
+ e^\rho\dot\rho^2\varphi^2
- {1\over 2} e^\rho\dot\phi^2\varphi^2
\nonumber \\
- e^\rho {\dot\phi^2\over\dot\rho} \varphi \dot\varphi 
&\!\!\!\!\!\!=\!\!\!\!\!\!&
- {1\over 2} e^\rho {\dot\phi^2\over\dot\rho} {d\over dt}\varphi^2 
= 
{d\over dt} \Bigl[ {1\over 2} e^\rho {\dot\phi^2\over\dot\rho} \varphi^2 \Bigr] 
+ {1\over 2} e^\rho \dot\phi^2 \varphi^2 
+ e^\rho {\dot\phi\ddot\phi\over\dot\rho} \varphi^2 
+ {1\over 4} e^\rho {\dot\phi^4\over\dot\rho^2} \varphi^2 . 
\nonumber 
\end{eqnarray}
In both of these equations the background equation $\ddot\rho = - {1\over 2}\dot\phi^2$ has again been freely applied.  Substituting these results back into the quadratic action produces
$$
S^{(2)} = {1\over 2} \int d^4x\, e^\rho \biggl\{ 
\dot\varphi^2 - e^{-2\rho} \partial_k\varphi\partial^k\varphi 
+ \dot\rho^2 \Bigl[ 2 + {5\over 2}{\dot\phi^2\over\dot\rho^2} 
+ {1\over 2} {\dot\phi^4\over\dot\rho^4} 
+ 2{\dot\phi\ddot\phi\over\dot\rho^3} 
- {1\over\dot\rho^2} {\delta^2 V\over\delta\phi^2}
 \Bigr] \varphi^2  
\biggr\} , 
$$
up to total derivatives.  The mass of the canonically normalized field can be inferred from the $\varphi^2$ term in the action to be 
$$
m^2 = - e^{2\rho} \dot\rho^2 \biggl[ 2 + {5\over 2}{\dot\phi^2\over\dot\rho^2} 
+ {1\over 2} {\dot\phi^4\over\dot\rho^4} 
+ 2{\dot\phi\ddot\phi\over\dot\rho^3} 
- {1\over\dot\rho^2} {\delta^2 V\over\delta\phi^2} \biggr] .
$$
While it might not yet be apparent, this is the same mass as was found before for $\varphi$.  Most of the terms can be immediately expressed in terms of the slow-roll parameters, 
$$
m^2 = - e^{2\rho} \dot\rho^2 \biggl[ 2 + 5\epsilon + 2\epsilon^2  
+ 4\epsilon\delta - {1\over\dot\rho^2} {\delta^2 V\over\delta\phi^2} \biggr] , 
$$
except for the derivative of the potential.  But by taking the time derivative of the background equation for the scalar field, this term can be expressed in terms of the slow-roll parameters too,
$$
{d\over dt}\biggl\{ 
{\delta V\over\delta\phi} = - \ddot\phi - 3\dot\rho\dot\phi \biggr\}
\quad\Rightarrow\quad
\dot\phi {\delta^2V\over\delta\phi^2} 
= - {d\over dt}\ddot\phi - 3\ddot\rho\dot\phi - 3\dot\rho\ddot\phi 
= - {d\over dt}\ddot\phi + {3\over 2}\dot\phi^3 - 3\dot\rho\ddot\phi . 
$$
The third derivative of $\phi$ is found by rearranging the definition of $\delta$ ($\ddot\phi = \dot\rho\dot\phi\delta$) and differentiating,
$$
{d\over dt} \ddot\phi 
= \ddot\phi\dot\rho\delta + \dot\phi\ddot\rho\delta + \dot\phi\dot\rho\dot\delta 
=\textstyle \dot\phi \bigl[ \dot\rho^2\delta^2 
- {1\over 2} \dot\phi^2\delta + \dot\rho\dot\delta \bigr] ,
$$
which means in turn that 
$$
\dot\phi {\delta^2V\over\delta\phi^2} =\textstyle 
\dot\phi \dot\rho^2 \bigl[ 3 \epsilon - 3 \delta - \delta^2 
+ \epsilon \delta - (\dot\delta/\dot\rho) \bigr] , 
$$
or 
$$
{1\over\dot\rho^2}{\delta^2V\over\delta\phi^2} = 
3 \epsilon - 3 \delta - \delta^2 
+ \epsilon \delta - {\dot\delta\over\dot\rho} , 
$$
Using these relations, the mass of the canonically normalized field becomes 
$$
m^2 = - e^{2\rho} \dot\rho^2 \biggl[ 2 + 2\epsilon + 3 \delta 
+ 2\epsilon^2 + 3\epsilon\delta + \delta^2 
+ {\dot\delta\over\dot\rho} \biggr] . 
$$
This still appears to be slightly different from the previous form for the canonically normalized field's mass.  But by differentiating $\epsilon$, 
$$
\dot\epsilon = {1\over 2} {d\over dt} {\dot\phi^2\over\dot\rho^2} 
= {\dot\phi\ddot\phi\over\dot\rho^2}
- {\dot\phi^2\over\dot\rho^3}\ddot\rho
= 2\epsilon\delta \dot\rho + 2 \epsilon^2 \dot\rho , 
$$
the mass can be put into its previous form, 
\begin{eqnarray}
m^2 &\!\!\!\!\!\!=\!\!\!\!\!\!& 
- e^{2\rho} \dot\rho^2 \biggl[ 2 + 2\epsilon + 3 \delta 
+ \epsilon\delta + \delta^2 
+ {\dot\epsilon\over\dot\rho} + {\dot\delta\over\dot\rho} \biggr] 
\nonumber \\
&\!\!\!\!\!\!=\!\!\!\!\!\!&
- e^{2\rho}\dot\rho^2 (2+\delta)(1+\epsilon+\delta) 
- e^{2\rho} \dot\rho (\dot\epsilon + \dot\delta) . 
\nonumber 
\end{eqnarray}

The relation between the canonically normalized field and the {\it general, coordinate-invariant\/} fluctuation is 
$$
\varphi = e^\rho \Bigl[ \psi + {\dot\phi\over\dot\rho} \zeta \Bigr] .
$$
This relation holds for {\it any\/} choice of the coordinates for the fluctuations about the background, $\{ \psi, \zeta, B, \Phi, \xi \}$---none of them are assumed to have been set equal to zero.  This general expression can be used for either of the two sets of coordinates that were chosen here:  $\psi=0$, as in the calculation of the three-point function, or $\zeta=0$, as in this appendix.

Returning to the cubic parts of the action, as will be shown next the field $\psi$ is related to $\zeta$ at leading order by 
$$
\psi = - {\dot\phi\over\dot\rho} \zeta + \cdots , 
$$
which means that $\psi\propto \sqrt{\epsilon}\zeta$.  The field that has been called $\chi$ here will further turn out to be exactly the same field $\chi$ that was defined earlier, which was shown to scale as $\epsilon$, when written in terms of $\zeta$.  

Looking at the terms in $S^{(3)}$ and using the fact that $\dot\phi/\dot\rho\sim\sqrt{\epsilon}$, it can be seen that the following terms are explicitly of order $\epsilon^2$, 
$$
- {1\over 4} {\dot\phi\over\dot\rho} \psi\!\!\psidot^2 
- {1\over 4} e^{-2\rho} {\dot\phi\over\dot\rho}\psi \partial_k\psi\partial^k\psi 
- \psidot\partial_k\psi\partial^k\chi 
+ \cdots . 
$$
These are the leading terms in the slow-roll limit.  The rest of the terms are third-order or higher.  This can be seen more or less explicitly for the following set, 
$$
{3\over 8} {\dot\phi^3\over\dot\rho} \psi^3
- {1\over 16} {\dot\phi^5\over\dot\rho^3}\psi^3
+ {1\over 4} {\dot\phi^3\over\dot\rho^2} \psi^2\!\!\psidot 
+ {1\over 4}{\dot\phi^2\over\dot\rho} \psi^2\partial_k\partial^k\chi  
- {1\over 4} {\dot\phi\over\dot\rho}\psi
   \bigl[ \partial_i\partial_j\chi\partial^i\partial^j\chi 
   - (\partial_k\partial^k\chi)^2 \bigr]
+ \cdots ,
$$
based on the scaling of the fields and the factors of $\dot\phi/\dot\rho$.  The terms involving the potential are also sub-leading in the slow-roll parameters.  The second-derivative term, 
$$
- {1\over 4}{\dot\phi\over\dot\rho}{\delta^2 V\over\delta\phi^2} \psi^3 ,
$$
already is suppressed by $\epsilon^2$ due to the ${\dot\phi\over\dot\rho}\psi^3$ factor; and, as was shown during the analysis of the quadratic terms, the second derivative of the potential is first order in $\epsilon$ and $\delta$.  For the third-derivative them, 
$$
- {1\over 6}{\delta^3 V\over\delta\phi^3} \psi^3,
$$
the $\psi^3$ again gives a factor of $\epsilon^{3/2}$.  The third derivative the potential can be expressed in terms of the slow-roll parameters by taking the time derivative of the expression for $\delta^2 V/\delta\phi^2$ derived above, 
$$
\dot\phi {\delta^3V\over\delta\phi^3} 
= \dot\rho^3 \biggl[ 12 \epsilon\delta 
+ 4 \epsilon\delta^2 
- 3 {\dot\delta\over\dot\rho} 
+ 2 {\epsilon\dot\delta\over\dot\rho} 
- 2 {\delta\dot\delta\over\dot\rho} 
- {\ddot\delta\over\dot\rho^2} \biggr] .
$$
The right side is manifestly second order in the slow-roll parameters.  Accounting for the $\dot\phi$ term on the left side, the third derivative of the potential must be at least ${3\over 2}$ order in the slow-roll parameters.  When combined with the $\epsilon^{3/2}$ from the $\psi^3$, the overall scaling of the corresponding term in the action is also at least third order in the slow-roll parameters.

In being able to make these statements about the scaling of the contributions to the cubic action, it was necessary to know something about the relation between the $\{\zeta=0, \delta\phi=\psi\not=0\}$ coordinates and the $\{\zeta\not=0, \delta\phi=0\}$ coordinates.  Even beyond this specific purpose, it is useful to know how to go between these two coordinates systems.  Therefore, the details of this change of coordinates, carried out to second order, will be derived next.
\vskip18truept

\noindent{\it B. Transforming between coordinate systems}
\vskip6truept

\noindent 
What is the transformation that connects these two choices for describing the small fluctuations?  It is determined, order by order, by making a small change in the coordinates, $x^\mu\to x^\mu + \delta x^\mu$, where $\delta x^\mu(t,\vec x)$ is of the same order as the fluctuations themselves.  For example, imagine starting in the coordinates that have been used so far in this appendix, where the scalar field is 
$$
\phi(t,\vec x) = \phi(t) + \psi(t,\vec x) . 
$$ 
In the other set of coordinates, which was best suited for describing the fluctuations once they have been stretched well outside the horizon, the scalar field is chosen to be given entirely by its background value, so $\psi(t,\vec x)$ is absent.  

Shifting the time coordinate by 
$$
t \to \tilde t = t + T(t,\vec x) 
$$
induces a corresponding change in the scalar field, 
$$
\phi(t,\vec x) \to \tilde\phi(\tilde t,\vec x) 
= \phi(t,\vec x) + (\tilde t-t) {\partial\phi\over\partial t}
+ {1\over 2} (\tilde t-t)^2 {\partial^2\phi\over\partial t^2} 
+ \cdots
$$
or 
$$\textstyle
\tilde\phi = \phi + \psi + \dot\phi T + T\!\!\psidot + {1\over 2} \ddot\phi T^2 
+ \hbox{3$^{\rm rd}$ order} . 
$$
In this last expression, the $\phi$ on the right side is really the background value, $\phi_0(t)$, though the subscripts are not being written explicitly.  The new coordinates are meant to be those where the scalar field has no quantum fluctuations.  Suppose that the field $T(t,\vec x)$ is written as a first order piece and a second order piece, as indicated by the corresponding subscripts, 
$$
T = T_1 + T_2 .
$$
Substituted into the expression for $\tilde\phi$,
$$\textstyle
\tilde\phi = \phi + \psi + \dot\phi T_1 + \dot\phi T_2 
+ T_1\!\!\psidot + {1\over 2} \ddot\phi T_1^2 + \cdots , 
$$
the first order piece is removed by choosing
$$
T_1 = - {1\over\dot\phi} \psi ,
$$
while the second order piece is removed by choosing 
$$
T_2 = {1\over\dot\phi^2} \psi \!\!\psidot 
- {1\over 2} {\ddot\phi\over\dot\phi^3} \psi^2 . 
$$
Thus, to remove the fluctuations from the scalar field, the function $T(t,\vec x)$ should be chosen to be
$$
T = - {1\over\dot\phi} \psi - {1\over 2} {\ddot\phi\over\dot\phi^3} \psi^2 
+ {1\over\dot\phi^2} \psi \!\!\psidot + \cdots . 
$$

The fluctuation $\zeta(t,\vec x)$ belongs to the spatial part of the metric, so the next step is to examine how this change of the coordinates generates $\zeta(t,\vec x)$.  From the general form of the metric, 
$$
ds^2 = \bigl[ N^2 - h_{ij}N^iN^j\bigr]\, dt^2 - 2N_i\, dtdx^i 
- h_{ij}\, dx^idx^j , 
$$
the change in the time coordinate causes it to become
$$
ds^2 = 
\tilde N^2\, d\tilde t^2 - \tilde h_{ij}\tilde N^i\tilde N^j\, d\tilde t^2 
- 2\tilde N_i\, d\tilde tdx^i - \tilde h_{ij}\, dx^idx^j .
$$
The change of coordinates also applies to the differentials, 
$$
d\tilde t = dt + \partial_i T\, dx^i ,
$$
so that the metric more explicitly is
\begin{eqnarray}
ds^2 &\!\!\!\!\!\!=\!\!\!\!\!\!&  
\bigl[ \tilde N^2 - \tilde h_{ij}\tilde N^i\tilde N^j \bigr]\, dt^2  
- 2\bigl[ \tilde N_i - \tilde N^2 \partial_iT 
 + \tilde h_{jk}\tilde N^j\tilde N^k\partial_iT \bigr]\, dtdx^i 
\nonumber \\
&&
- \bigl[ \tilde h_{ij} + \tilde N_i\, \partial_jT + \tilde N_j\, \partial_iT 
- \tilde N^2\, \partial_iT\, \partial_jT
+ \tilde h_{kl}\tilde N^k\tilde N^l\partial_iT\, \partial_jT
\bigr]\, dx^idx^j ,
\nonumber 
\end{eqnarray}
and in particular, the spatial part of the metric in the new coordinates is 
$$
h_{ij} = e^{2\rho(\tilde t)} \delta_{ij} 
+ \tilde N_i\, \partial_jT + \tilde N_j\, \partial_iT 
- \partial_iT\, \partial_jT
+ \cdots 
$$
to second order.  A factor of $e^{2\rho(\tilde t)}$ can also be extracted from the second and the third terms by noting that 
$$
\tilde N_i = \tilde h_{ij} \tilde N^j 
= e^{2\rho(\tilde t)} \delta_{ij} \delta^{jk}\partial_k\tilde\chi 
= e^{2\rho(\tilde t)} \partial_i\chi + \cdots , 
$$
so that 
$$
h_{ij} = e^{2\rho(\tilde t)} \bigl[ \delta_{ij} 
+ \partial_i\chi\partial_jT + \partial_iT\partial_j\chi
- e^{-2\rho(\tilde t)}\partial_iT\, \partial_jT \bigr]
+ \cdots . 
$$
For most of the terms, which are already second order, the exponent can be replaced with $e^{2\rho(\tilde t)} = e^{2\rho(t)}$ directly; but for the first term, this exponential factor needs to be expanded in a Taylor series about $\tilde t = t + T$, 
$$
e^{2\rho(\tilde t)} 
= e^{2\rho(t)} \bigl[ 1 + 2\dot\rho T + \ddot\rho T^2 + 2\dot\rho^2 T^2 + \cdots
\bigr] .
$$
Doing so leads to the following expression for the metric in the new coordinates, 
$$
h_{ij} = e^{2\rho(t)} \Bigl\{ 
\delta_{ij} \bigl[ 1 + 2\dot\rho T + \ddot\rho T^2 + 2\dot\rho^2 T^2 \bigr] 
+ \partial_i\chi\partial_jT + \partial_iT\partial_j\chi
- e^{-2\rho(t)}\partial_iT\, \partial_jT \Bigr\}
+ \cdots .
$$

The goal is to arrive at a coordinate system where the spatial part of the metric is of the form that was used throughout the calculation of the three-point function,
$$
h_{ij} = e^{2\rho(t)+2\zeta(t,\vec x)}\delta_{ij} .
$$
The current form of the metric has not quite reached this stage yet---for example, there is no obvious reason that the term 
$$
\pi_{ij} \equiv \partial_i\chi\partial_jT + \partial_iT\partial_j\chi
- e^{-2\rho(t)}\partial_iT\, \partial_jT 
$$
should automatically be proportional to $\delta_{ij}$.  This means that although there was no $\xi(t,\vec x)$ field in the original metric, it could have been generated again in the process of changing the time coordinate,
$$
h_{ij} = e^{2\rho} \bigl[ e^{2\zeta}\, \delta_{ij} + \partial_i\partial_j\xi 
\bigr] .
$$

So far, only the time coordinate has been transformed.  There is still the freedom to transform the spatial coordinate by a small amount too, 
$$
x^i \to \tilde x^i  = x^i + \epsilon^i(t,\vec x) ,
$$
where the index of $\epsilon^i(t,\vec x)$ is raised or lowered by $\epsilon^i = \delta^{ij}\epsilon_j$.  In terms of differentials,
$$
d\tilde x^i = dx^i + {\partial\epsilon^i\over\partial x^j}\, dx^j + \cdots .
$$
The time-derivative part would not alter the spatial parts of the metric and is therefore not needed for computing $\zeta(t,\vec x)$, so it has not been written explicitly here.  The part that is to be cancelled, which was called `$\pi_{ij}$', is already a second-order effect.  Correspondingly, this change in the spatial coordinates will have no effect on the $\pi_{ij}$---it will look the same in the $x^i$ and the $\tilde x^i$ coordinates up to still higher-order corrections.  Moreover, since $\epsilon^i$ is meant to cancel a second-order term, $\epsilon^i$ itself should be second-order.

Changing the spatial coordinates does affect the diagonal part of the spatial metric, 
\begin{eqnarray}
\delta_{ij}\, d\tilde x^i d\tilde x^j &\!\!\!\!\!\!=\!\!\!\!\!\!& 
\delta_{ij}\, \bigl[ dx^i + \partial_k\epsilon^i\, dx^k \bigr]
\bigl[ dx^j + \partial_l\epsilon^j\, dx^l \bigr]
\nonumber \\
&\!\!\!\!\!\!=\!\!\!\!\!\!& 
\delta_{ij}\, dx^i dx^j 
+ \delta_{kj}\, \partial_i\epsilon^k\, dx^i dx^j 
+ \delta_{ik}\, \partial_j\epsilon^k\, dx^i dx^j 
+ \cdots 
\nonumber \\
&\!\!\!\!\!\!=\!\!\!\!\!\!& 
\bigl[ \delta_{ij} + \partial_i\epsilon_j + \partial_j\epsilon_i + \cdots 
\bigr]\, dx^i dx^j . 
\nonumber 
\end{eqnarray}
After this transformation of the spatial coordinates, the spatial part of the metric becomes, 
$$
h_{ij} = e^{2\rho} \Bigl\{ 
\delta_{ij} \bigl[ 1 + 2\dot\rho T + \ddot\rho T^2 + 2\dot\rho^2 T^2 \bigr] 
+ \partial_i\epsilon_j + \partial_j\epsilon_i
+ \pi_{ij} \Bigr\}
+ \cdots .
$$
Now everything is finally being evaluated in the $(t,\vec x)$ coordinates.  Concentrating on the parts of the metric that transform as spatial scalars, this new metric must be generally of the form 
$$
h_{ij} = e^{2\rho+2\zeta}\, \delta_{ij} + e^{2\rho}\, \partial_i\partial_j\xi ,
$$
where the $\xi$ contribution is traceless, $\partial_k\partial^k\xi = 0$.  If it had not been, the trace could always have been removed and regrouped with the $\zeta$ part.

What is $\zeta(t,\vec x)$ in terms of $\psi(t,\vec x)$?  The relation between the two can be determined even without knowing the detailed form of $\epsilon^i(t,\vec x)$.  To find this relation, first equate the two forms of the metric,
$$
e^{2\rho + 2\zeta}\, \delta_{ij} + e^{2\rho}\, \partial_i\partial_j\xi 
= e^{2\rho} \Bigl\{ 
\delta_{ij} \bigl[ 1 + 2\dot\rho T + \ddot\rho T^2 + 2\dot\rho^2 T^2 \bigr] 
+ \partial_i\epsilon_j + \partial_j\epsilon_i
+ \pi_{ij} \Bigr\} ,
$$
cancel the appearance of the scale factor, 
$$
e^{2\zeta}\, \delta_{ij} + \partial_i\partial_j\xi 
= \delta_{ij} \bigl[ 1 + 2\dot\rho T + \ddot\rho T^2 + 2\dot\rho^2 T^2 \bigr] 
+ \partial_i\epsilon_j + \partial_j\epsilon_i
+ \pi_{ij} ,
$$
and expand the left side to second order, to obtain
$$
\bigl[ 1 + 2\zeta + 2\zeta^2 \bigr] \, \delta_{ij} + \partial_i\partial_j\xi 
= \delta_{ij} \bigl[ 1 + 2\dot\rho T + \ddot\rho T^2 + 2\dot\rho^2 T^2 \bigr] 
+ \partial_i\epsilon_j + \partial_j\epsilon_i
+ \pi_{ij} . 
$$
It is a bit easier to follow what is happening order by order if both $\zeta$ and $T$ are broken into terms of a specific order, indicated by a subscript, 
$$
\zeta = \zeta_1 + \zeta_2 + \cdots 
\qquad\hbox{and}\qquad
T = T_1 + T_2 + \cdots , 
$$
as was done for earlier for $T$ alone.  Then,
$$
\bigl[ 1 + 2\zeta_1 + 2\zeta_2 + 2\zeta_1^2 \bigr] \, \delta_{ij} 
+ \partial_i\partial_j\xi 
= \delta_{ij} \bigl[ 1 + 2\dot\rho T_1 + 2\dot\rho T_2 + \ddot\rho T_1^2 + 2\dot\rho^2 T_1^2 \bigr] 
+ \partial_i\epsilon_j + \partial_j\epsilon_i
+ \pi_{ij} . 
$$
The zeroth-order terms plainly match, and there is only one first-order term on either side---remember that $\epsilon^i$ and $\pi_{ij}$ are both second order---
$$
\zeta_1 = \dot\rho T_1 = - {\dot\rho\over\dot\phi} \psi .
$$
This result, turned around,
$$
\psi = - {\dot\phi\over\dot\rho} \zeta + \cdots , 
$$
is already sufficient to prove the earlier claim that $\psi\propto \sqrt{\epsilon}\zeta$, which was used to establish that the cubic action in $\psi$ was of order $\epsilon^2$.

Substituting $\zeta_1=\dot\rho T_1$ into this equation leaves just,
$$
2\zeta_2\, \delta_{ij} + \partial_i\partial_j\xi 
= \delta_{ij} \bigl[ 2\dot\rho T_2 + \ddot\rho T_1^2 \bigr] 
+ \partial_i\epsilon_j + \partial_j\epsilon_i
+ \pi_{ij} . 
$$
The $\partial_i\partial_j\xi$ part on the left is supposed to be traceless, $\partial_k\partial^k\xi=0$.  There are two independent ways to use this fact to remove the $\xi$ term from the left side.  Obviously, taking the trace of the equation is one way to do so,
$$
6\zeta_2 + \partial_k\partial^k\xi = 6\zeta_2
= 3 \bigl[ 2\dot\rho T_2 + \ddot\rho T_1^2 \bigr] 
+ 2\partial_k\epsilon^k 
+ \pi_k^k , 
$$
while another is to differentiate the equation twice using the operator $\partial^i\partial^j$, 
$$
2\partial_k\partial^k\zeta_2 + \partial_i\partial_j\partial^i\partial^j\xi 
= 2\partial_k\partial^k\zeta_2 
= \partial_k\partial^k \bigl[ 2\dot\rho T_2 + \ddot\rho T_1^2 \bigr] 
+ 2\partial_k\partial^k\partial_l\epsilon_l 
+ \partial^i\partial^j\pi_{ij} . 
$$
Applying the inverse Laplacian, $\partial^{-2}$, to both sides results in 
$$
2\zeta_2 
= 2\dot\rho T_2 + \ddot\rho T_1^2 + 2\partial_k\epsilon^k 
+ \partial^{-2}[\partial^i\partial^j\pi_{ij}] . 
$$
Both equations have the same appearance of $\partial_k\epsilon^k$, 
\begin{eqnarray}
6\zeta_2
&\!\!\!\!\!\!=\!\!\!\!\!\!&
3 \bigl[ 2\dot\rho T_2 + \ddot\rho T_1^2 \bigr] 
+ 2\partial_k\epsilon^k + \pi_k^k 
\nonumber \\
2\zeta_2
&\!\!\!\!\!\!=\!\!\!\!\!\!&
2\dot\rho T_2 + \ddot\rho T_1^2 + 2\partial_k\epsilon^k 
+ \partial^{-2}[\partial^i\partial^j\pi_{ij}] , 
\nonumber 
\end{eqnarray}
so by taking their difference, the second-order part of $\zeta$ has been determined,
$$\textstyle 
\zeta_2 = \dot\rho T_2 + {1\over 2} \ddot\rho T_1^2 + {1\over 4} \pi_k^k 
- {1\over 4}\partial^{-2}[\partial^i\partial^j\pi_{ij}] . 
$$
Using the definition of $\pi_{ij}$, the quantities appearing in this expression are
\begin{eqnarray}
\pi_k^k &\!\!\!\!\!\!=\!\!\!\!\!\!& 
2\partial_k\chi\partial^kT - e^{-2\rho(t)}\partial_kT\, \partial^kT 
\nonumber \\
\partial^i\partial^j\pi_{ij}
&\!\!\!\!\!\!=\!\!\!\!\!\!&
2 \partial^i\partial^j [\partial_i\chi\partial_jT] 
- e^{-2\rho(t)}\partial^i\partial^j [\partial_iT\, \partial_jT] . 
\nonumber 
\end{eqnarray}

Thus, to second order in the fluctuations, the relation of the $\zeta(t,\vec x)$ fluctuation in the coordinates used for the calculation of the three-point function and the fluctuation $\psi(t,\vec x)$ in the scalar field is given by 
\begin{eqnarray}
\zeta &\!\!\!\!\!\!=\!\!\!\!\!\!&\textstyle 
\dot\rho T + {1\over 2} \ddot\rho T^2 
+ {1\over 2}\partial_k\chi\partial^kT 
- {1\over 4} e^{-2\rho(t)}\partial_kT\, \partial^kT 
\nonumber \\
&&\textstyle
- {1\over 2}\partial^{-2} \partial^i\partial^j [\partial_i\chi\partial_jT] 
+ {1\over 4}e^{-2\rho(t)} \partial^{-2} \partial^i\partial^j [\partial_iT\, \partial_jT]
\nonumber 
\end{eqnarray}
where 
\begin{eqnarray}
T &\!\!\!\!\!\!=\!\!\!\!\!\!& 
- {1\over\dot\phi} \psi - {1\over 2} {\ddot\phi\over\dot\phi^3} \psi^2 
+ {1\over\dot\phi^2} \psi \!\!\psidot + \cdots 
\nonumber \\
\partial_k\partial^k\chi 
&\!\!\!\!\!\!=\!\!\!\!\!\!& 
- {1\over 2} {\dot\phi^2\over\dot\rho^2} 
{d\over dt} \biggl\{ {\dot\rho\over\dot\phi}\psi \biggr\} .
\nonumber 
\end{eqnarray}
Upon substituting the expression for $T(t,\vec x)$ into the relation for $\zeta(t,\vec x)$ directly, it becomes 
\begin{eqnarray}
\zeta &\!\!\!\!\!\!=\!\!\!\!\!\!& 
- {\dot\rho\over\dot\phi} \psi 
- {1\over 2} {\dot\rho\ddot\phi\over\dot\phi^3} \psi^2 
+ {\dot\rho\over\dot\phi^2} \psi \!\!\psidot 
+ {1\over 2} {\ddot\rho\over\dot\phi^2} \psi^2
- {1\over 2} {1\over\dot\phi} \partial_k\chi\partial^k\psi 
- {1\over 4} {e^{-2\rho(t)}\over\dot\phi^2} \partial_k\psi \partial^k\psi 
\nonumber \\
&&
+ {1\over 4} {e^{-2\rho(t)}\over\dot\phi^2} \partial^{-2} \partial_k\partial_l (\partial^k\psi\partial^l\psi)  
+ {1\over 2}{1\over\dot\phi} \partial^{-2} \partial_k\partial_l (\partial^k\psi\partial^l\chi) . 
\nonumber 
\end{eqnarray}

Finally, notice what happens when the first-order relation for $\psi = - {\dot\phi\over\dot\rho}\zeta$ is substituted into the expression for $\chi$,
$$
\partial_k\partial^k\chi 
= - {1\over 2} {\dot\phi^2\over\dot\rho^2} 
{d\over dt} \biggl\{ {\dot\rho\over\dot\phi}\psi \biggr\} 
= {1\over 2} {\dot\phi^2\over\dot\rho^2} 
{d\over dt} \biggl\{ {\dot\rho\over\dot\phi}{\dot\phi\over\dot\rho}\zeta \biggr\} 
= {1\over 2} {\dot\phi^2\over\dot\rho^2} \dot\zeta . 
$$
This is exactly the expression for $\chi(t,\vec x)$ as it was originally defined when solving the constraint equations in terms of $\zeta$.
\vskip36truept

\centerline{\bf\large AFTERWORD}
\vskip9truept

These notes are not meant to provide a complete introduction to the theory of inflation nor are they intended to treat all of the ways for generating non-Gaussianities through inflation.  For the former purpose there are many textbooks and reviews already available [5--7], while the latter subject is discussed elsewhere [8].  Rather, these notes are meant to explain the derivation of an important prediction of the simplest inflationary picture in complete detail.  Nonetheless, for students and researchers who are already familiar with the basic idea of inflation, the material presented here might serve as a useful pedagogical introduction to how inflation generates the primordial fluctuations in the space-time metric, and how to derive the two-point and three-point functions for these fluctuations in the simplest class of inflationary models.

The first parts of these notes were completed while at the Discovery Center at the Niels Bohr Institute; these earlier parts were supported by the Niels Bohr International Academy and the Discovery Center at the Niels Bohr Institute.  In the course of moving from one institution to another, further progress on these notes was interrupted and for a long while they were left in an unfinished form.  Work on them was only taken up again much later at Carnegie Mellon University in the course of delivering a long series of informal lectures on Maldacena's calculation of the three-point function.

I am grateful to the Physics Department of Carnegie Mellon University for supporting the completion of these notes.  But I should especially like to thank Nishant Agarwal and Tereza Vardanyan for their curiosity and eagerness, and for their ability to endure these lectures to the very end.
\vfill\eject

\centerline{\bf\large REFERENCES}
\vskip12truept

\renewcommand{\theenumi}{[\arabic{enumi}]}
\renewcommand{\labelenumi}{\theenumi}

\begin{enumerate}
\item J.~M.~Maldacena, ``Non-Gaussian features of primordial fluctuations in single field inflationary models,'' JHEP {\bf 0305}, 013 (2003).
\item R.~L.~Arnowitt, S.~Deser and C.~W.~Misner, ``The dynamics of general relativity,'' gr-qc/0405109.
\item E.~Komatsu {\it et al.}, ``Seven-Year Wilkinson Microwave Anisotropy Probe Observations:  Cosmological Interpretation,'' astro-ph.CO/1001.4538.
\item J.~S.~Schwinger, ``Brownian motion of a quantum oscillator,'' J.\ Math.\ Phys.\  {\bf 2}, 407 (1961); 
K.~T.~Mahanthappa, ``Multiple production of photons in quantum electrodynamics,'' Phys.\ Rev.\  {\bf 126}, 329 (1962); 
P.~M.~Bakshi and K.~T.~Mahanthappa, ``Expectation value formalism in quantum field theory. 1.,'' J.\ Math.\ Phys.\  {\bf 4}, 1 (1963); 
P.~M.~Bakshi and K.~T.~Mahanthappa, ``Expectation value formalism in quantum field theory. 2.,'' J.\ Math.\ Phys.\  {\bf 4}, 12 (1963).
L.~V.~Keldysh, ``Diagram technique for nonequilibrium processes,'' Zh.\ Eksp.\ Teor.\ Fiz.\  {\bf 47}, 1515 (1964) [Sov.\ Phys.\ JETP {\bf 20}, 1018 (1965)].
\item A.~D.~Linde, {\it Particle Physics and Inflationary Cosmology\/}, Contemporary concepts in physics, 5 (Harwood, Chur, Switzerland) (1990), 
hep-th/0503203.
\item V.~F.~Mukhanov, H.~A.~Feldman and R.~H.~Brandenberger, ``Theory of cosmological perturbations,'' Phys.\ Rept.\  {\bf 215}, 203 (1992).
\item S.~Weinberg, ``Cosmology,'' Oxford University Press, Oxford (2008).
\item N.~Bartolo, E.~Komatsu, S.~Matarrese and A.~Riotto, ``Non-Gaussianity from inflation: Theory and observations,'' Phys.\ Rept.\  {\bf 402}, 103 (2004) astro-ph/0406398; 
N.~Bartolo, S.~Matarrese and A.~Riotto, ``Non-Gaussianity and the Cosmic Microwave Background Anisotropies,''
astro-ph.CO/1001.3957.

\end{enumerate}

\end{document}